\documentclass{aastex631}

\usepackage{graphicx}
\let\splitbox\relax

\usepackage{array}
\usepackage{multirow}

\usepackage{adjustbox}

\usepackage{hyperref}
\usepackage{nameref}
\usepackage[percent]{overpic}

\begin{document}

\title{Investigating FRB~20240114A with FAST: Morphological Classification and Drifting Rate Measurements in a Burst-Cluster Framework}

\author[0009-0002-3020-9123]{Long-Xuan Zhang}
\altaffiliation{These authors contributed equally to this work.}
\affiliation{School of Physics, Huazhong University of Science and Technology, Wuhan, 430074. China}

\author{Shiyan Tian}
\altaffiliation{These authors contributed equally to this work.}
\affiliation{School of Physics, Huazhong University of Science and Technology, Wuhan, 430074. China}

\author[0000-0002-7949-3906]{Junyi Shen}
\affiliation{School of Physics, Huazhong University of Science and Technology, Wuhan, 430074. China}

\author[0009-0005-8586-3001]{Jun-Shuo Zhang}
\affiliation{National Astronomical Observatories, Chinese Academy of Sciences, Beijing 100101, China}
\affiliation{University of Chinese Academy of Sciences, Beijing 100049, China}

\author[0000-0002-6423-6106]{Dejiang Zhou}
\affiliation{National Astronomical Observatories, Chinese Academy of Sciences, Beijing 100101, China}

\author[0009-0004-8180-3055]{Lin Zhou}
\affiliation{School of Physics, Huazhong University of Science and Technology, Wuhan, 430074. China}

\author[0009-0000-0635-5679]{Po Ma}
\affiliation{School of Physics, Huazhong University of Science and Technology, Wuhan, 430074. China}

\author[0000-0002-9332-5562]{Tian-Cong Wang}
\affiliation{School of Physics and Astronomy, Beijing Normal University, Beijing 100875, China}
\affiliation{Institute for Frontiers in Astronomy and Astrophysics, Beijing Normal University, Beijing 102206, China}

\author[0000-0002-7420-9988]{Dengke Zhou}
\affiliation{Research Center for Astronomical Computing, Zhejiang Laboratory, Hangzhou 311121, China}

\author{Jinlin Han}
\affiliation{National Astronomical Observatories, Chinese Academy of Sciences, Beijing 100101, China}
\affiliation{University of Chinese Academy of Sciences, Beijing 100049, China}
\affiliation{State Key Laboratory of Radio Astronomy and Technology, Beijing 100101, China}

\author[0000-0003-4137-4247]{Yunpeng Men}
\affiliation{Max-Planck-Institut für Radioastronomie, Auf dem Hügel 69, Bonn, D-53121, Germany}

\author{Fayin Wang}
\affiliation{School of Astronomy and Space Science, Nanjing University, Nanjing 210023, China}
\affiliation{Key Laboratory of Modern Astronomy and Astrophysics (Nanjing University), Ministry of Education, Nanjing 210093, China}

\author{Jiarui Niu}
\affiliation{National Astronomical Observatories, Chinese Academy of Sciences, Beijing 100101, China}

\author[0000-0002-3386-7159]{Pei Wang*}
\email{wangpei@nao.cas.cn}
\affiliation{National Astronomical Observatories, Chinese Academy of Sciences, Beijing 100101, China}
\affiliation{Institute for Frontiers in Astronomy and Astrophysics, Beijing Normal University, Beijing 102206, China}
\affiliation{State Key Laboratory of Radio Astronomy and Technology, Beijing 100101, China}

\author[0000-0001-5105-4058]{Weiwei Zhu*}
\email{zhuww@nao.cas.cn}
\affiliation{National Astronomical Observatories, Chinese Academy of Sciences, Beijing 100101, China}
\affiliation{Institute for Frontiers in Astronomy and Astrophysics, Beijing Normal University, Beijing 102206, China}
\affiliation{State Key Laboratory of Radio Astronomy and Technology, Beijing 100101, China}

\author[0000-0002-9725-2524]{Bing Zhang*}
\email{bzhang1@hku.hk}
\affiliation{The Hong Kong Institute for Astronomy and Astrophysics, the University of Hong Kong, Pokfulam, Hong Kong, P. R. China}
\affiliation{Department of Physics, the University of Hong Kong, Pokfulam, Hong Kong, P. R. China}
\affiliation{The Nevada Center for Astrophysics, University of Nevada, Las Vegas, Las Vegas, NV 89154, USA}
\affiliation{Department of Physics and Astronomy, University of Nevada, Las Vegas, Las Vegas, NV 89154, USA}

\author[0000-0003-3010-7661]{Di Li*}
\email{dili@tsinghua.edu.cn}
\affiliation{New Cornerstone Science Laboratory, Department of Astronomy, Tsinghua University, Beijing 100084, China}
\affiliation{National Astronomical Observatories, Chinese Academy of Sciences, Beijing 100101, China}
\affiliation{State Key Laboratory of Radio Astronomy and Technology, Beijing 100101, China}

\author[0000-0002-5400-3261]{Yuan-Chuan Zou*}
\email{zouyc@hust.edu.cn}
\affiliation{School of Physics, Huazhong University of Science and Technology, Wuhan, 430074. China}
\affiliation{Purple Mountain Observatory, Chinese Academy of Sciences, Nanjing 210023, China}

\author[0000-0001-9036-8543]{Wei-Yang Wang}
\affiliation{University of Chinese Academy of Sciences, Beijing 100049, China}

\author[0000-0001-6374-8313]{Yuan-Pei Yang}
\affiliation{South-Western Institute for Astronomy Research, Key Laboratory of Survey Science of Yunnan Province, Yunnan University, Kunming 650500, China}

\author[0000-0001-6021-5933]{Qin Wu}
\affiliation{School of Astronomy and Space Science, Nanjing University, Nanjing 210023, China}

\author{He Gao}
\affiliation{School of Physics and Astronomy, Beijing Normal University, Beijing 100875, China}
\affiliation{Institute for Frontiers in Astronomy and Astrophysics, Beijing Normal University, Beijing 102206, China}

\author{Ke-Jia Lee}
\affiliation{Department of Astronomy, School of physics, Peking University, Beijing 100871, China}
\affiliation{National Astronomical Observatories, Chinese Academy of Sciences, Beijing 100101, China}
\affiliation{Yunnan Astronomical Observatories, Chinese Academy of Sciences, Kunming 650216, China}
\affiliation{Beijing Laser Acceleration Innovation Center, Huairou, Beijing 101400, China}

\author[0000-0002-9642-9682]{Jia-Wei Luo}
\affiliation{College of Physics and Hebei Key Laboratory of Photophysics Research and Application, Hebei Normal University, Shijiazhuang 050024, China}
\affiliation{Shijiazhuang Key Laboratory of Astronomy and Space Science, Hebei Normal University, Shijiazhuang 050024, China}

\author[0000-0002-4300-121X]{Rui Luo}
\affiliation{Department of Astronomy, School of Physics and Materials Science, Guangzhou University, Guangzhou 510006, China}

\author[0000-0002-9390-9672]{Chao-Wei Tsai}
\affiliation{National Astronomical Observatories, Chinese Academy of Sciences, Beijing 100101, China}
\affiliation{Institute for Frontiers in Astronomy and Astrophysics, Beijing Normal University,  Beijing 102206, China}
\affiliation{University of Chinese Academy of Sciences, Beijing 100049, China}
\affiliation{State Key Laboratory of Radio Astronomy and Technology, Beijing 100101, China}

\author{Lin Lin}
\affiliation{School of Physics and Astronomy, Beijing Normal University, Beijing 100875, China}

\author{Wanjin Lu}
\affiliation{National Astronomical Observatories, Chinese Academy of Sciences, Beijing 100101, China}
\affiliation{University of Chinese Academy of Sciences, Beijing 100049, China}

\author[0000-0001-5649-2591]{Jintao Xie}
\affiliation{School of Computer Science and Engineering, Sichuan University of Science and Engineering, Yibin 644000, China}

\author[0000-0001-9956-6298]{Jianhua Fang}
\affiliation{Research Center for Astronomical Computing, Zhejiang Laboratory, Hangzhou 311121, China}

\author{Jinhuang Cao}
\affiliation{National Astronomical Observatories, Chinese Academy of Sciences, Beijing 100101, China}
\affiliation{University of Chinese Academy of Sciences, Beijing 100049, China}

\author{Chen-Chen Miao}
\affiliation{College of Physics and Electronic Engineering, Qilu Normal University, Jinan 250200, China}

\author{Yuhao Zhu}
\affiliation{National Astronomical Observatories, Chinese Academy of Sciences, Beijing 100101, China}
\affiliation{University of Chinese Academy of Sciences, Beijing 100049, China}

\author[0009-0000-4795-8767]{Yunchuan Chen}
\affiliation{Research Center for Astronomical Computing, Zhejiang Laboratory, Hangzhou 311121, China}

\author{Yong-Kun Zhang}
\affiliation{National Astronomical Observatories, Chinese Academy of Sciences, Beijing 100101, China}

\author{Shuo Cao}
\affiliation{Yunnan Astronomical Observatories, Chinese Academy of Sciences, Kunming 650216, China}
\affiliation{University of Chinese Academy of Sciences, Beijing 100049, China}

\author{Zi-Wei Wu}
\affiliation{National Astronomical Observatories, Chinese Academy of Sciences, Beijing 100101, China}

\author[0000-0002-4327-711X]{Chunfeng Zhang}
\affiliation{National Astronomical Observatories, Chinese Academy of Sciences, Beijing 100101, China}

\author{Silu Xu}
\affiliation{National Astronomical Observatories, Chinese Academy of Sciences, Beijing 100101, China}
\affiliation{University of Chinese Academy of Sciences, Beijing 100049, China}

\author[0009-0000-6108-2730]{Huaxi Chen}
\affiliation{Research Center for Astronomical Computing, Zhejiang Laboratory, Hangzhou 311121, China}

\author{Xiang-Lei Chen}
\affiliation{National Astronomical Observatories, Chinese Academy of Sciences, Beijing 100101, China}

\author[0000-0002-6165-0977]{Xianghan Cui}
\affiliation{National Astronomical Observatories, Chinese Academy of Sciences, Beijing 100101, China}

\author{Yi Feng}
\affiliation{Research Center for Astronomical Computing, Zhejiang Laboratory, Hangzhou 311121, China}
\affiliation{Institute for Astronomy, School of Physics, Zhejiang University, Hangzhou 310027, China}

\author{Yu-Xiang Huang}
\affiliation{Yunnan Astronomical Observatories, Chinese Academy of Sciences, Kunming 650216, China}

\author[0000-0002-1056-5895]{Weicong Jing}
\affiliation{National Astronomical Observatories, Chinese Academy of Sciences, Beijing 100101, China}

\author{Dong-Zi Li}
\affiliation{Department of Astrophysical Sciences, Princeton University, Princeton, NJ 89154, USA}

\author{Jian Li}
\affiliation{Department of Astronomy, University of Science and Technology of China, Hefei 230026, China}
\affiliation{School of Astronomy and Space Science, University of Science and Technology of China, Hefei 230026, China}

\author{Ye Li}
\affiliation{Purple Mountain Observatory, Chinese Academy of Sciences, Nanjing 210023, China}

\author[0000-0001-6651-7799]{Chen-Hui Niu}
\affiliation{Institute of Astrophysics, Central China Normal University, Wuhan 430079, China}

\author[0000-0001-7199-2906]{Yong-Feng Huang}
\affiliation{School of Astronomy and Space Science, Nanjing University, Nanjing 210023, China}
\affiliation{Key Laboratory of Modern Astronomy and Astrophysics (Nanjing University), Ministry of Education, Nanjing 210093, China}

\author{Qingyue Qu}
\affiliation{National Astronomical Observatories, Chinese Academy of Sciences, Beijing 100101, China}
\affiliation{University of Chinese Academy of Sciences, Beijing 100049, China}

\author[0000-0003-4721-4869]{Yuanhong Qu}
\affiliation{The Nevada Center for Astrophysics, University of Nevada, Las Vegas, Las Vegas, NV 89154, USA}
\affiliation{Department of Physics and Astronomy, University of Nevada, Las Vegas, Las Vegas, NV 89154, USA}

\author[0000-0002-9434-4773]{Bojun Wang}
\affiliation{National Astronomical Observatories, Chinese Academy of Sciences, Beijing 100101, China}

\author{Yi-Dan Wang}
\affiliation{National Astronomical Observatories, Chinese Academy of Sciences, Beijing 100101, China}
\affiliation{University of Chinese Academy of Sciences, Beijing 100049, China}

\author[0000-0001-7746-9462]{Suming Weng}
\affiliation{National Key Laboratory of Dark Matter Physics, School of Physics and Astronomy, Shanghai Jiao Tong University, Shanghai 200240, China}
\affiliation{Laboratory for Laser Plasmas and Collaborative Innovation Centre of IFSA, Shanghai Jiao Tong University, Shanghai 200240, China} 

\author[0000-0002-6299-1263]{Xuefeng Wu}
\affiliation{Purple Mountain Observatory, Chinese Academy of Sciences, Nanjing 210023, China}

\author{Heng Xu}
\affiliation{National Astronomical Observatories, Chinese Academy of Sciences, Beijing 100101, China}

\author[0000-0002-5799-9869]{Shihan Yew}
\affiliation{National Key Laboratory of Dark Matter Physics, School of Physics and Astronomy, Shanghai Jiao Tong University, Shanghai 200240, China}
\affiliation{Laboratory for Laser Plasmas and Collaborative Innovation Centre of IFSA, Shanghai Jiao Tong University, Shanghai 200240, China}

\author{Aiyuan Yang}
\affiliation{National Astronomical Observatories, Chinese Academy of Sciences, Beijing 100101, China}
\affiliation{State Key Laboratory of Radio Astronomy and Technology, Beijing 100101, China}

\author{Wenfei Yu}
\affiliation{Shanghai Astronomical Observatory, Chinese Academy of Sciences, Shanghai 200030, China}

\author{Lei Zhang}
\affiliation{National Astronomical Observatories, Chinese Academy of Sciences, Beijing 100101, China}
\affiliation{Centre for Astrophysics and Supercomputing, Swinburne University of Technology, Hawthorn, VIC 3122, Australia}

\author{Rushuang Zhao}
\affiliation{Guizhou Provincial Key Laboratory of Radio Astronomy and Data Processing, Guizhou Normal University, Guiyang 550001, China}




\begin{abstract}
This study investigates the morphological classification and drifting rate measurement of the repeating fast radio burst (FRB) source FRB~20240114A using the Five-hundred-meter Aperture Spherical Telescope (FAST). Detected on January 14, 2024, FRB~20240114A exhibited an exceptionally high burst rate, revealing unique properties. Through observational campaigns over several months, we selected a dataset comprising 3,203 bursts (2,109 burst-clusters) during a continuous monitoring session (15,780 seconds) on March~12, 2024. Improving upon previous work, we clarify the definitions of sub-bursts, bursts and burst-clusters. Using an average dispersion measures (DM) of 529.2 pc cm$^{-3}$, we classified the burst-clusters into Downward Drifting, Upward Drifting, No Drifting, No Evidence for Drifting, Not-Clear, and Complex burst-clusters. Among the 978 burst-clusters that exhibit drifting behavior, 233 (23.82\%) show upward drifting. Additionally, if 142 upward drifting single-component burst-clusters are excluded, upward drifting double- and multi-component burst-clusters still account for 10.89\% of the 836 burst-clusters exhibiting drifting behavior, equating to 91 burst-clusters. Furthermore, if only upward drifting burst-clusters with consecutive time intervals (or upward drifting bursts) are considered, only 9 bursts remain. Drifting rate comparisons with other physical quantities reveal that the drifting rate increases with peak frequency for single-component burst-clusters with drifting behavior. Moreover, in single-component burst-clusters, those with upward drifting exhibit smaller effective widths, bandwidths, and fluxes than their downward drifting counterparts. A Kolmogorov-Smirnov test further indicates that upward drifting burst-clusters possess longer consecutive time intervals than downward drifting ones, suggesting distinct underlying physical mechanisms.
\end{abstract}


\keywords{
    \href{https://astrothesaurus.org/uat/2008}{Radio transient sources (2008)},
    \href{https://astrothesaurus.org/uat/1339}{Radio bursts (1339)},
    \href{https://astrothesaurus.org/uat/1338}{Radio astronomy (1338)},
    \href{https://astrothesaurus.org/uat/740}{High time resolution astrophysics (740)},
    \href{https://astrothesaurus.org/uat/739}{High energy astrophysics (739)}
}


\section{Introduction}

\label{sec:intro}

Fast radio bursts (FRBs) are brief radio flashes, typically lasting milliseconds \citep{2007Sci...318..777L}. They exhibit high dispersion measures (DMs), exceeding the maximum estimated from the electron column density model \citep{cordes2003ne2001inewmodelgalactic, 2017ApJ...835...29Y} for the Milky Way, making almost all of them considered extragalactic in origin \citep{doi:10.1126/science.1236789, annurev-astro-091918-104501, 2019A&ARv..27....4P, 2020Natur.587...45Z, 2023RvMP...95c5005Z}. Since the discovery of the first event, FRB~20010724 \citep{2007Sci...318..777L}, hundreds of FRB sources have been discovered, with a small number of sources observed emitting repeated bursts. After the discovery of the first repeater, FRB~20121102A \citep{2014ApJ...790..101S}, a large number of repeated bursts have been detected from 72 repeating FRB sources\footnote{\url{https://blinkverse.zero2x.org/}}, up to September 2025 \citep{2019ApJ...885L..24C, 2020ApJ...891L...6F, 2021ApJS..257...59C, 2023ApJ...947...83C}. These FRBs exhibit high brightness temperatures ($T_b > 10^{35}$ K), indicating that their emission mechanisms must be coherent \citep{2023RvMP...95c5005Z}. With the advent of new-generation radio telescopes, such as the Canadian Hydrogen Intensity Mapping Experiment (CHIME; \citeauthor{2018ApJ...863...48C} \citeyear{2018ApJ...863...48C}), the Green Bank Telescope (GBT; \citeauthor{2020AAS...23624305W} \citeyear{2020AAS...23624305W}), the Australian Square Kilometer Array Pathfinder (ASKAP; \citeauthor{2022arXiv220808245K} \citeyear{2022arXiv220808245K}), the Karoo Array Telescope (MeerKAT; \citeauthor{2016mks..confE...1J} \citeyear{2016mks..confE...1J}), the Very Large Array (VLA; \citeauthor{2011ApJ...739L...1P} \citeyear{2011ApJ...739L...1P}), the Low-Frequency Array (LOFAR; \citeauthor{2013A&A...556A...2V} \citeyear{2013A&A...556A...2V}), the Westerbork Synthesis Radio Telescope (WSRT; \citeauthor{2022A&A...658A.146V} \citeyear{2022A&A...658A.146V}), the Kunming 40-Meter Radio Telescope (KM40M; \citeauthor{2010RAA....10..805H} \citeyear{2010RAA....10..805H}) and the Shanghai Tianma Radio Telescope (TMRT; \citeauthor{2024AstTI...1..239L} \citeyear{2024AstTI...1..239L}) which offer higher time resolution and broader fields of view, the detection of FRBs has become more frequent and precise, further accelerating the development of this field. In particular, China's Five-Hundred-Meter Aperture Spherical Radio Telescope (FAST; \citet{2011IJMPD..20..989N, 2019SCPMA..6259502J, 2020Innov...100053Q}), with its extremely high sensitivity, has become an important tool for FRB observations and has provided a wealth of valuable data.

Although most FRBs still appear as one-off burst events, the study of repeating FRBs provides in-depth insight into repeated burst-clusters and drives the exploration of their potential origins (magnetars, black holes, or other extreme astrophysical environments). For further discussion on the historical development and physical mechanisms of FRBs, refer to relevant reviews \citep{2020Natur.587...45Z,2023RvMP...95c5005Z, 2019ARA&A..57..417C, 2019A&ARv..27....4P, 2019PhR...821....1P}. In particular, observations of repeating FRBs have consistently reported a phenomenon often called the ``sad trombone'' effect. This pattern involves sub-bursts that appear at progressively lower frequencies over time. It was initially highlighted in FRB~20121102A \citep{Hessels_2019} and has since been documented in additional sources \citep{2019ApJ...885L..24C,2020Natur.582..351C,2020ApJ...891L...6F}. Later studies confirmed that this downward drifting is widespread and can be detected at various observing frequencies \citep{2021MNRAS.508.5354H,2023MNRAS.519..666J, 2022RAA....22l4001Z, 2022RAA....22l4002Z}. Detailed analyses have shown that multiple sub-bursts can each shift to lower frequencies as they evolve \citep{2021MNRAS.505.3041P,2021ApJ...923....1P}, and recent work demonstrated that this effect extends to low radio bands of around a few hundred megahertz \citep{2020ApJ...891L...6F, 2021Natur.596..505P}. These phenomena have motivated diverse theoretical explanations, which are elaborated in the following discussion.

The apparent downward drifting feature can be readily interpreted within the framework of magnetospheric models invoking the so-called radius-to-frequency mapping \citep{2019ApJ...876L..15W,osti_1802960,2022ApJ...925...53Z}. In particular, if FRBs are produced by bunches via curvature radiation or inverse Compton scattering, the line of sight would usually sweep the emitters at lower altitudes earlier and those at higher altitudes later \citep{2019ApJ...876L..15W,2022ApJ...925...53Z,2024ApJ...972..124Q} so that lower-frequency emission arrives later. In order to interpret upward drifting features, one needs to invoke different injection times of bunches from different field lines \citep{2020ApJ...899..109W}. This is a less natural assumption, and therefore, such events are expected to be less common than downward drifting. Within far-away models that invoke the synchrotron maser in relativistic shocks, the downward drifting feature can be accounted for if the shock undergoes deceleration so that there is continuous decay of the spectral break \citep{10.1093/mnras/stz700,2022ApJ...925..135M}. 
Independently, studies incorporating relativistic bulk motion and Dicke's superradiance have demonstrated that such kinematic effects can predict inverse correlations between drifting rates and sub-burst durations \citep{2020MNRAS.498.4936R}.


Although these emission mechanisms address the intrinsic properties of FRB, propagation effects through the interstellar and circum-burst media introduce additional complexities: scintillation in turbulent media modulates burst morphology \citep{1990ARA&A..28..561R}; temporal variations in the DM of a repeating FRB could potentially indicate changes in the electron density along the propagation path or in the immediate vicinity of the source \citep{Zhao:2021vns}; while plasma lensing from electron density gradients \citep{2017ApJ...842...35C} can mimic downward drifting via chromatic deflection. \citet{2021arXiv210713549T} proposed that a post-dispersion modulating screen imprints temporally sharp broadband structures, with subsequent dispersion in the circum-burst plasma generating the observed drift. Their model further posits that emission is beamed away from Earth, requiring chromatic deflection by plasma prisms ($\alpha \propto \nu^{-2}$) to redirect radiation into our line of sight, a mechanism that simultaneously explains the narrow observed bandwidths despite the broadband intrinsic emission. However, the overwhelming prevalence of downward drifting events, combined with rare upward drifting bursts \citep{2021ApJ...923....1P}, suggests that a single physical framework may be insufficient to explain the full diversity of the observed drifting phenomena.

Temporal analyses across the FRB population reveal two key patterns: (1) bimodal waiting-time distributions in repeaters, suggesting distinct triggering regimes \citep{2021Natur.598..267L} or potentially linked to starquake-aftershock sequences \citep{2025arXiv250216626L,2025arXiv250812567Q}; (2) periodic activity windows, such as FRB~20180916B's 16.35-day cycle, indicative of orbital/precessional modulation \citep{2020Natur.582..351C}. Recently, a potential second-scale periodicity has been reported in the repeating FRB~20201124A, with periods on the order of 1.7 seconds detected on two separate observation days. This may suggest a possible intrinsic periodicity associated with the rotation of the central engine, in contrast to longer-term orbital or precessional cycles; although no periodic signal was found in other datasets \citep{du2025secondscaleperiodicityactiverepeating}. Notably absent are sub-second periodicities matching magnetar spin periods, with upper limits ($P < 10$ ms) ruling out canonical pulsar-like beam sweeping \citep{2022RAA....22l4004N,2024ApJ...977..129D}. For FRB~20240114A, 
comprehensive timing analysis based on FAST data is underway to search for potential periodicities \citep{2025arXiv250714708Z}, a study motivated by its recent detection by CHIME, which revealed an exceptional burst rate (\(>100\,\mathrm{hr}^{-1}\)) and a high dispersion measure (DM = 527.7 pc cm\(^{-3}\)) \citep{2023ApJ...947...83C,2024ATel16420....1S,2025arXiv250513297S}, as well as the observed near-complete linear polarization (\(>95\%\)) \citep{2024arXiv241010172X}, a signature that is consistent with coherent curvature radiation in ordered magnetospheric fields \citep{2019ApJ...876L..15W}. If the downward drifting and upward drifting are partly caused by incorrect DMs, proper dedispersion correction would clarify these patterns and strengthen the two temporal pillars outlined above. \cite{2025arXiv250403569H} reported observations of the repeating FRB~20240114A using the KM40M and FAST, revealing a flattened high-energy tail in its burst energy distribution that aligns with the properties of non-repeating FRBs, supporting a potential common origin between repeating and one-off FRBs, while \cite{2025ApJS..278...49X} demonstrated using the GBT that FRB~20240114A exhibits the highest burst rate ever recorded with (sub)-100-meter radio telescopes (264\,hr$^{-1}$) and extreme linear polarization ($>90\%$ in 81\% of bursts). In addition, long-term simultaneous monitoring at 2.25/8.60 GHz with TMRT detected 155 bursts at 2.25 GHz but none at 8.60 GHz, placing strong constraints on the high-frequency activity of FRB~20240114A \citep{2025ApJ...992..185W}.

In this work, we focus solely on the data obtained from the FAST on March 12, 2024, which provided the longest-duration monitoring and the highest burst rate of FRB~20240114A observed to date. The observation lasted 15{,}780 seconds, during which a large number of bursts were detected. These abundant data allow us to investigate the properties of repeating bursts in considerable detail. This paper is organized as follows. Section~\ref{sec:preprocessing} introduces the data preprocessing procedure and the calculation of various physical quantities. Section~\ref{sec:morphology} focuses on the morphological classification, providing definitions for sub-burst, burst, and burst-cluster, along with the concepts of consecutive and intermittent time intervals. Section~\ref{sec:drift} presents drifting rate measurements, discusses their correlation with other physical quantities, and highlights the distinct recurrence patterns between upward and downward drifting events, suggesting that multi-mechanism models may be necessary to fully understand FRB mechanisms.

\section{Observations and burst-cluster Detection}

\label{sec:preprocessing}

Following its discovery by CHIME on January 14, 2024, FRB~20240114A was subsequently targeted for intensive follow-up observations with FAST. The observations were carried out using the center beam of the L-band 19-beam receiver \citep{2018IMMag..19..112L, 2020RAA....20...64J}, initially pointed to R.A.\,=\,21:27:39.888, decl.\,=\,+04:21:00.36, as reported by CHIME \citep{2024ATel16420....1S}. On 2024 February~15, the pointing was calibrated to the position R.A.\,=\,21:27:39.84, decl.\,=\,+04:19:46.3, localized by MeerKAT \citep{10.1093/mnras/stae2013}. After obtaining the new coordinates, the subsequent observations with FAST were updated accordingly. The monitoring campaign spanned from January 28, 2024, to August 30, 2024, comprising 59 separate observing sessions and generating a cumulative on-source integration time of 34.7 hours. Data acquisition was performed in pulsar searching mode, with four polarization channels (XX, YY, XY*, and X*Y) sampled at a time resolution of 49.152\,$\mu$s by the digital backend. A total of 4096 frequency channels, each with a bandwidth of about 0.122\,MHz, were used to fully cover the 1.0--1.5\,GHz range. All data were recorded in 8-bit PSRFITS format \citep{2004PASA...21..302H}, with each FITS file containing about 6.442\,s of data. The detection threshold $7\sigma$ of FAST was measured as $0.015~\mathrm{Jy~ms}$ for bursts with a width of $1~\mathrm{ms}$ \citep{2021Natur.598..267L}. This sensitivity represents at least a threefold improvement compared to previous Arecibo observations \citep{7911064}, enabling the detection of fainter bursts.

These high time-resolution observations provide a rich dataset for investigating the burst properties and temporal evolution of FRB~20240114A. At the beginning of each session, calibration signals consisting of a 10\,K periodic noise were injected for one minute before each session to calibrate the system temperature ($T_{\rm sys}$). The temporal variation in background was taken into account by considering the telescope gain dependence on the zenith angle and frequency \citep{2020RAA....20...64J}, enabling us to convert the  antenna temperature (K) to flux density (Jy). We then computed each burst isotropic-equivalent energy via
\begin{equation}
E = 10^{39} \,\mathrm{erg} \cdot \frac{4\pi}{1+z} 
    \cdot \left(\frac{D_L}{10^{28}\,\mathrm{cm}}\right)^2 
    \cdot \left(\frac{F_{\nu}}{\mathrm{Jy}\cdot\mathrm{ms}}\right) 
    \cdot \left(\frac{\Delta\nu}{\mathrm{GHz}}\right),
\end{equation}
where \(D_L \approx 630.72\)~Mpc and \(z = 0.13\) for FRB~20240114A \citep{2024ATel16613....1B}.

For additional details on the calibration procedures and a comprehensive overview of all 59 observing sessions, including the data collected during the March~12, 2024 session, we refer the reader to our first paper in this series \citep{2025arXiv250714707Z}. In the present work, we focus solely on the data collected during the March~12, 2024 session, with the burst timestamps obtained from the analysis pipeline described in that paper.

\subsection{Data Analysis and Burst Detection}

The observation on March~12, 2024, session yielded the highest burst rate, with 3203 bursts (corresponding to 2109 burst-clusters in total; for detailed definitions of burst-cluster, sub-burst, and burst, see Section~\ref{sec:morphology}) detected in 15,780~seconds (average $\sim730.72$~bursts~hr$^{-1}$). Using the burst timestamps from \cite{2025arXiv250714707Z}, referenced to 1.5 GHz, we grouped the bursts whose adjacent timestamps were separated by less than 400~ms into individual groups. We used the \texttt{replot\_fil} function from the \textsc{TransientX}\footnote{\url{https://github.com/ypmen/TransientX}} software \citep{2024A&A...683A.183M}, producing a single time-frequency waterfall plot per group (or per burst-cluster). During processing, we applied a time downsampling factor of 4 (resulting in a time resolution of $49.152\ \mu\text{s} \times 4 = 196.608\ \mu\text{s}$), while no frequency downsampling was performed. The plot range extended to multiples (20$\times$, 50$\times$, 100$\times$, etc.) of the burst effective width, with the specific multiple chosen based on the temporal span of the whole burst-cluster. The corresponding archival data for each burst-cluster were saved individually for further analysis. In this procedure, we adopted the average DM derived from all bursts detected on March~12,, as described in Section \ref{DM_Optimized_for_Structure}.

Furthermore, to mitigate RFI and obtain cleaner data for subsequent analysis, we enabled the \texttt{--zdot} and \texttt{--kadane} options during the \texttt{replot\_fil} process. The \texttt{--zdot} option was used to remove zero-DM RFI, typically caused by terrestrial radio sources and unrelated to genuine astrophysical signals. The \texttt{--kadane} option applied the Kadane algorithm, which uses a maximum subarray summation method to detect and remove high-power anomalous signals. This method efficiently identifies and removes RFI, minimizing the impact of spurious signals for subsequent analysis. Details about the algorithms can be found in \cite{2019MNRAS.488.3957M, 2023A&A...679A..20M}.

\subsection{DM Optimized for Structure}
\label{DM_Optimized_for_Structure}

The radio signal experiences dispersion as it propagates through the plasma. This dispersive delay is inversely proportional to the square of the radio frequency and directly proportional to the dispersion measure (DM), which quantifies the total plasma content along the line
of sight from the source to the observer:
\begin{equation}
    \mathrm{DM}=\int n {\rm d}l.
\end{equation}
\noindent where $n$ is the free electron density, and $dl$ is the differential path length.

The FRB search algorithm explores a range of DM values to compensate for the delay. According to \citet{Hessels_2019}, the DM of the FRB is determined based on either the maximum signal-to-noise ratio (S/N) or the most detailed temporal structure of the burst. As described by the propagation scenario of \citet{2021arXiv210713549T}, the intrinsic burst emission passes through an inhomogeneous modulation region near the FRB source, where short-duration intensity dropouts or narrow absorption features are imprinted simultaneously across the full observing band onto the already dispersed pulse. Assuming no significant source-inherent drifting features exist, the modulated temporal structure imprinted by the screen can be optimally resolved by adopting a structure-maximizing DM, whereas the S/N-maximizing DM corresponds to the value that yields the highest integrated burst significance, though this choice may smear out complex temporal substructures or multiple components. Conversely, when propagation effects are negligible, the de-dispersed dynamic spectrum morphology directly manifests the source's intrinsic radiation mechanism. The interpretation becomes more nuanced when both propagation-induced modulation and source-inherent drifting features coexist.

\begin{figure}[ht]
    \centering
    \begin{minipage}{0.45\textwidth}  
        \centering
        \includegraphics[width=\textwidth]{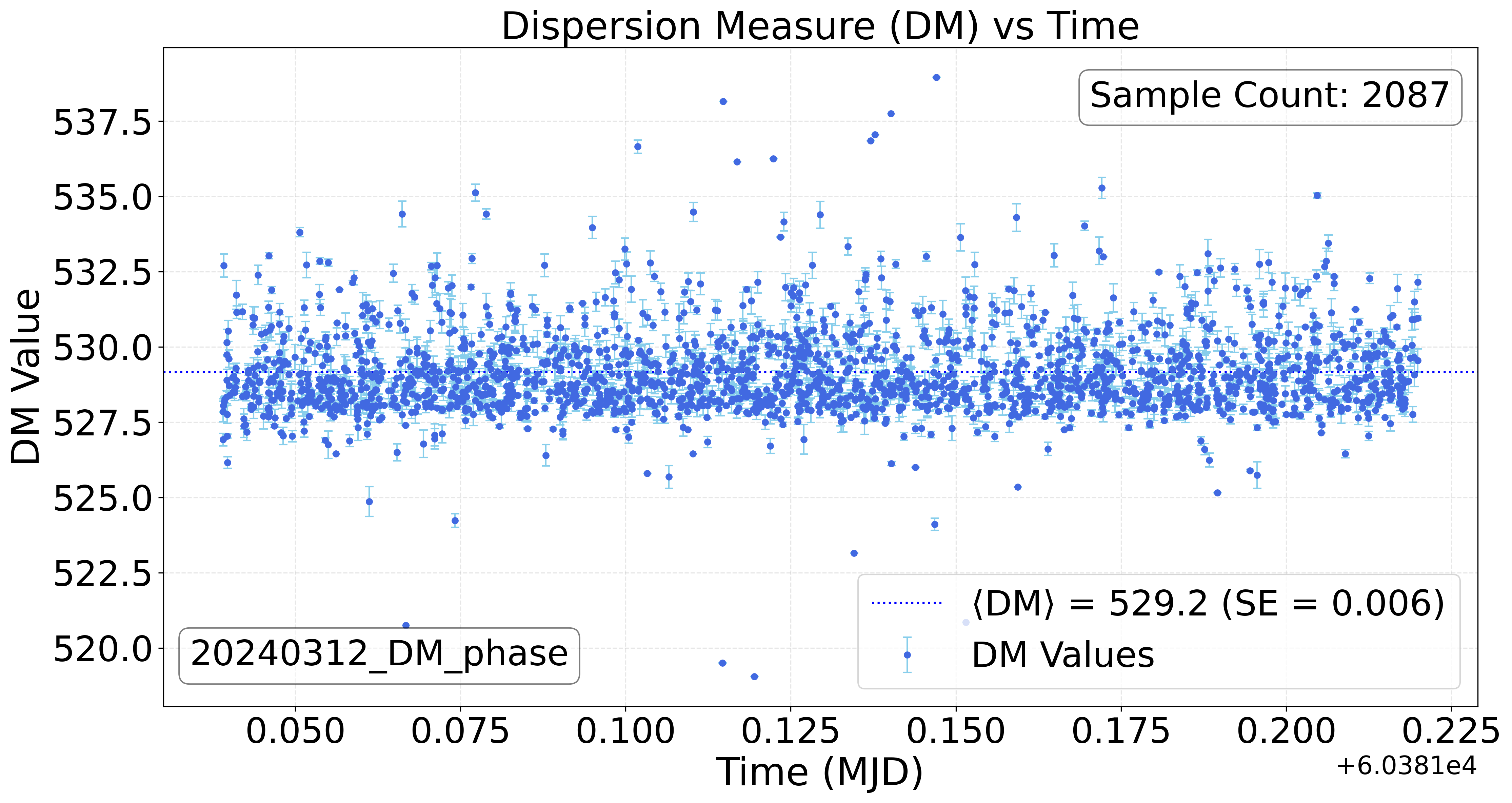}
        \label{fig:fig00}
    \end{minipage}  
    \hspace{0.0\textwidth}  
    \begin{minipage}{0.45\textwidth}  
        \centering
        \includegraphics[width=\textwidth]{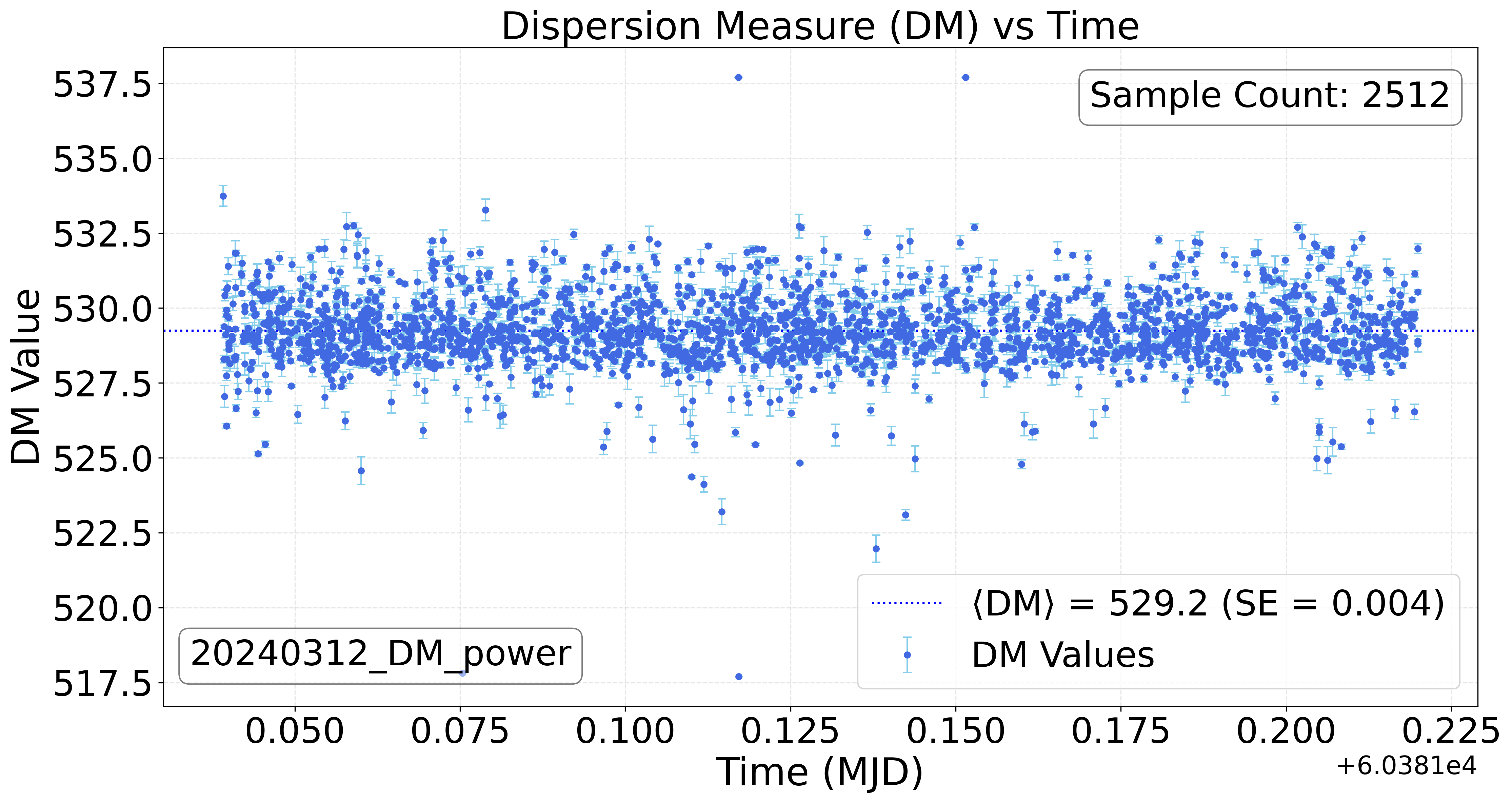}
        \label{fig:fig000}
    \end{minipage}
    \caption{
        Comparison of two DM measurement methods, \texttt{DM\_phase} and \texttt{DM\_power}, applied to the data from March~12, 2024 (20240312). 
        The left panel shows results from \texttt{DM\_phase}, while the right panel corresponds to \texttt{DM\_power}, both yielding consistent results. The dotted line indicates the average DM value, denoted by ⟨DM⟩, which is determined to be $529.2~\text{pc~cm}^{-3}$ for both \texttt{DM\_phase} and \texttt{DM\_power}, with the standard error (SE) reflecting the variability in the measurements. 
        Only data points with error bars $< 1~\text{pc~cm}^{-3}$ (i.e., a one-sided error bar of $0.5~\text{pc~cm}^{-3}$) are plotted, selected from the original 3203 bursts, with the number of precise data points noted in Sample Count.
    }
    \label{fig1}
\end{figure}

\begin{figure}[ht]
    \centering
    \begin{minipage}{0.45\textwidth}
        \centering
        \includegraphics[width=\textwidth]{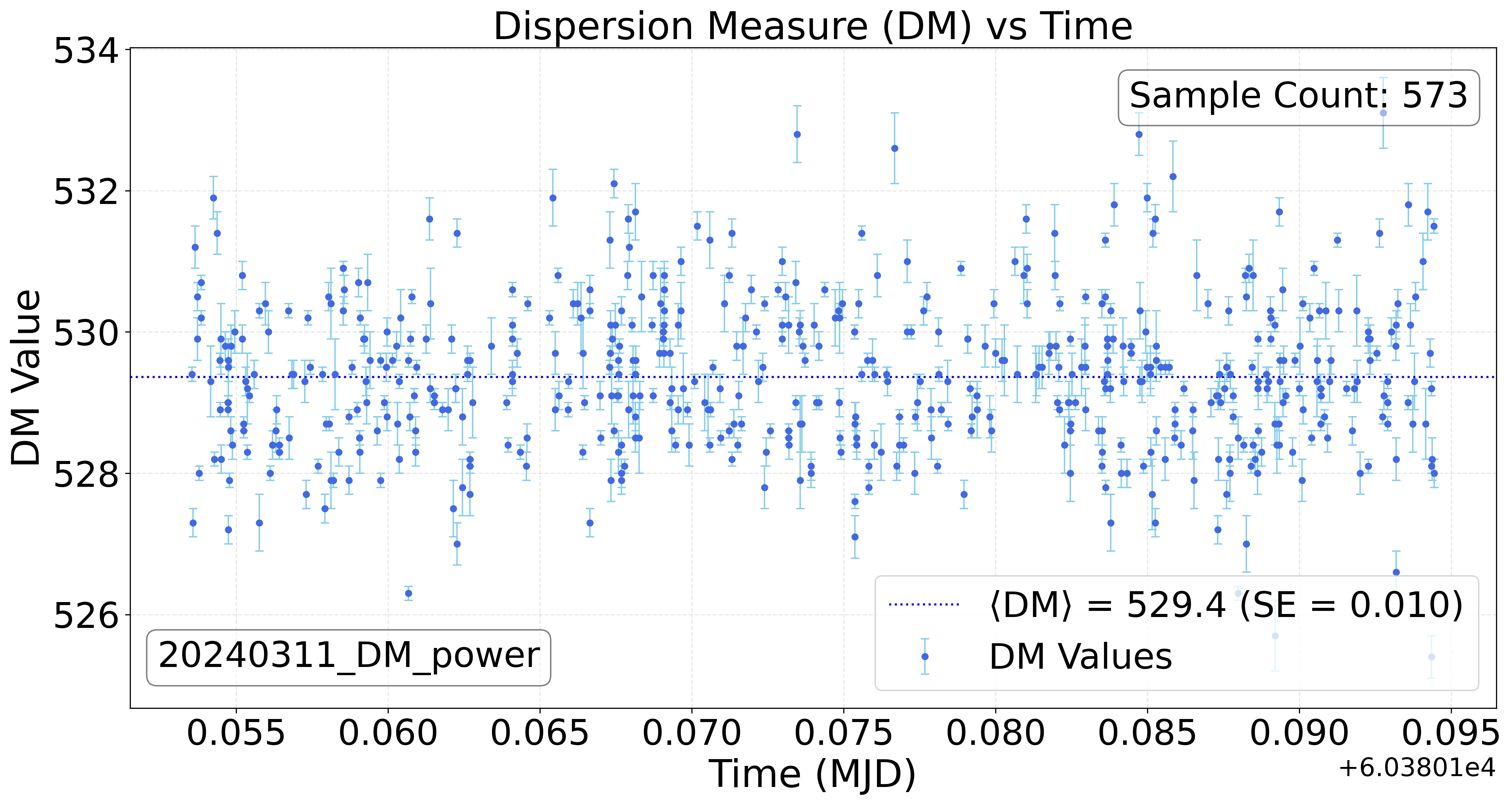}
        \label{fig:fig1}
    \end{minipage}  
    \hspace{0.0\textwidth}
    \begin{minipage}{0.45\textwidth}
        \centering
        \includegraphics[width=\textwidth]{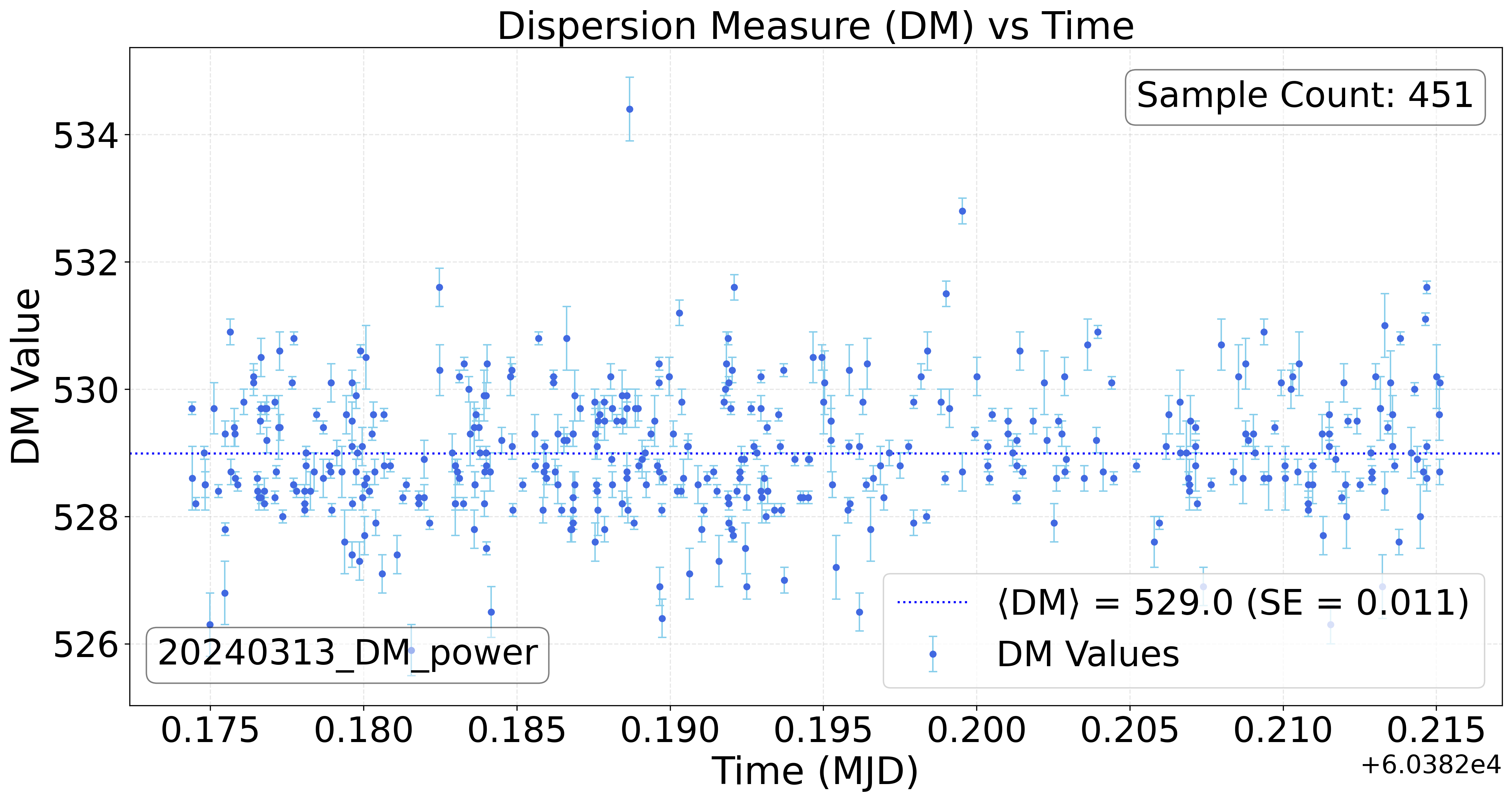}
        \label{fig:fig2}
    \end{minipage}
    \caption{Both figures compare the \texttt{DM\_power} results for two data sets: March~11 2024 session (20240311) and March~13 2024 session (20240313), obtained using the \texttt{DM\_power} software. The average DM values for these two days are shown for comparison. Interestingly, the DM value for the March~12, 2024 (20240312) data lies precisely between the values of the March~11 and March~13 datasets, suggesting that the DM for this day is consistent with a midpoint between the two neighboring observations.}
    \label{fig2}
\end{figure}

For observations on the March~12, 2024 session, we utilize the \texttt{DM\_power}\footnote{\url{https://github.com/hsiuhsil/DM-power}} \citep{2022arXiv220813677L} and \texttt{DM\_phase}\footnote{\url{https://github.com/danielemichilli/DM_phase}} \citep{2019ascl.soft10004S} packages, which optimally weight the pulse structure at each Fourier frequency. The results of both methods for March~12, 2024 (15,780 seconds of data) are consistent with $529.2~\text{pc~cm}^{-3}$, as shown in Figure~\ref{fig1}. Additionally, we present data from March~11 and March~13, each representing 1,200 seconds of observation, with DM values of $529.4~\mathrm{pc\,cm^{-3}}$ and $529.0~\mathrm{pc\,cm^{-3}}$, respectively, measured using the \texttt{DM\_power} package, as shown in Figure~\ref{fig2}. 
The average DM of March~12, 2024, with a value of $529.2~\text{pc~cm}^{-3}$, is between the values of March~11, 2024, and March~13, 2024.
According to \citet{2021arXiv210713549T}, this variation may indicate that the amount of plasma along the line of sight between the modulation screen and the Earth has a long-term variation.
This comparison demonstrates that the use of the daily average DM for March~12, 2024, for plotting, classification, and drifting rate measurement could be reliable.
We choose DM$=529.2~\text{pc~cm}^{-3}$ for the entire analysis in this work for simplicity and consistency.

\section{burst-cluster Morphological Classification} 
\label{sec:morphology}

\begin{table}[ht]
    \centering
    \caption{The morphological classification results after de-dispersion using the average DM value (529.2 pc cm$^{-3}$). U and D denote upward and downward, respectively. H, M, L, and W represent high, middle, low, and wide frequency bands (based on the 1-1.5 GHz observation range of the FAST telescope). The numbers 1, 2, and m indicate single-, double-, and multi-component burst-clusters, respectively. C denotes complex morphology, NE represents burst-clusters with no evidence of drifting, and Not-Clear indicates cases with low signal-to-noise ratio or severe RFI, making it impossible to determine their morphology. ND stands for no drifting.}
    \label{tab:stat_table}
    \begin{tabular}{lccc@{\hspace{2em}}lccc}
    \hline
    \hline
    \multicolumn{4}{c}{Downward (D=745)} & \multicolumn{4}{c}{Upward (U=233)} \\
    \hline
    Band & 1 & 2 & m & Band & 1 & 2 & m \\
    \hline
    H & 122 & 64 & 5 & H & 27 & 16 & 2 \\
    M & 35 & 24 & 1 & M & 42 & 5 & 0 \\
    L & 162 & 127 & 11 & L & 73 & 42 & 2 \\
    W & 55 & 119 & 30 & W & 0 & 23 & 1 \\
    \hline
    \multicolumn{8}{c}{Other morphological classifications} \\
    \hline
    C & NE-H & NE-L & Not-Clear & \multicolumn{4}{c}{No drifting (ND=720)} \\
    \cline{5-8}
    224 & 52 & 135 & 449 & & 1 & 2 & m \\
    & & & & & 670 & 49 & 1 \\
    \hline
    \end{tabular}
\end{table}

\begin{figure}[ht]
    \centering
    \includegraphics[width=\textwidth]{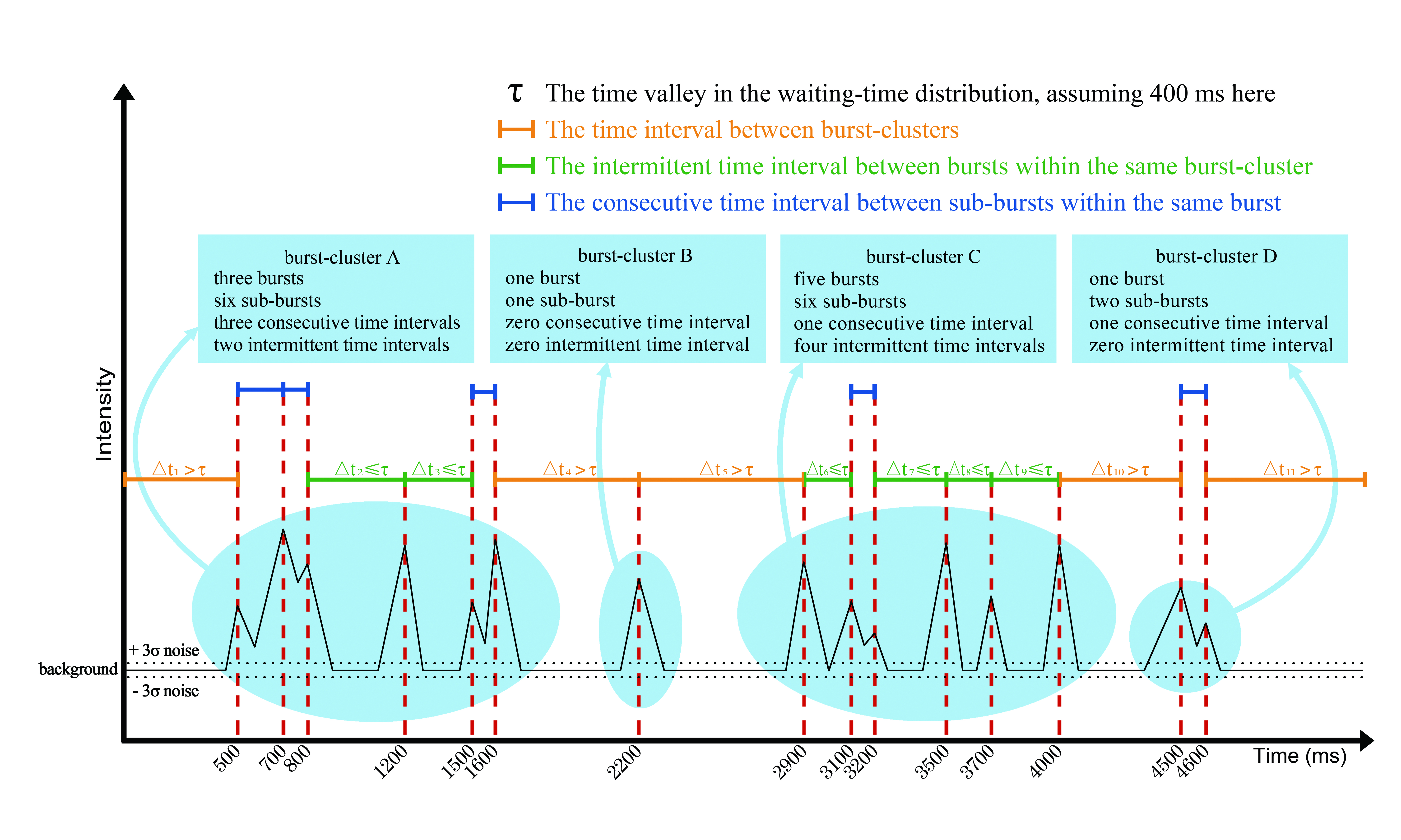}
    \caption{Illustration of the definitions for burst-clusters, bursts, sub-bursts, as well as consecutive and intermittent time intervals. $\tau$ represents the time valley in the waiting-time distribution, assuming 400 ms here. Burst-cluster A consists of three bursts, six sub-bursts, three consecutive time intervals, and two intermittent time intervals. Burst-cluster B, C, and D are similar, as shown in the figure. In this observation session, the total counts of burst-clusters, bursts, sub-bursts, consecutive time intervals, and intermittent time intervals are four, ten, fifteen, five, and six, respectively. The total time scale in the figure is in the order of seconds. Since the bursts are distributed on the millisecond scale and are far apart, the figure employs an exaggerated representation to make the millisecond-scale bursts more visible.
    }
    \label{burst_definition}
\end{figure}

In this section, we present our morphological classification method for bursts, which is adapted from the framework of \citet{2022RAA....22l4001Z} with modifications. However, it should be noted that there is a major caveat regarding our DM averaging approach, which may affect the identification of drifting burst-clusters, particularly for single-component burst-clusters where the observed drifting may be an artifact rather than genuine frequency evolution. A follow-up study using precise DM values for all FAST observations of FRB~20240114A is currently underway to address this limitation. Our morphological classification criteria are defined as follows (for a more detailed schematic illustration, see Figure~\ref{burst_definition}):
\begin{enumerate}
    \item \textbf{Sub-Burst}: One of the quasi-connected components in the dynamic spectrum, where the emission intensity between adjacent sub-bursts remains above the background level and exhibits a distinguished peak (visually separable from background fluctuations) in the de-dispersed burst profile. These sub-bursts constitute elements within a burst. In particular, a component with only a single peak, standing alone, is also called a sub-burst.
    
    \item \textbf{Burst}: A collection of sub-bursts, which may consist of only one sub-burst.
    
    \item \textbf{Burst-Cluster}: A collection of bursts (excluding those labeled as Not-Clear), with adjacent components (or peaks) separations less than the time threshold $\tau$, which is the valley derived from the bimodal waiting-time distribution. In particular, a single burst is still classified as a burst-cluster if no other bursts are detected within the $\tau$-duration intervals immediately preceding and following it. Here, the time threshold $\tau$ is 400 ms for FRB~20240114A, as derived from the bimodal waiting-time distribution shown in \citet{2025arXiv250714707Z} (see their Extended Data Fig.~2 for the distribution). Each entry excluding Not-Clear events in Table~\ref{tab:stat_table} represents one such burst-cluster.
\end{enumerate}

The counts for single-, double-, and multi-component burst-clusters (1, 2, and m) in Table~\ref{tab:stat_table} only consider the number of sub-bursts of same burst-cluster. In contrast, the 1, 2, and m counts defined by \cite{2022RAA....22l4001Z} specifically refer to single-, double-, and multi-component bursts (rather than burst-clusters), focusing on the drifting of sub-bursts within the same burst. Therefore, a single-component burst-cluster contains only one burst, and this burst only contains one sub-burst. A double-component burst-cluster may consist of either one burst containing two sub-bursts, or two bursts each containing one sub-burst. By extension, a multi-component burst-cluster can manifest as either one burst with multiple sub-bursts, or multiple bursts containing varied numbers of sub-bursts. 

Table \ref{tab:stat_table} displays the de-dispersed morphological classifications that were derived by applying the mean DM of 529.2 pc cm$^{-3}$. In the table, D represents the downward drifting and U represents the upward drifting. The codes H, M, L, and W classify emissions with a bandwidth relative to the FAST observing band (1.0-1.5 GHz). M (Middle): Emission is within the 1.0-1.5 GHz band. H (High): Emission extends above the 1.5 GHz upper boundary. L (Low): Emission extends below the 1.0 GHz lower boundary. W (Wide): Emission extends beyond both the 1.0 GHz and 1.5 GHz boundaries. The numbers 1, 2, and m represent single-, double-, and multi-component burst-cluster, respectively. C indicates complex morphology, NE-H stands for no-evidence high frequency band, NE-L stands for no-evidence low frequency band, and Not-Clear indicates cases where the signal-to-noise ratio is too low or the radio frequency interference (RFI) is too severe. ND represents no drifting.

Then, we further distinguish two types of time intervals (for a more detailed schematic illustration, see Figure~\ref{burst_definition}): 
\begin{enumerate}
    \item \textbf{Consecutive Time Interval}: The time intervals between sub-bursts within the same burst, where the emission intensity between these sub-bursts remains above the background level. Clearly, for a burst containing only one sub-burst, the concept of consecutive time intervals does not apply.
    
    \item \textbf{Intermittent Time Interval}: The intervals between bursts within the same burst-cluster, where the emission intensity between these bursts returns to the background level. Clearly, for a burst-cluster containing only one burst, the concept of intermittent time intervals does not apply.
\end{enumerate}

According to the definitions described above, as shown in Figure~\ref{burst_definition}, burst-cluster A consists of three bursts (including two intermittent time intervals). The first burst contains three sub-bursts (including two consecutive time intervals), the second burst contains only one sub-burst (with no consecutive time intervals), and the third burst contains two sub-bursts (including one consecutive time interval). In total, this results in six sub-bursts, three consecutive time intervals, and two intermittent time intervals. Similarly, burst-cluster B contains one burst, one sub-burst, zero consecutive time intervals, and zero intermittent time intervals. Burst-cluster C contains five bursts, six sub-bursts, one consecutive time interval, and four intermittent time intervals. Notably, in burst-cluster C, although the first two sub-bursts are closely connected, the emission intensity between them returns to the background level; thus according to our definition, these two sub-bursts belong to different bursts. Burst-cluster D contains one burst, two sub-bursts, one consecutive time interval, and zero intermittent time intervals. Therefore, in this observation session, the overall counts of burst-clusters, bursts, sub‐bursts, consecutive time intervals, and intermittent time intervals are four, ten, fifteen, five and six, respectively. For a more comprehensive analysis, the distribution of consecutive and intermittent time intervals for the 978 burst-clusters exhibiting drifting is presented in Section \ref{Analysis_of_consecutive_and_intermittent}.

\subsection{Downward Drifting burst-clusters}

\begin{figure*}[ht]
\centering
\begin{tabular}{ccc}
\adjustbox{valign=t}{\includegraphics[width=0.3\textwidth]{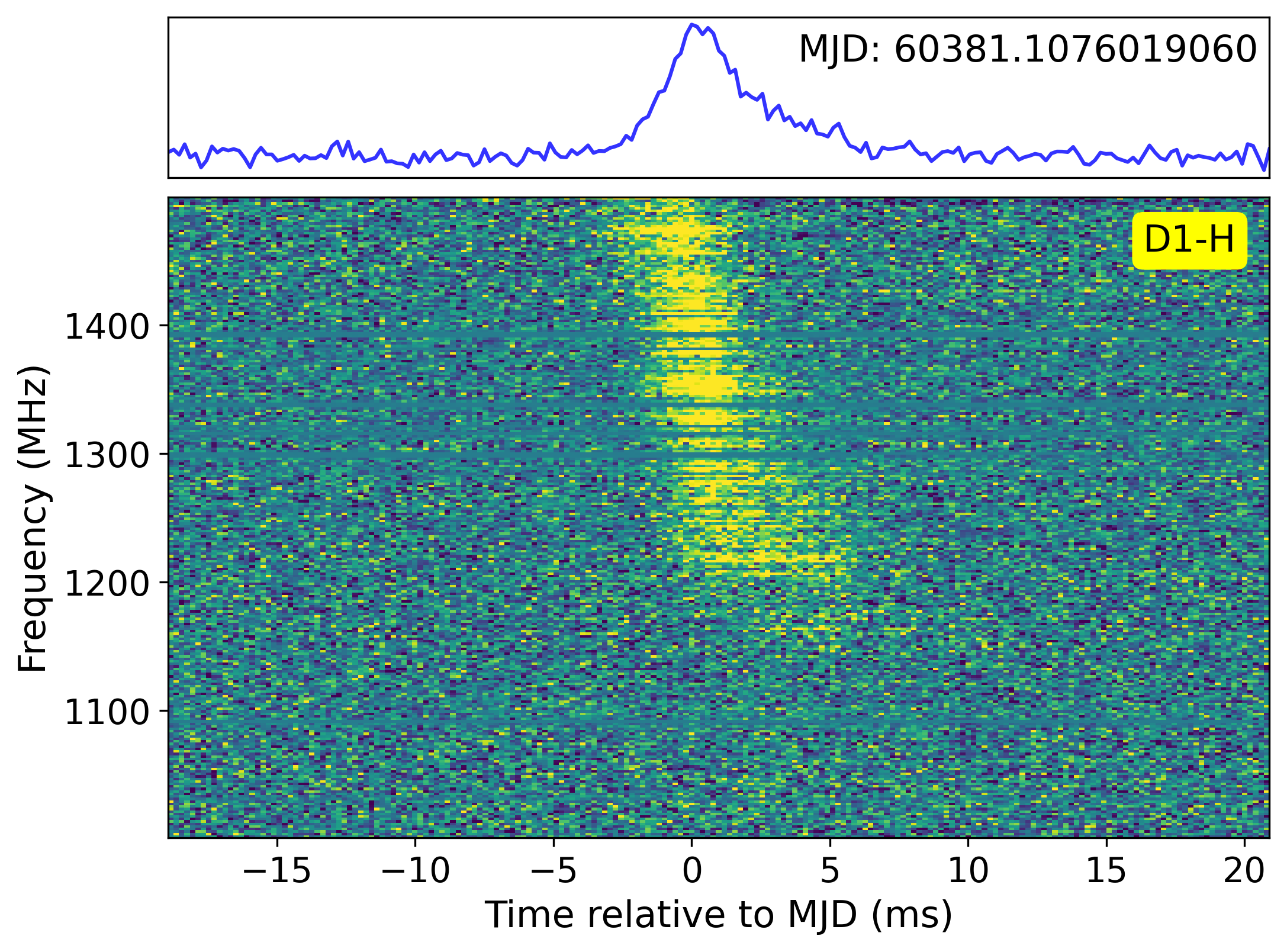}} & 
\adjustbox{valign=t}{\includegraphics[width=0.3\textwidth]{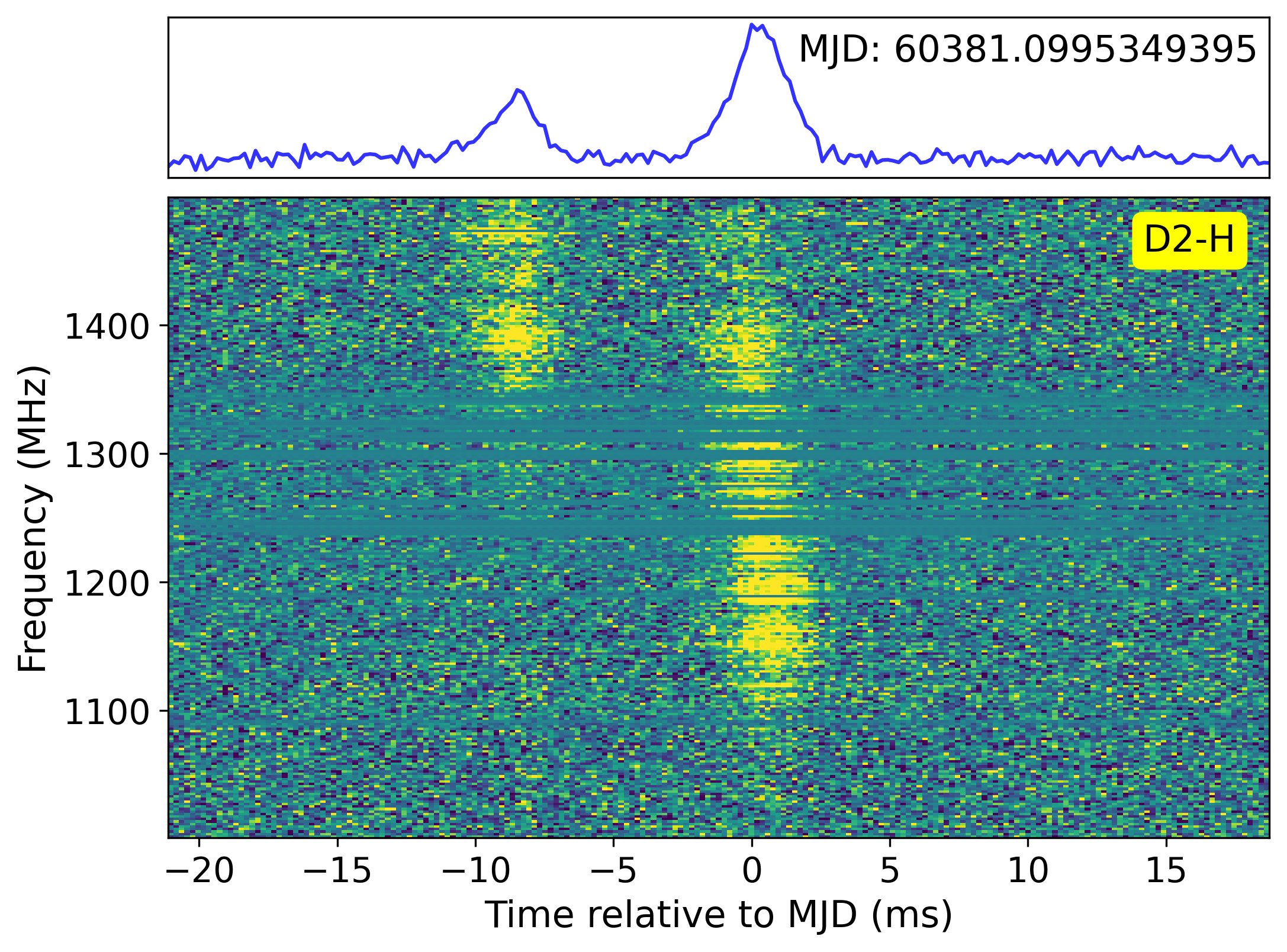}} & 
\adjustbox{valign=t}{\includegraphics[width=0.3\textwidth]{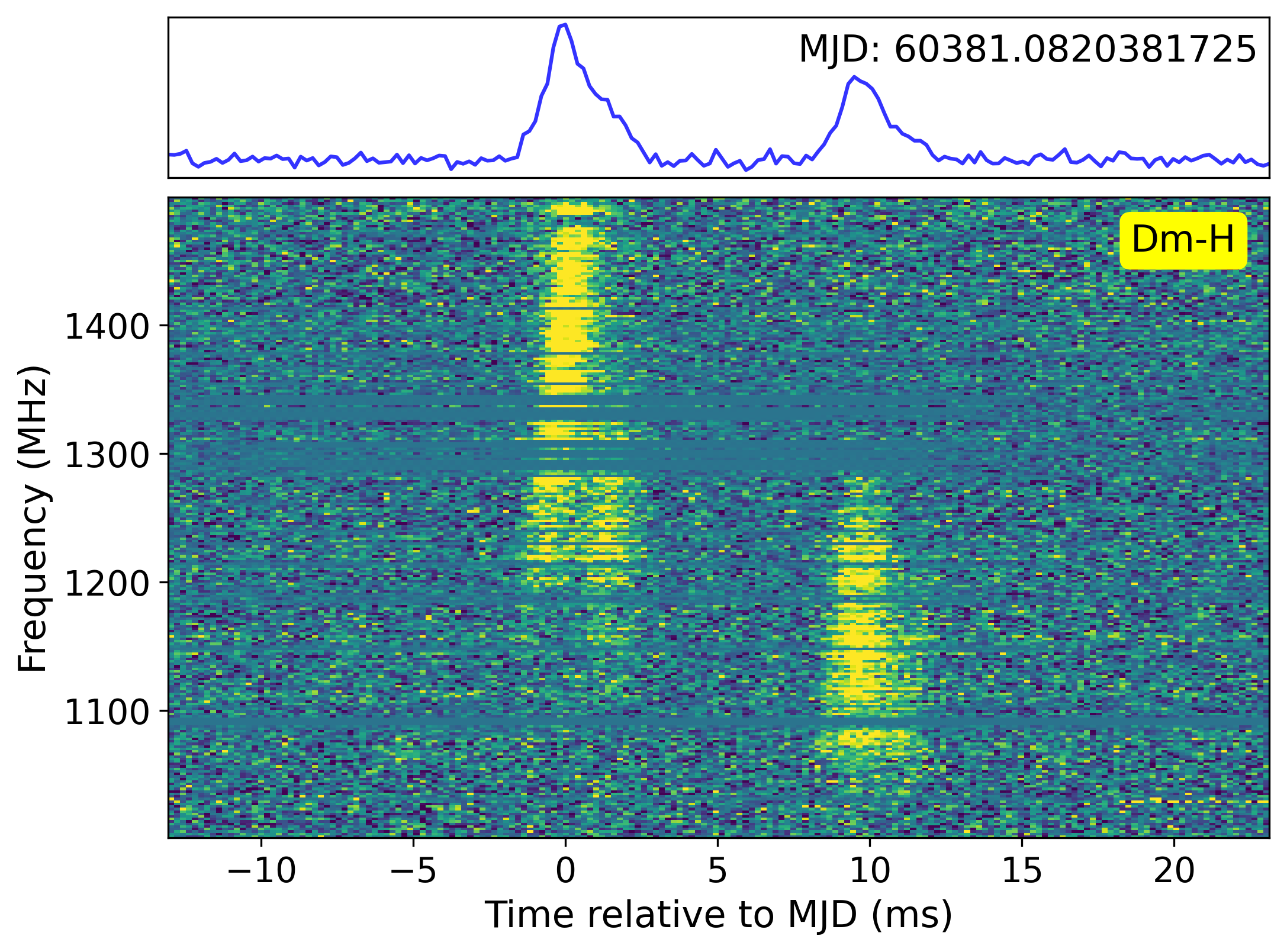}} \\[2mm]
\adjustbox{valign=t}{\includegraphics[width=0.3\textwidth]{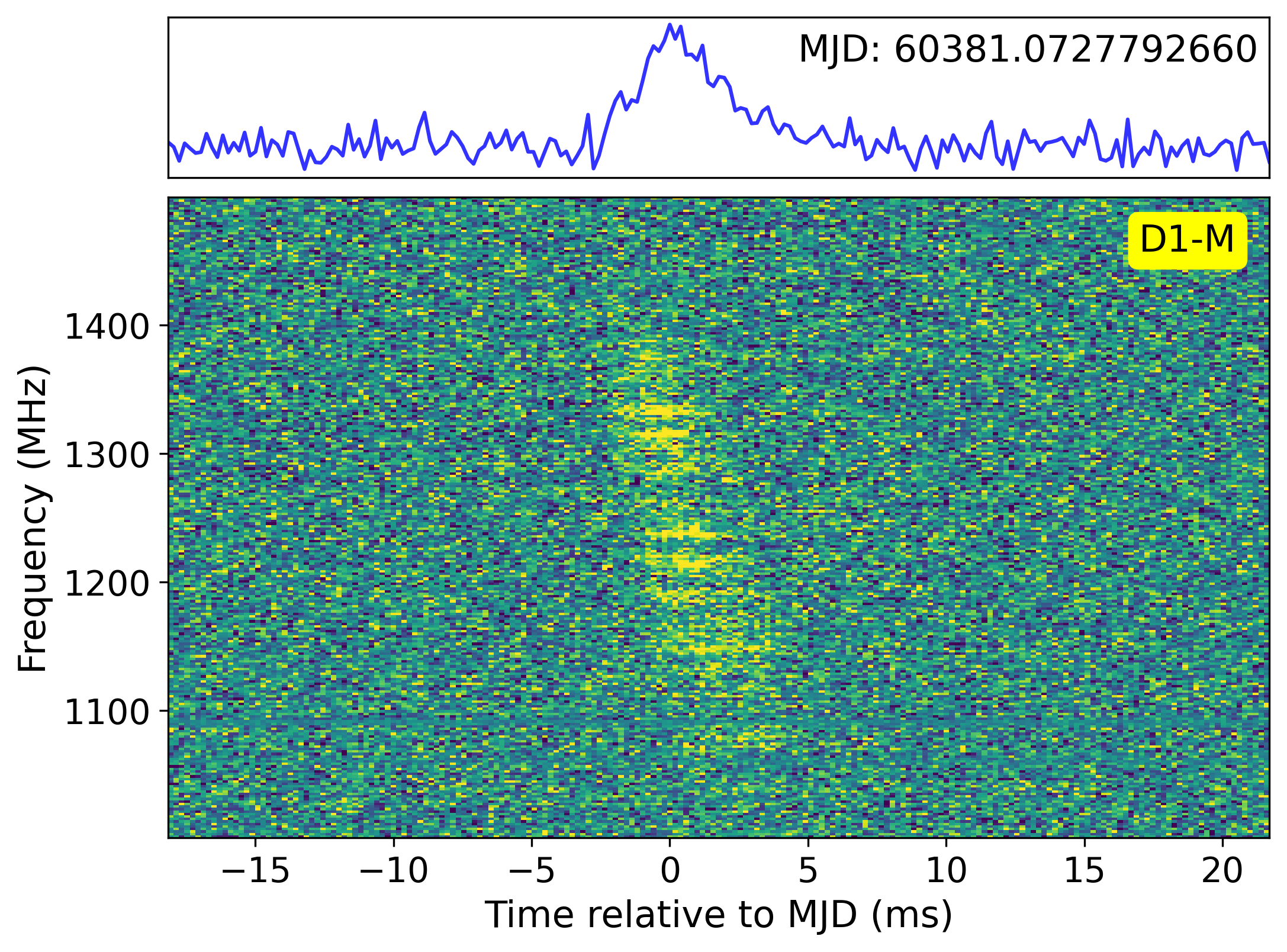}} & 
\adjustbox{valign=t}{\includegraphics[width=0.3\textwidth]{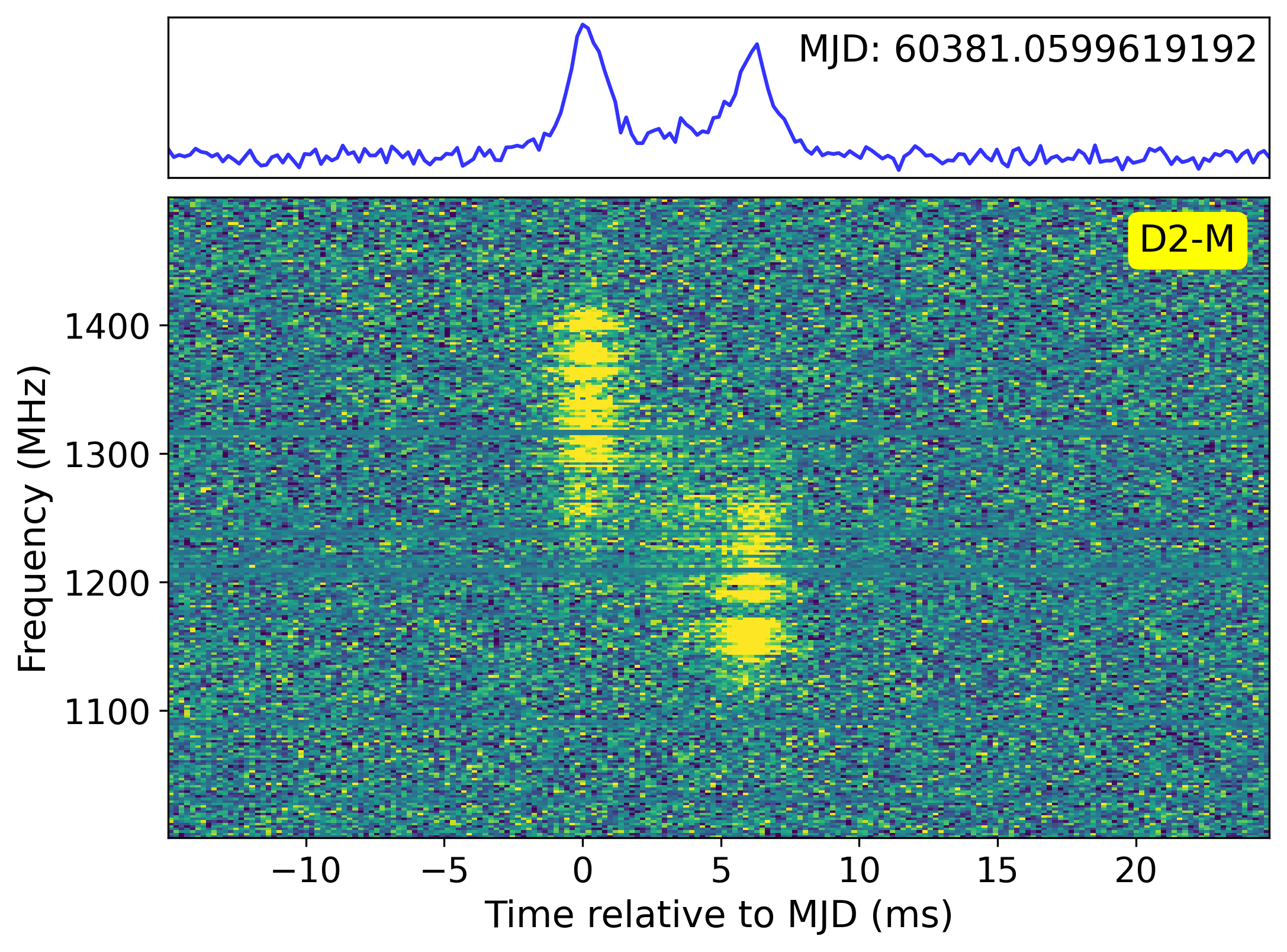}} & 
\adjustbox{valign=t}{\includegraphics[width=0.3\textwidth]{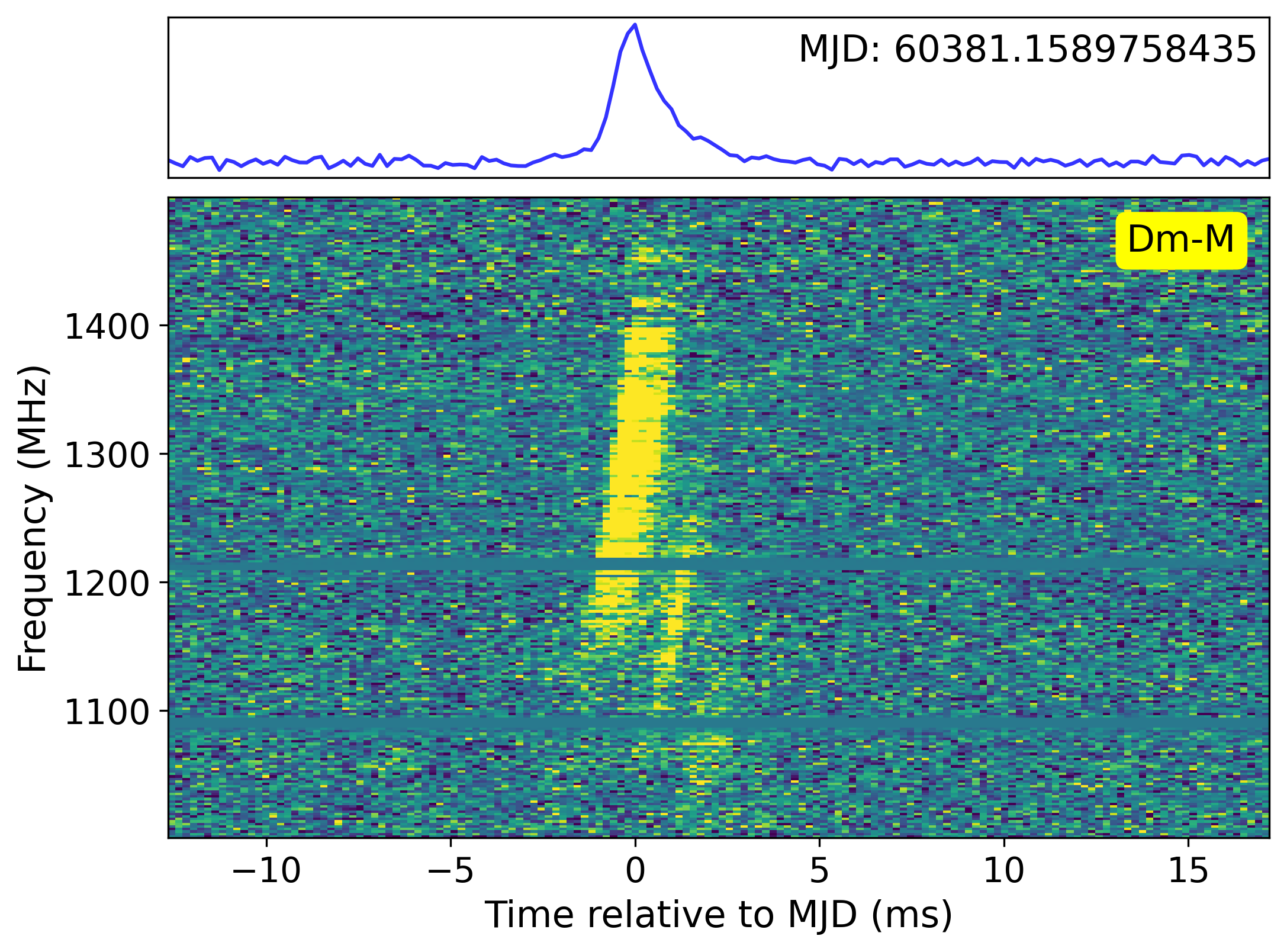}} \\[2mm]
\adjustbox{valign=t}{\includegraphics[width=0.3\textwidth]{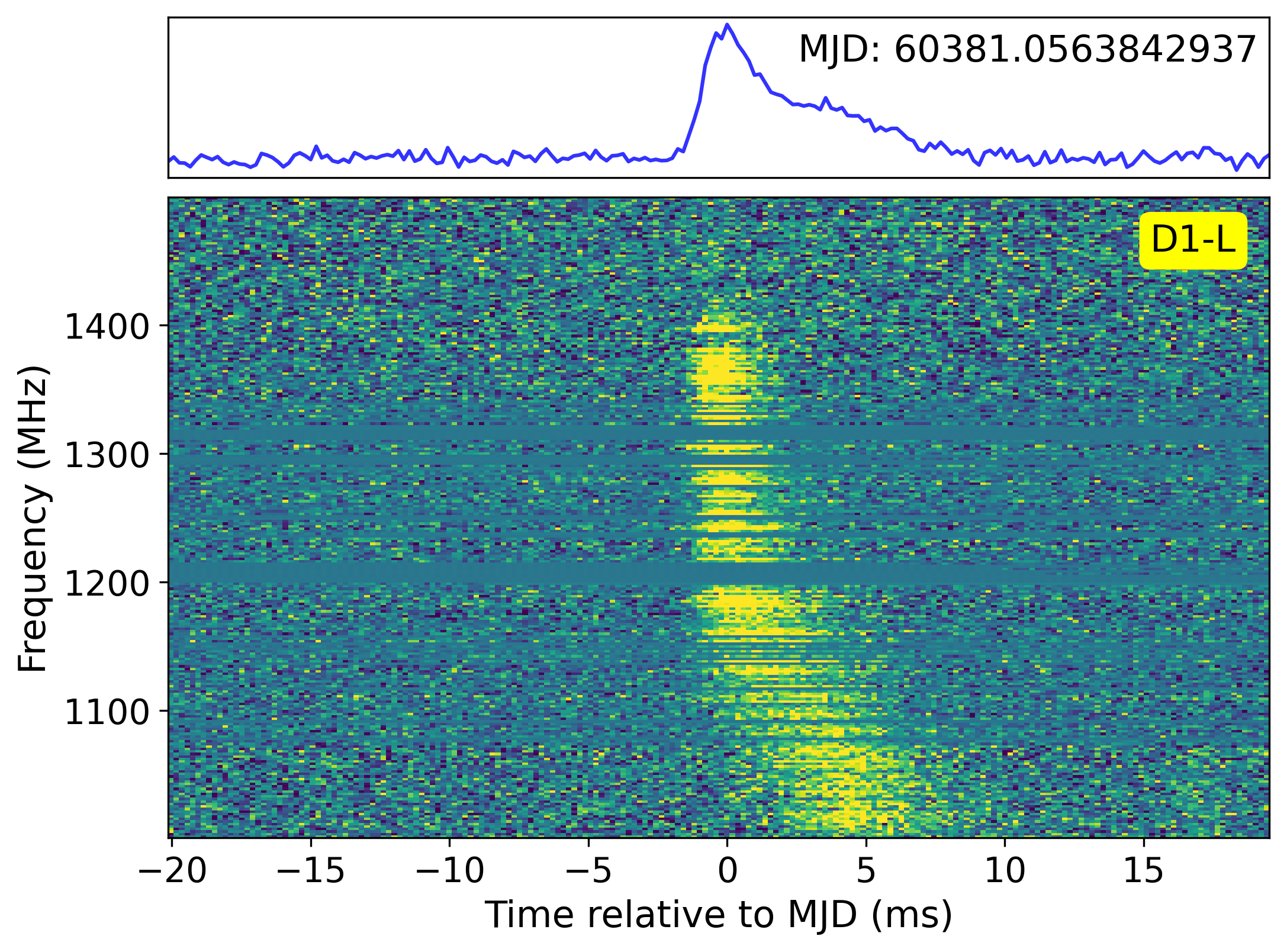}} & 
\adjustbox{valign=t}{\includegraphics[width=0.3\textwidth]{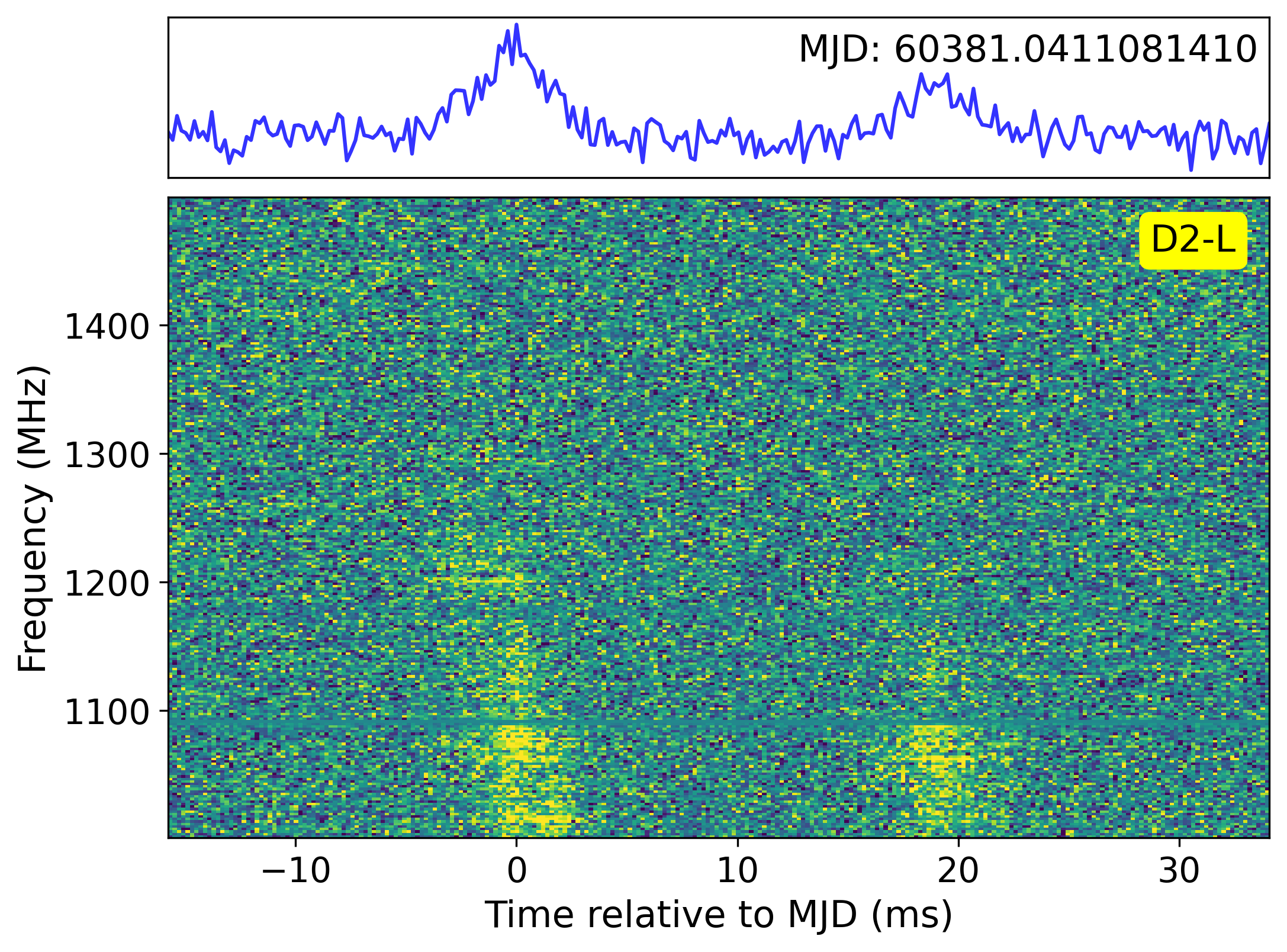}} & 
\adjustbox{valign=t}{\includegraphics[width=0.3\textwidth]{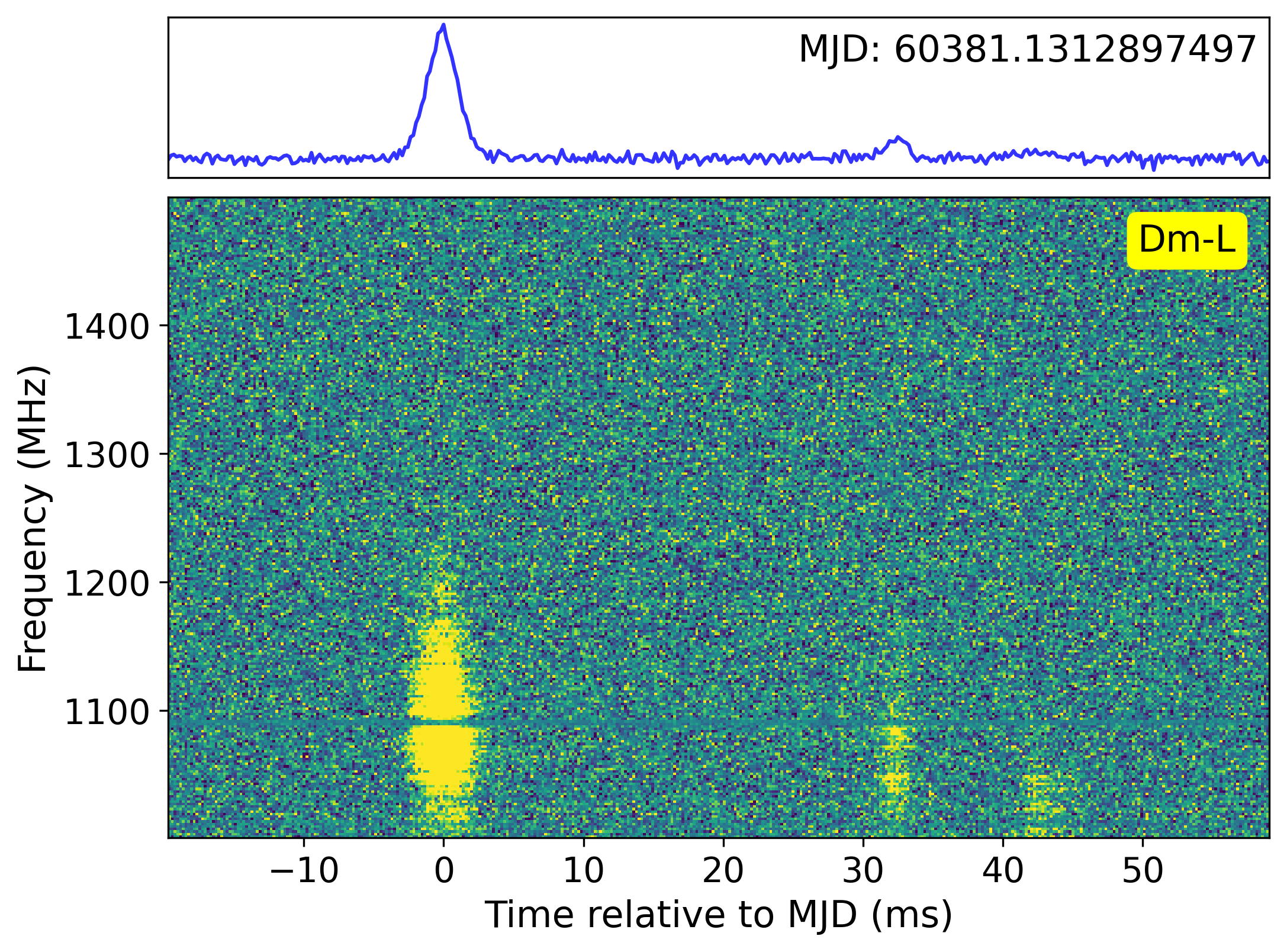}} \\[2mm]
\adjustbox{valign=t}{\includegraphics[width=0.3\textwidth]{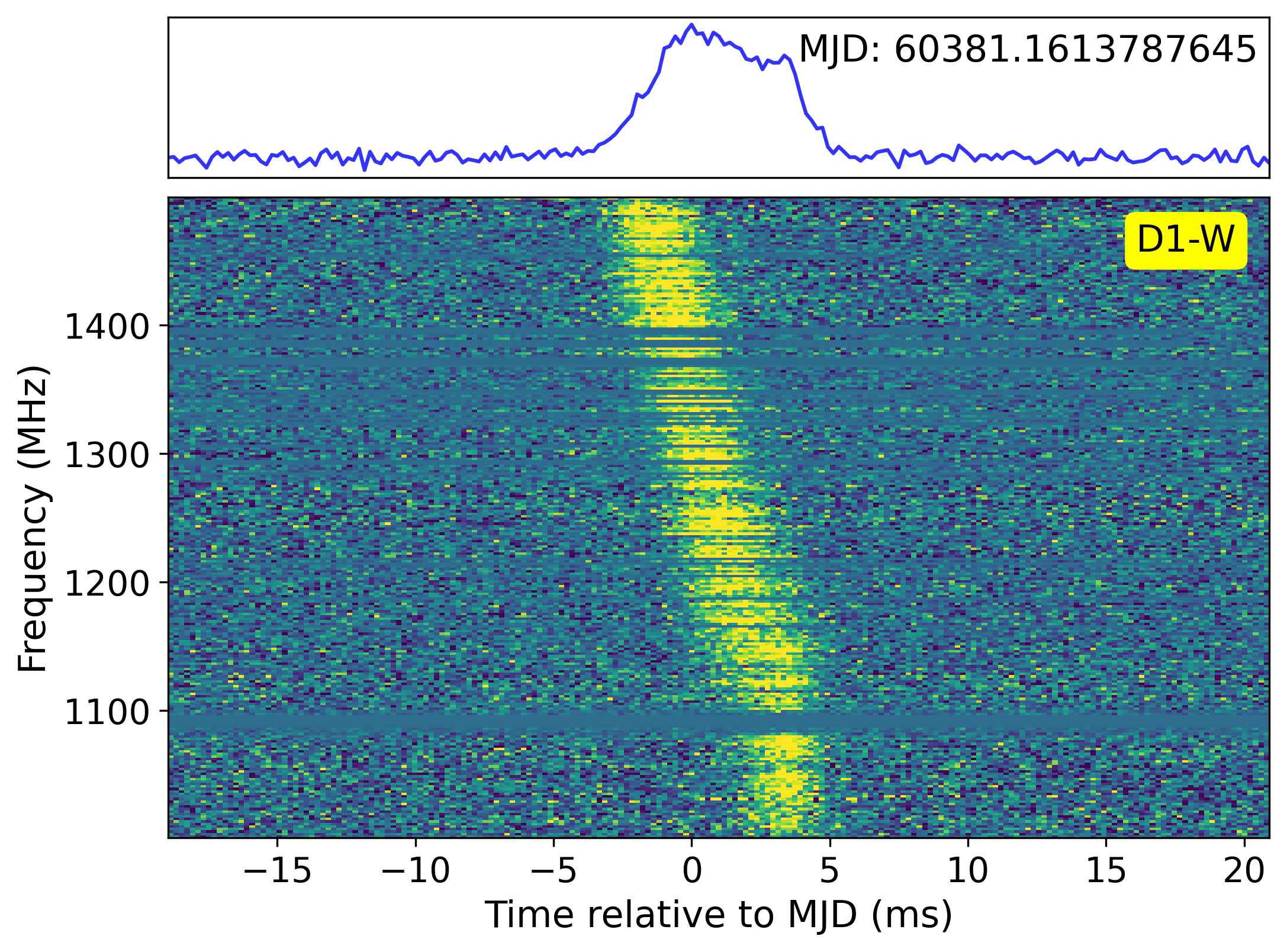}} & 
\adjustbox{valign=t}{\includegraphics[width=0.3\textwidth]{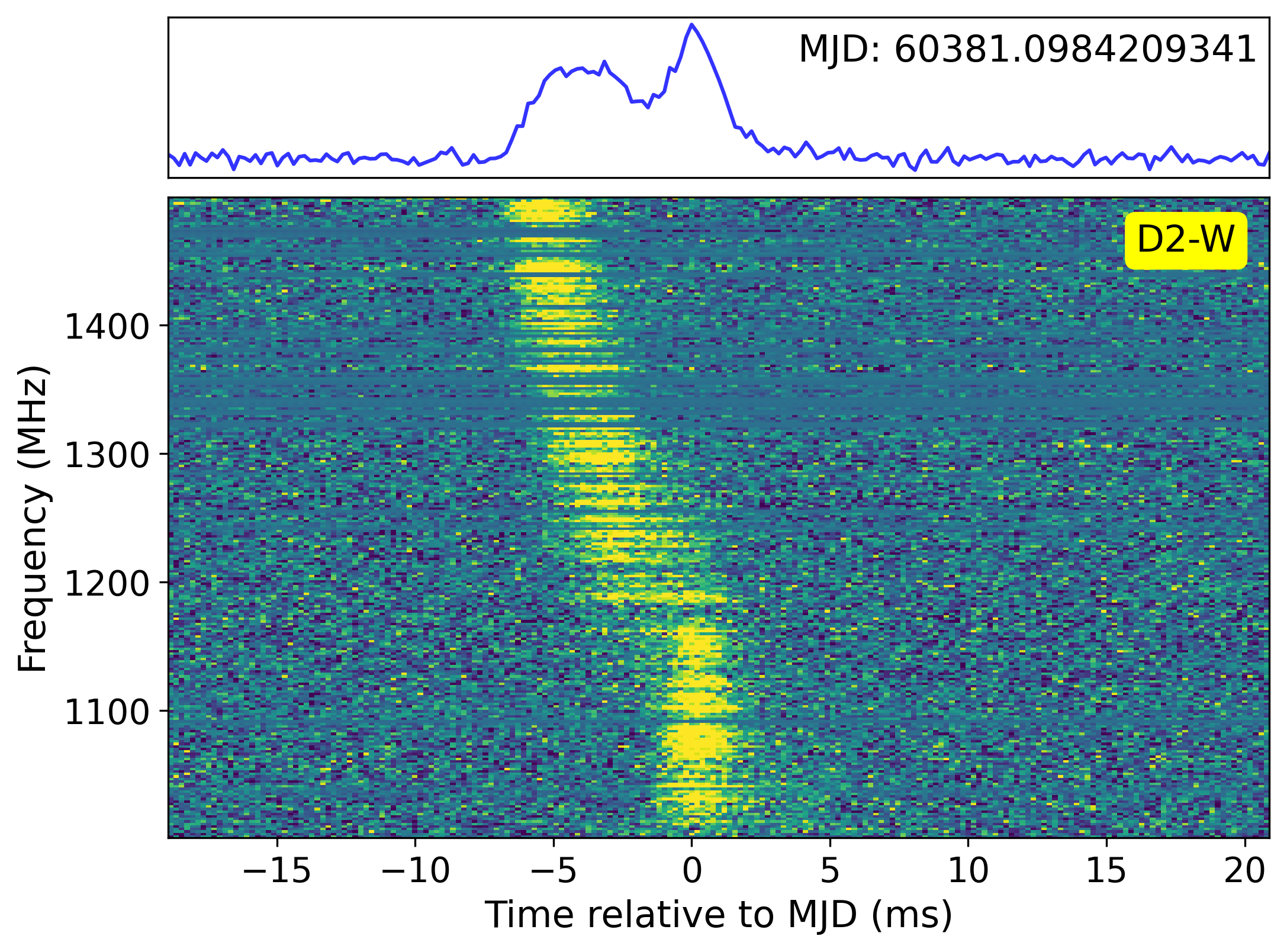}} & 
\adjustbox{valign=t}{\includegraphics[width=0.3\textwidth]{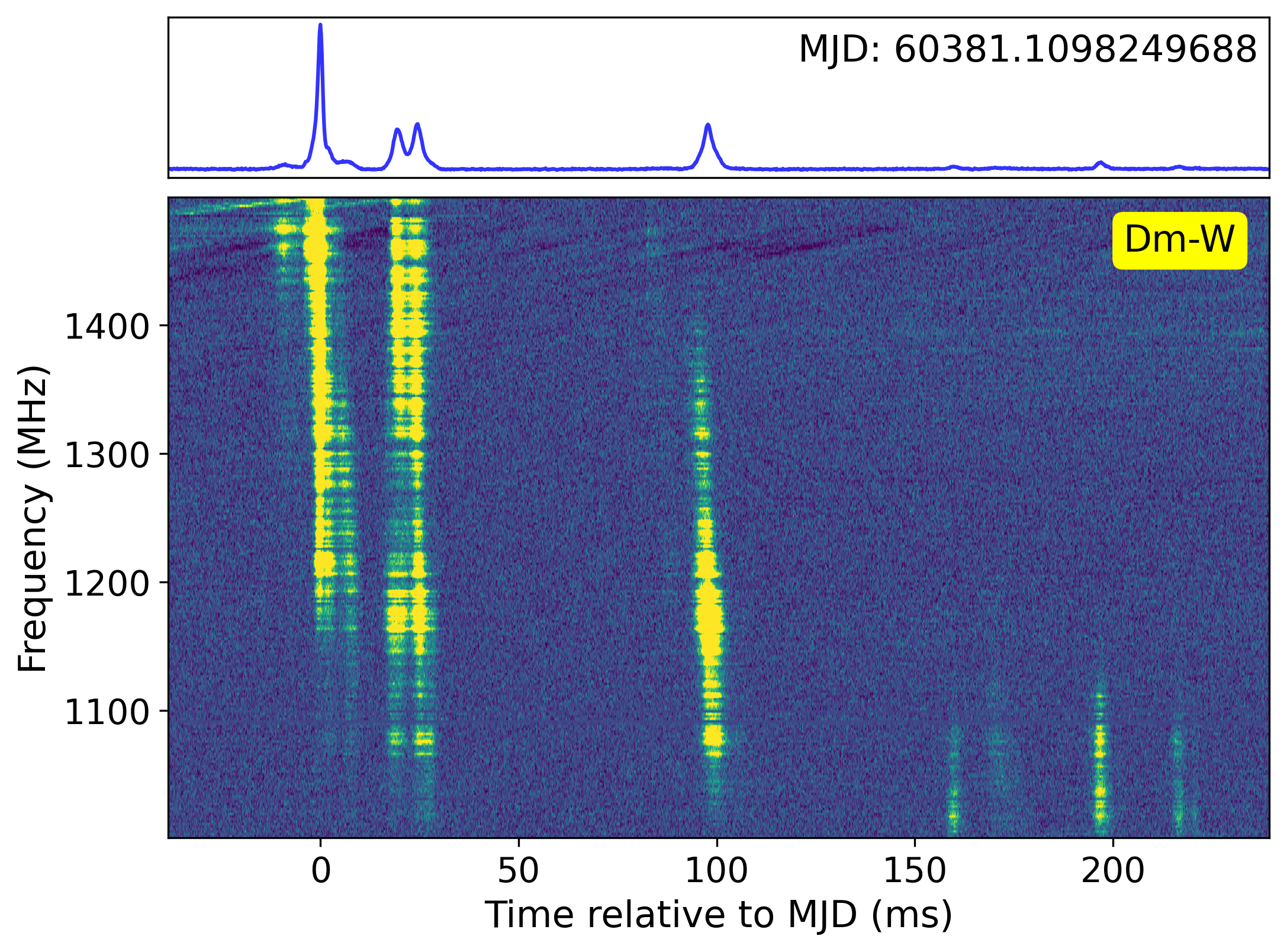}} \\
\end{tabular}
\caption{The typical dynamic spectra of downward drifting burst-clusters. Each burst-cluster is labeled as Dx-y, where x denotes the number of components. If the number of components exceeds two, x is noted as m. The y value indicates the observation band of the burst-cluster: L (low frequency), M (middle frequency), H (high frequency), W (wide frequency). The plots are arranged in a grid of four rows and three columns, with x corresponding to each column and y corresponding to each row.} 
\label{downward_drifting_burst-clusters}
\end{figure*}

Downward drifting burst-clusters refer to cases where the higher-frequency components of the burst-cluster arrive earlier than the lower-frequency ones. These burst-clusters are categorized based on the number of components (sub-bursts) observed in their dynamic spectrum and the position of the frequency bands: high-frequency band (H), middle-frequency band (M), low-frequency band (L), and wide-frequency band (W), as recorded by the FAST. For single-component burst-clusters, this implies that the higher-frequency sub-band arrives earlier than the lower-frequency one. In the case of double-component burst-clusters, the sub-burst with the higher peak frequency is detected earlier than the one with the lower peak frequency. For multi-component burst-clusters, the sub-burst with the highest peak frequency is the first to arrive, while the sub-burst with the lowest peak frequency arrives last, which facilitates the calculation of the drifting rate. A representation of these burst-clusters can be found in Figure~\ref{downward_drifting_burst-clusters}. 

As shown in Table \ref{tab:stat_table}, a total of 745 downward drifting burst-clusters were detected. Among them, there were 364 single-component burst-clusters (48.9\%). It should be noted that since we used an average DM value rather than the precise DM for each individual burst-cluster, the apparent drifting in single-component burst-clusters could be an artifact of this DM averaging approach, and the true number of genuinely drifting single-component burst-clusters could be much smaller than what we report here. Additionally, there were 334 double-component burst-clusters (44.8\%), and 47 multi-component burst-clusters (6.3\%). The distribution across frequency bands is as follows: 181 burst-clusters (24.3\%) in the high-frequency band (H), 60 burst-clusters (8.1\%) in the middle-frequency band (M), 300 burst-clusters (40.3\%) in the low-frequency band (L), and 204 burst-clusters (27.4\%) in the wide-frequency band (W). The number of consecutive time intervals for downward drifting burst-clusters was 184, and the number of intermittent time intervals for downward drifting burst-clusters was 265.

The prevalent downward drifting can be explained by multiple physical frameworks. The most natural interpretation invokes radius to frequency mapping in magnetospheric models  \citep{2019ApJ...876L..15W,osti_1802960,2022ApJ...925...53Z}. The curvature radiation model suggests that relativistic particle bunches moving along neutron star open field lines naturally exhibit downward drifting, as emission regions at higher altitudes encounter larger curvature radii \citep{2019ApJ...876L..15W, 2020ApJ...899..109W}. Separately, relativistic kinematics introduces a distinct mechanism: bulk motion combined with Dicke's superradiance predicts an inverse correlation between sub-burst drifting rates and durations through Doppler-shifted narrowband emission \citep{2020MNRAS.498.4936R}. Propagation effects also contribute significantly; chromatic deflection by circum-burst plasma prisms \citep{2021arXiv210713549T} can redirect low-frequency components into the observer's line of sight, mimicking downward drifting. Synchrotron maser emission from decelerating shocks \citep{10.1093/mnras/stz700,2022ApJ...925..135M} provides an alternative mechanism where shock cooling drives $\nu_c \propto t^{-\beta}$ decay, consistent with the majority of drifting events. Taken together, these scenarios indicate that downward drifting is a natural and robust outcome of several distinct physical processes, though distinguishing their relative contributions remains challenging.

\subsection{Upward Drifting burst-clusters}

\begin{figure}[ht]
\centering
\setlength{\tabcolsep}{6pt}

\begin{tabular}{ccc}
\begin{minipage}[c]{0.3\textwidth}\centering
\includegraphics[width=\linewidth]{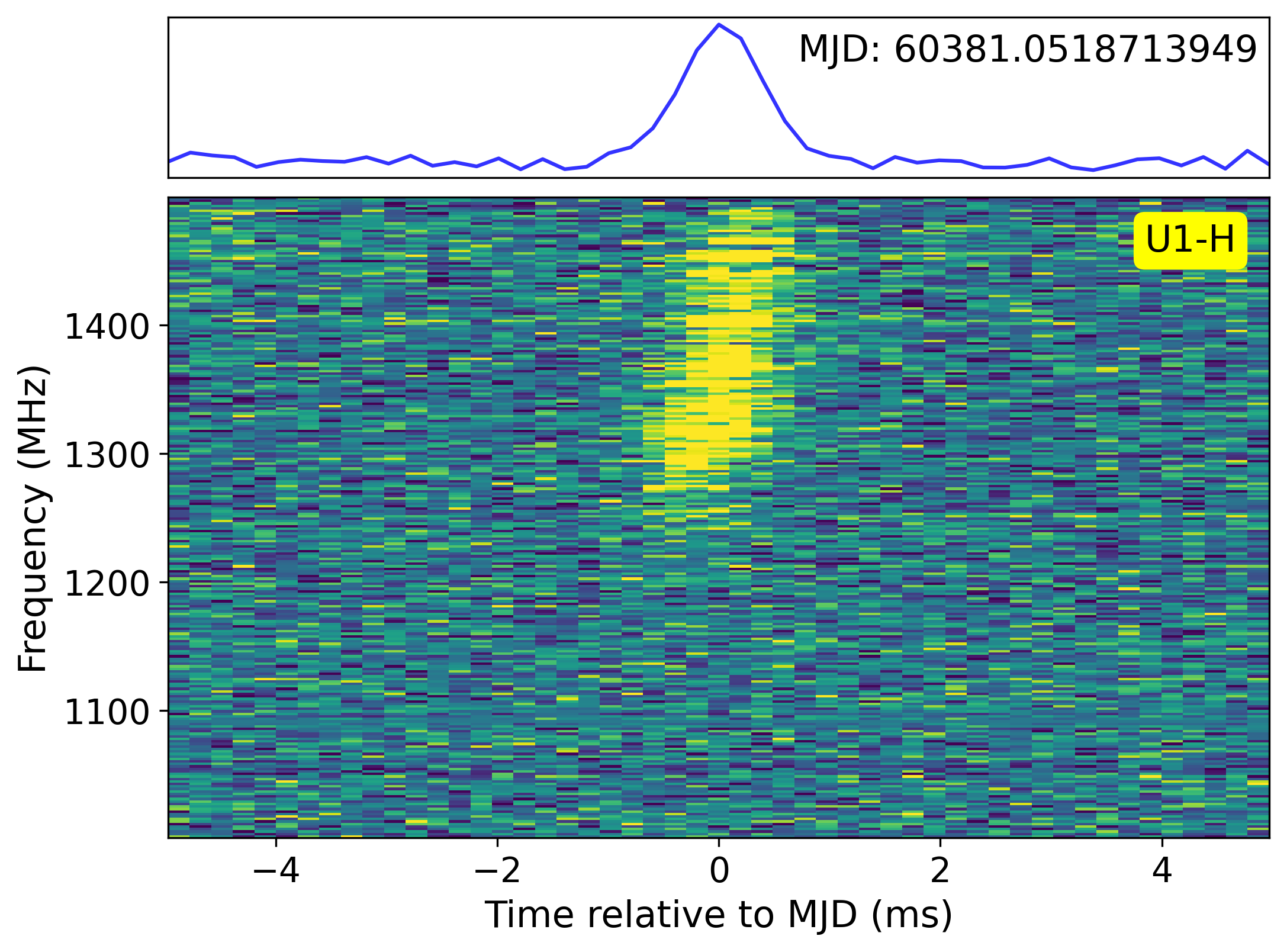}
\end{minipage} &
\begin{minipage}[c]{0.3\textwidth}\centering
\includegraphics[width=\linewidth]{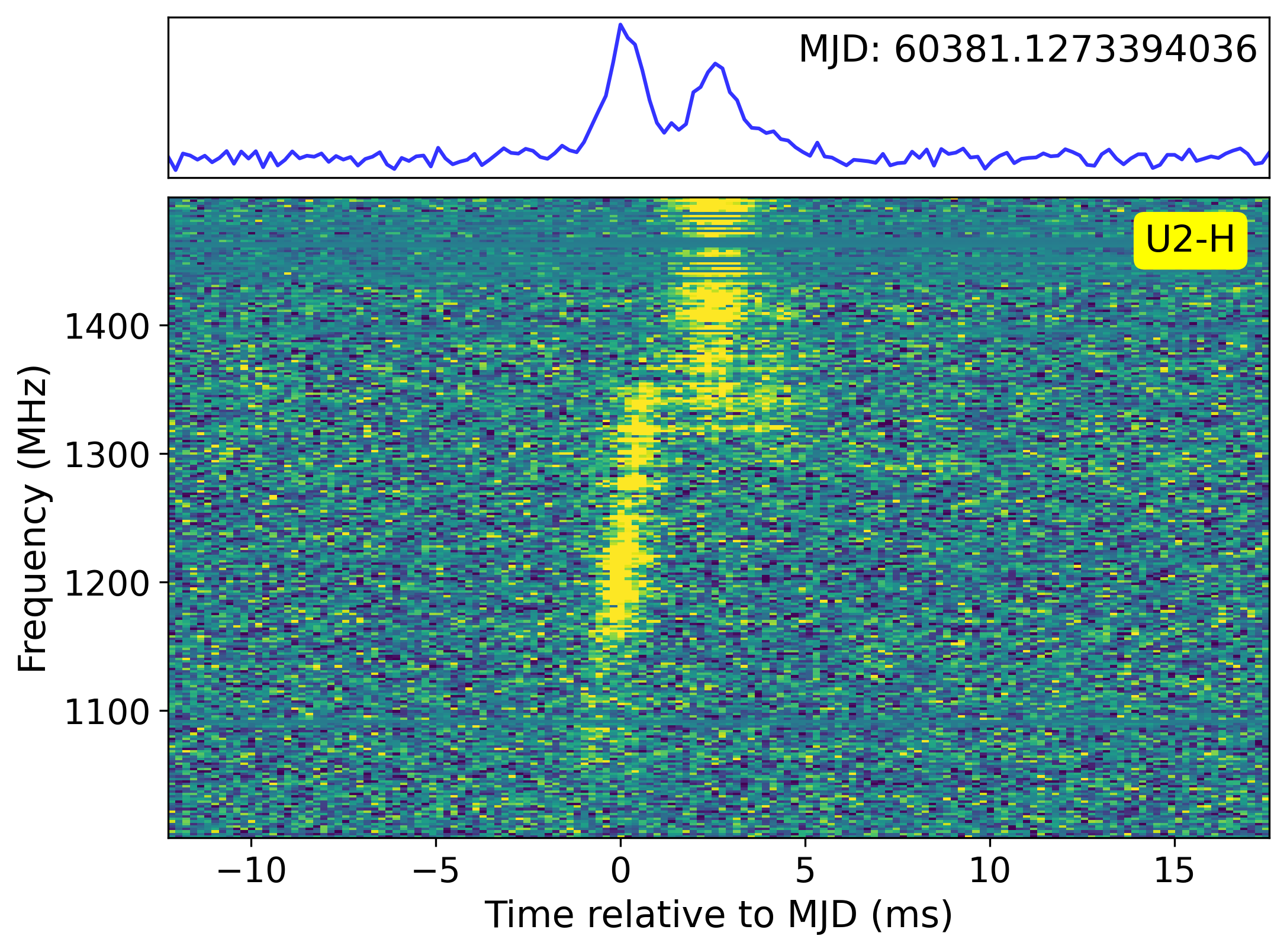}
\end{minipage} &
\begin{minipage}[c]{0.3\textwidth}\centering
\includegraphics[width=\linewidth]{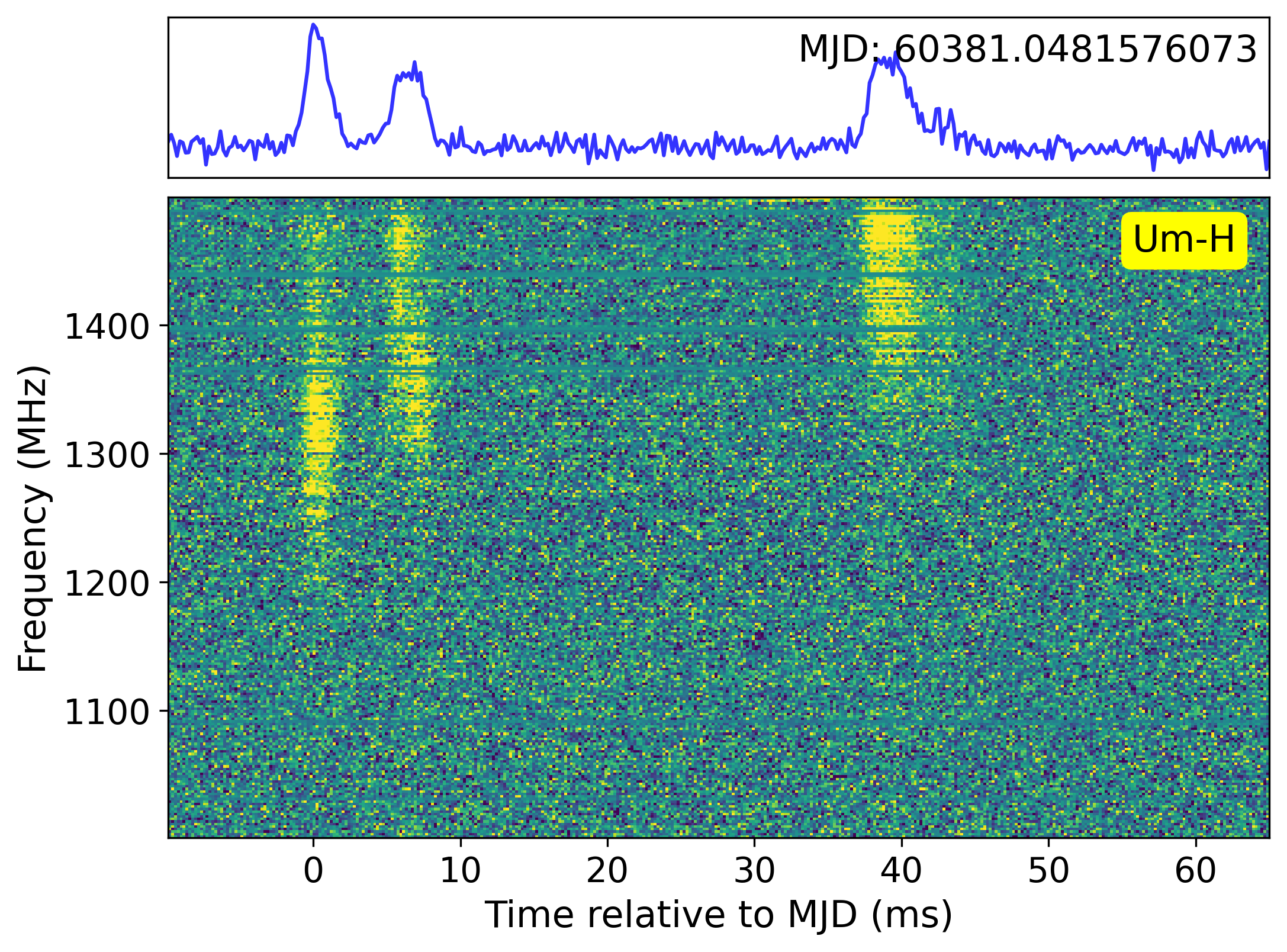}
\end{minipage} \\[2pt]

\begin{minipage}[c]{0.3\textwidth}\centering
\includegraphics[width=\linewidth]{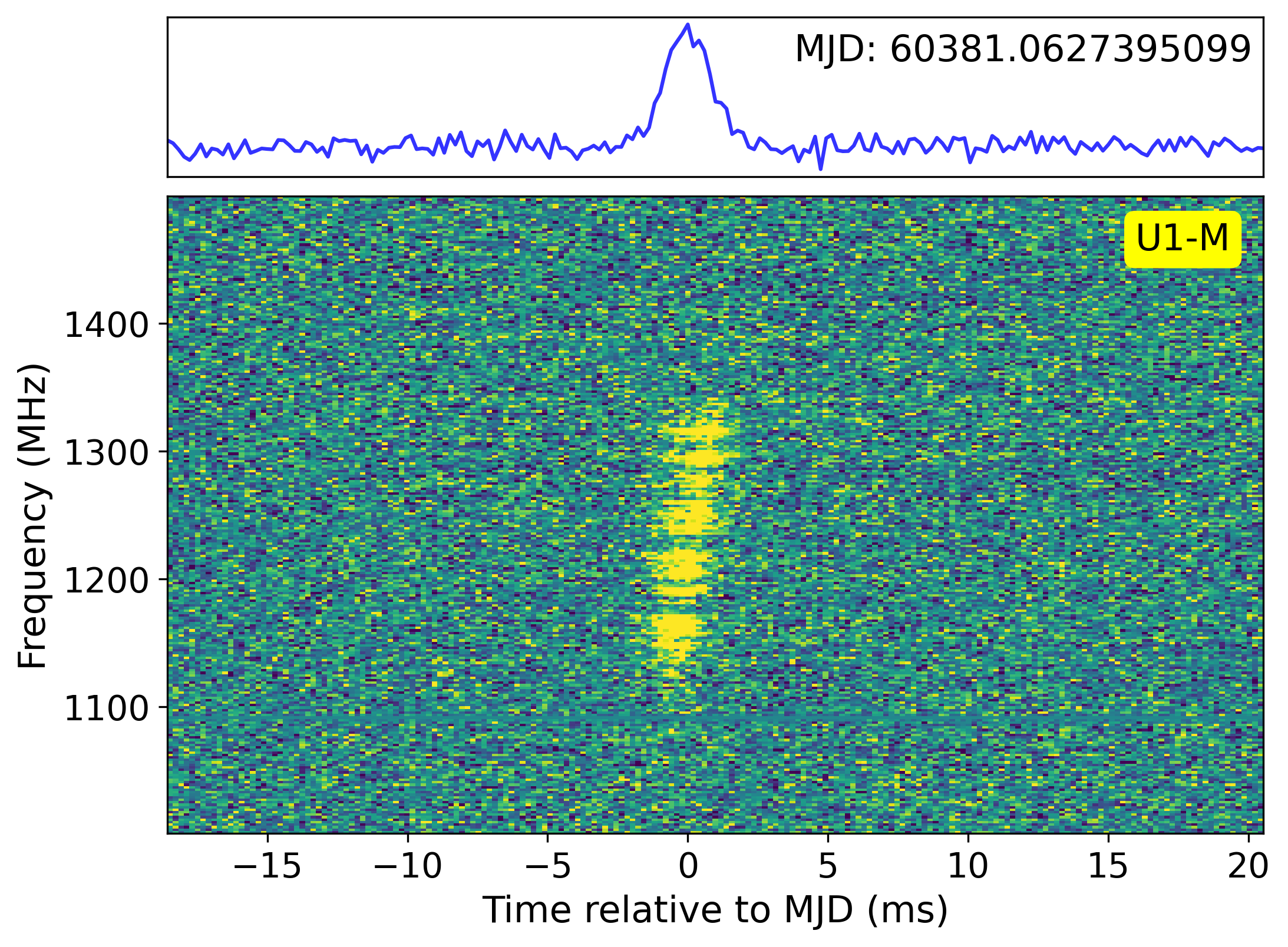}
\end{minipage} &
\begin{minipage}[c]{0.3\textwidth}\centering
\includegraphics[width=\linewidth]{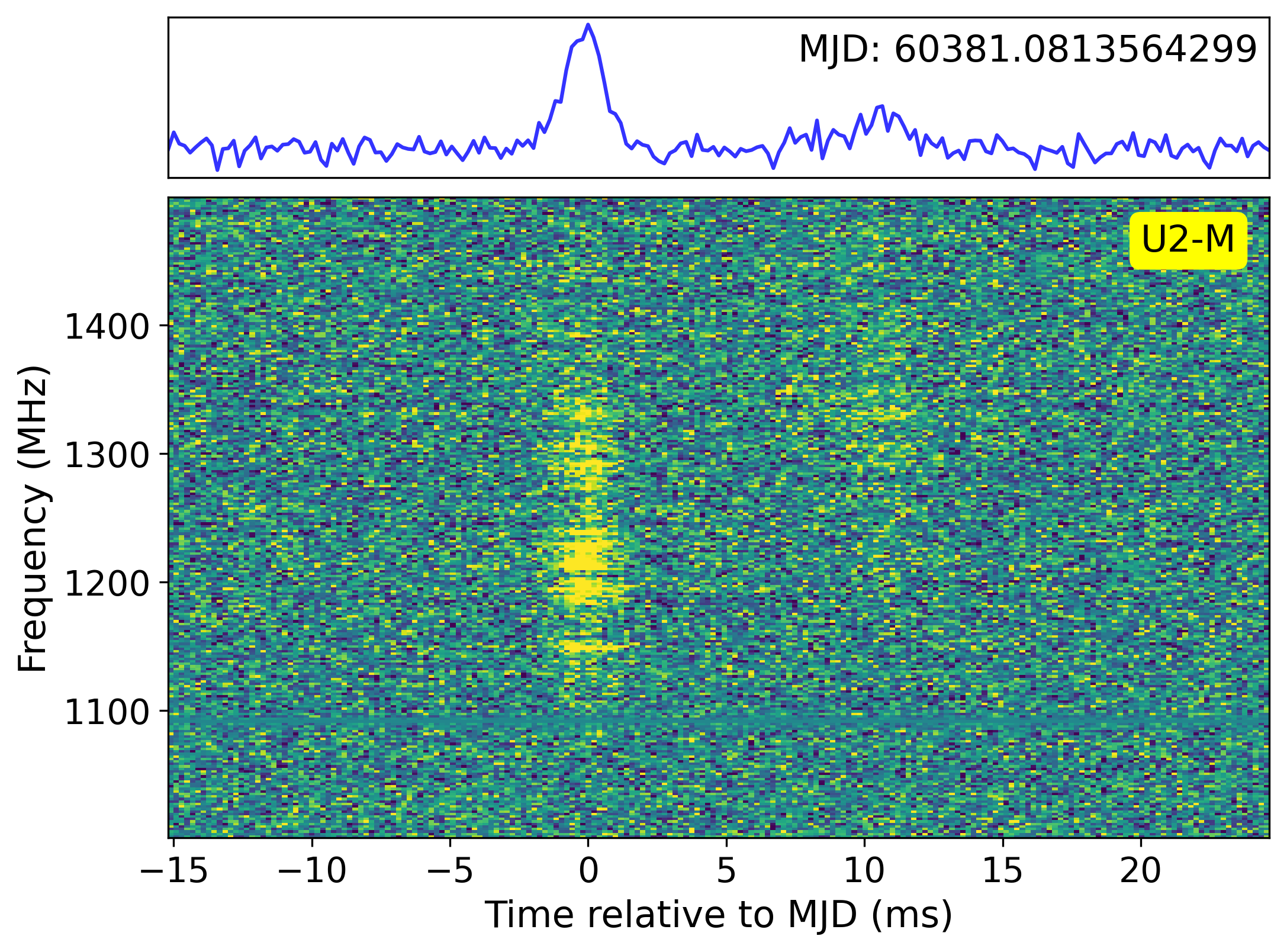}
\end{minipage} &
\begin{minipage}[c]{0.3\textwidth}\centering
\textbf{None (Um-M)}
\end{minipage} \\[2pt]

\begin{minipage}[c]{0.3\textwidth}\centering
\includegraphics[width=\linewidth]{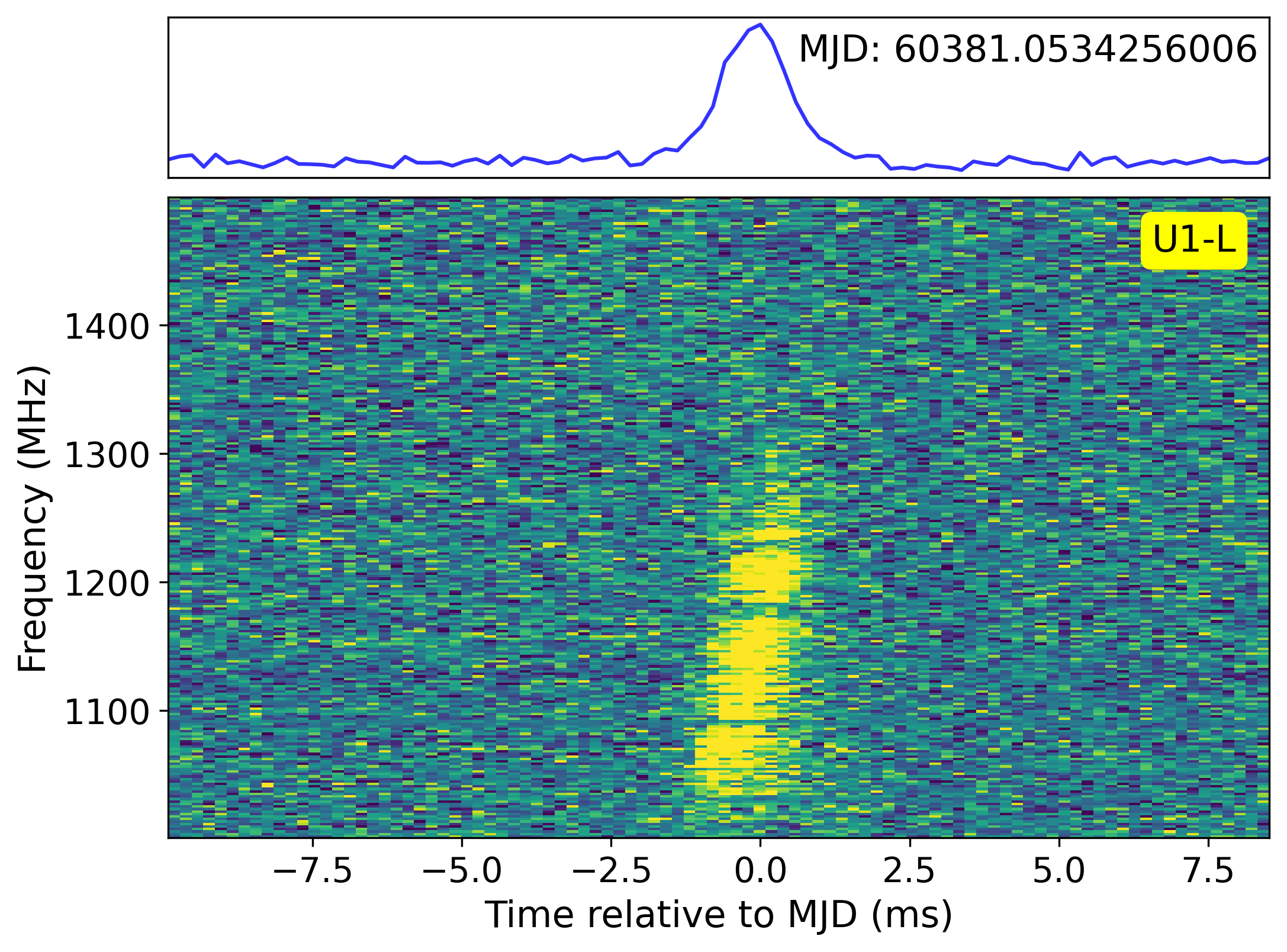}
\end{minipage} &
\begin{minipage}[c]{0.3\textwidth}\centering
\includegraphics[width=\linewidth]{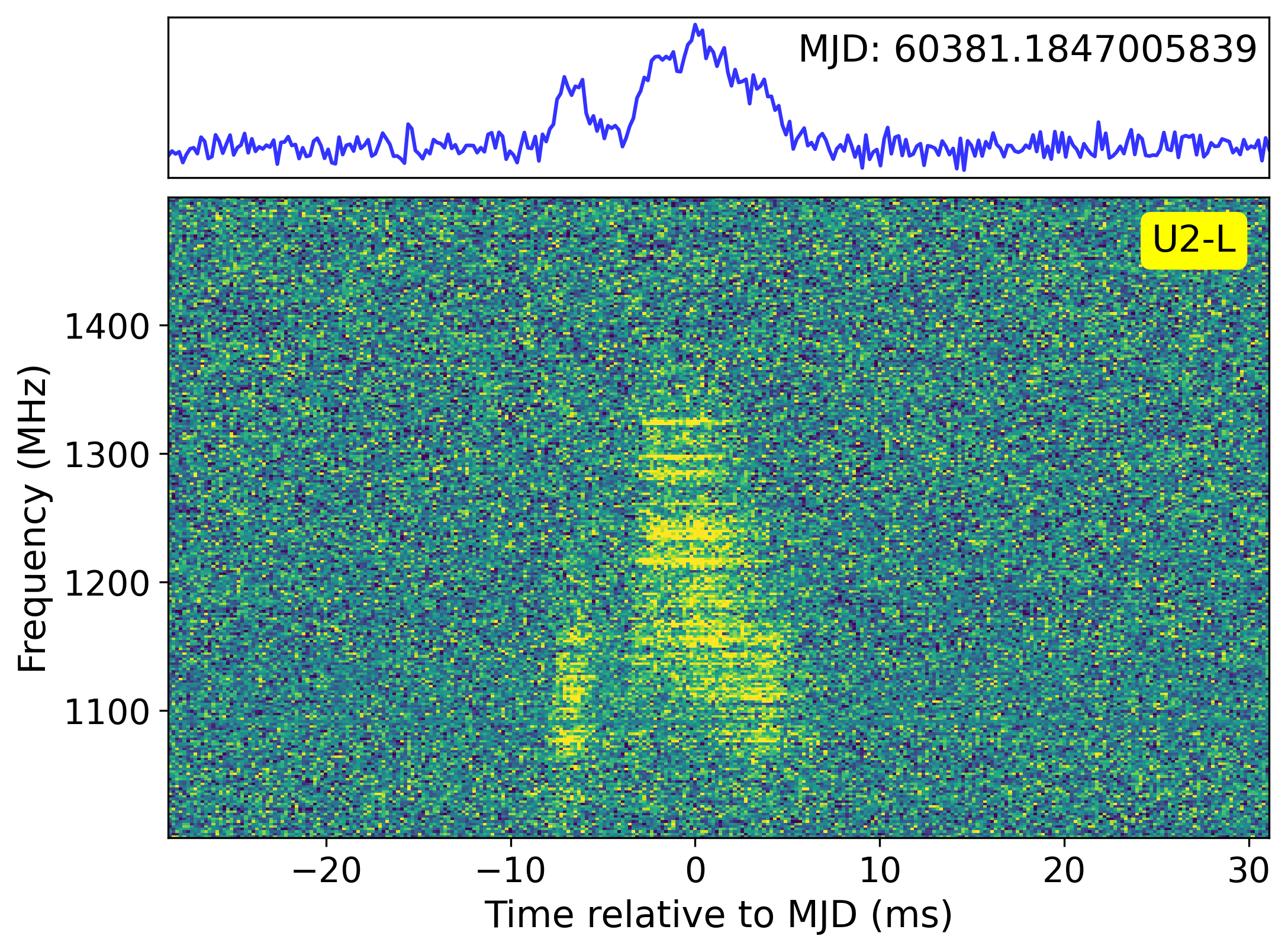}
\end{minipage} &
\begin{minipage}[c]{0.3\textwidth}\centering
\includegraphics[width=\linewidth]{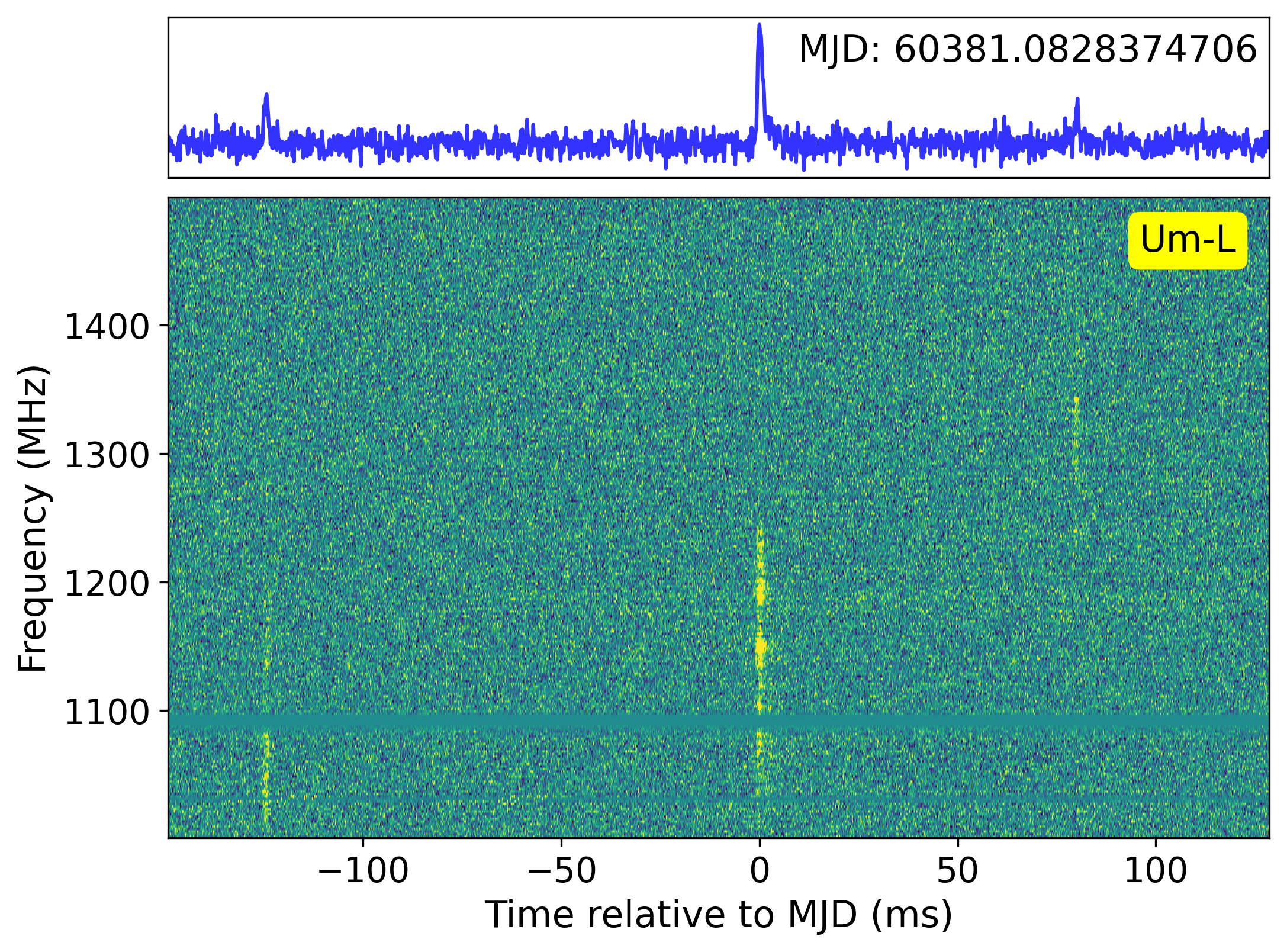}
\end{minipage} \\[2pt]

\begin{minipage}[c]{0.3\textwidth}\centering
\textbf{None (U1-W)}
\end{minipage} &
\begin{minipage}[c]{0.3\textwidth}\centering
\includegraphics[width=\linewidth]{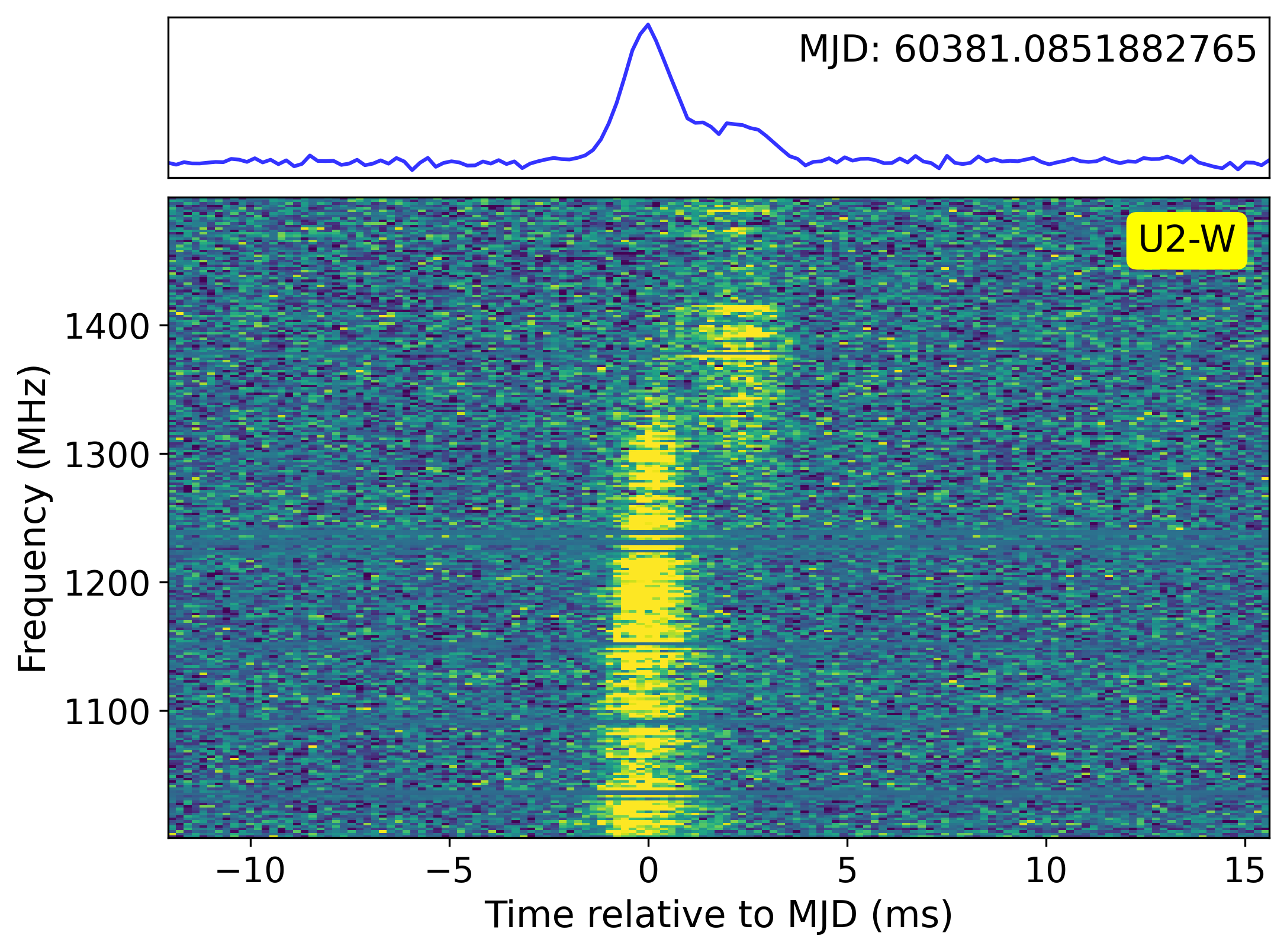}
\end{minipage} &
\begin{minipage}[c]{0.3\textwidth}\centering
\includegraphics[width=\linewidth]{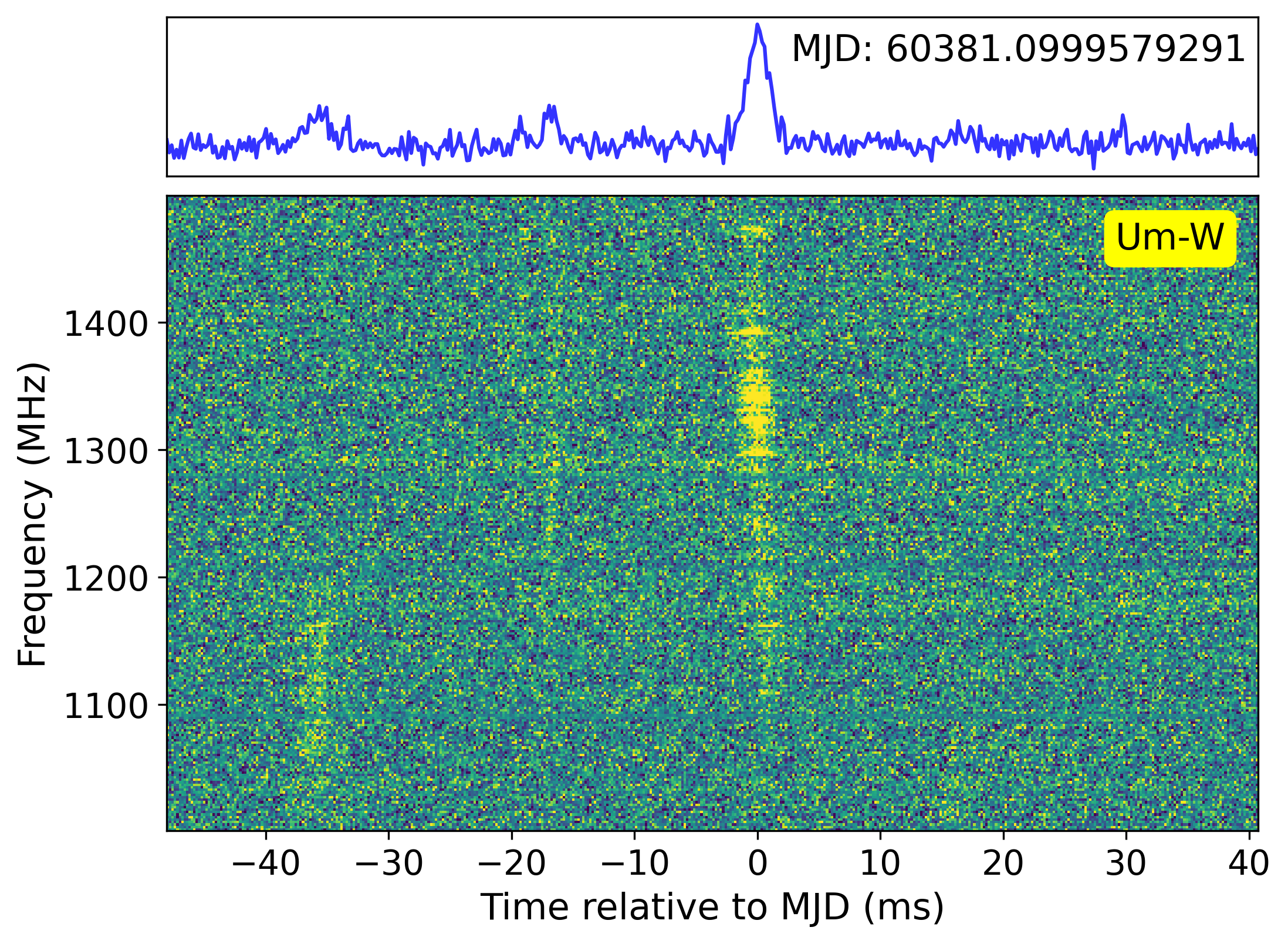}
\end{minipage} \\
\end{tabular}

\caption{The typical dynamic spectra of upward drifting burst-clusters. Each burst-cluster is labeled as Ux-y, where x denotes the number of components (labeled as 'm' if the number exceeds two), and y indicates the observation band: L (low frequency), M (middle frequency), H (high frequency), W (wide frequency). The plots are arranged in a grid of four rows and three columns, with x corresponding to each column and y corresponding to each row. Notably, our sample contains no events classified as Um-M or U1-W.}
\label{upward_drifting_burst-clusters}
\end{figure}

\begin{figure*}[ht]
\centering
\begin{tabular}{ccc}
\adjustbox{valign=t}{\includegraphics[width=0.3\textwidth]{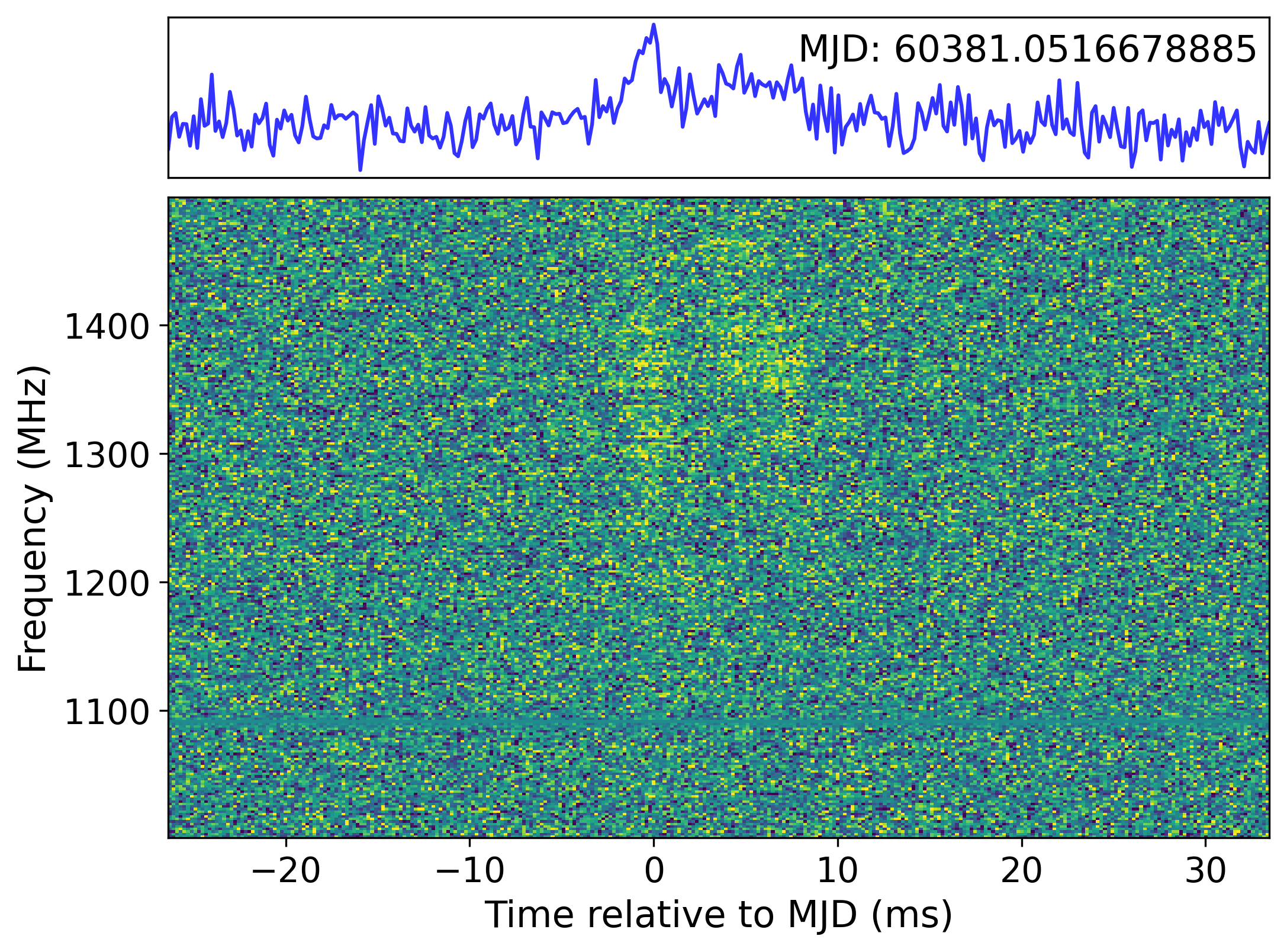}} & 
\adjustbox{valign=t}{\includegraphics[width=0.3\textwidth]{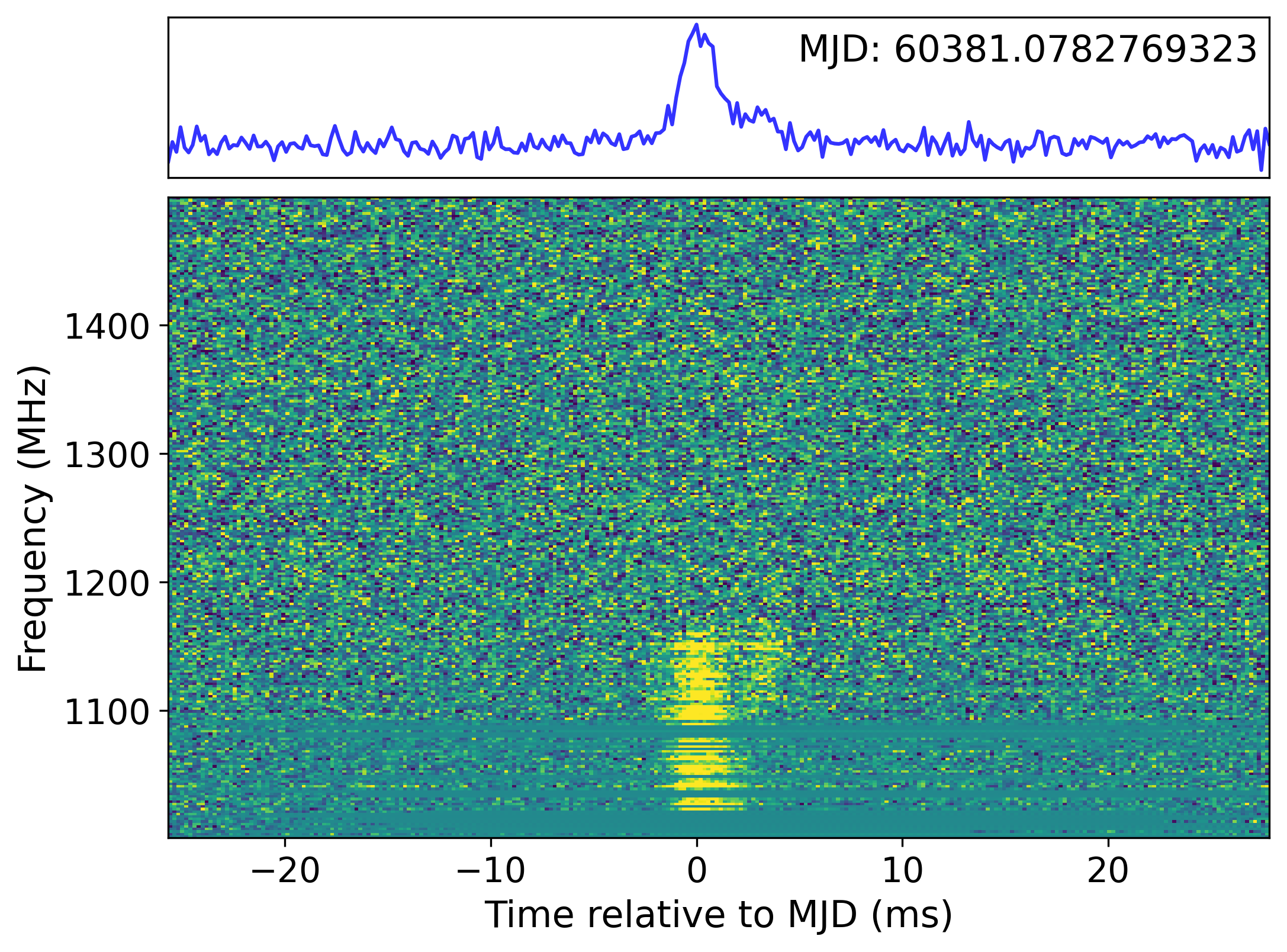}} & 
\adjustbox{valign=t}{\includegraphics[width=0.3\textwidth]{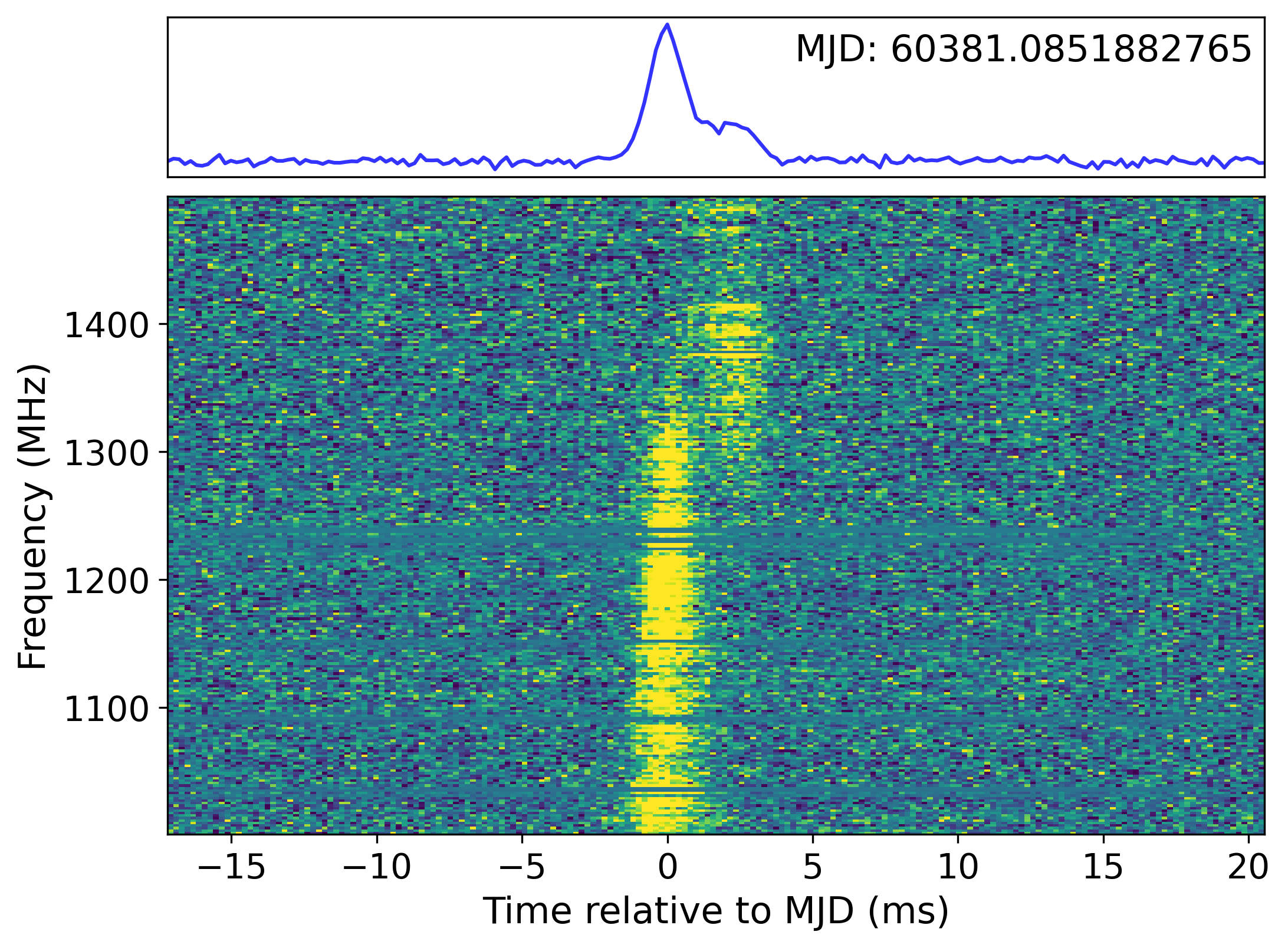}} \\[2mm]

\adjustbox{valign=t}{\includegraphics[width=0.3\textwidth]{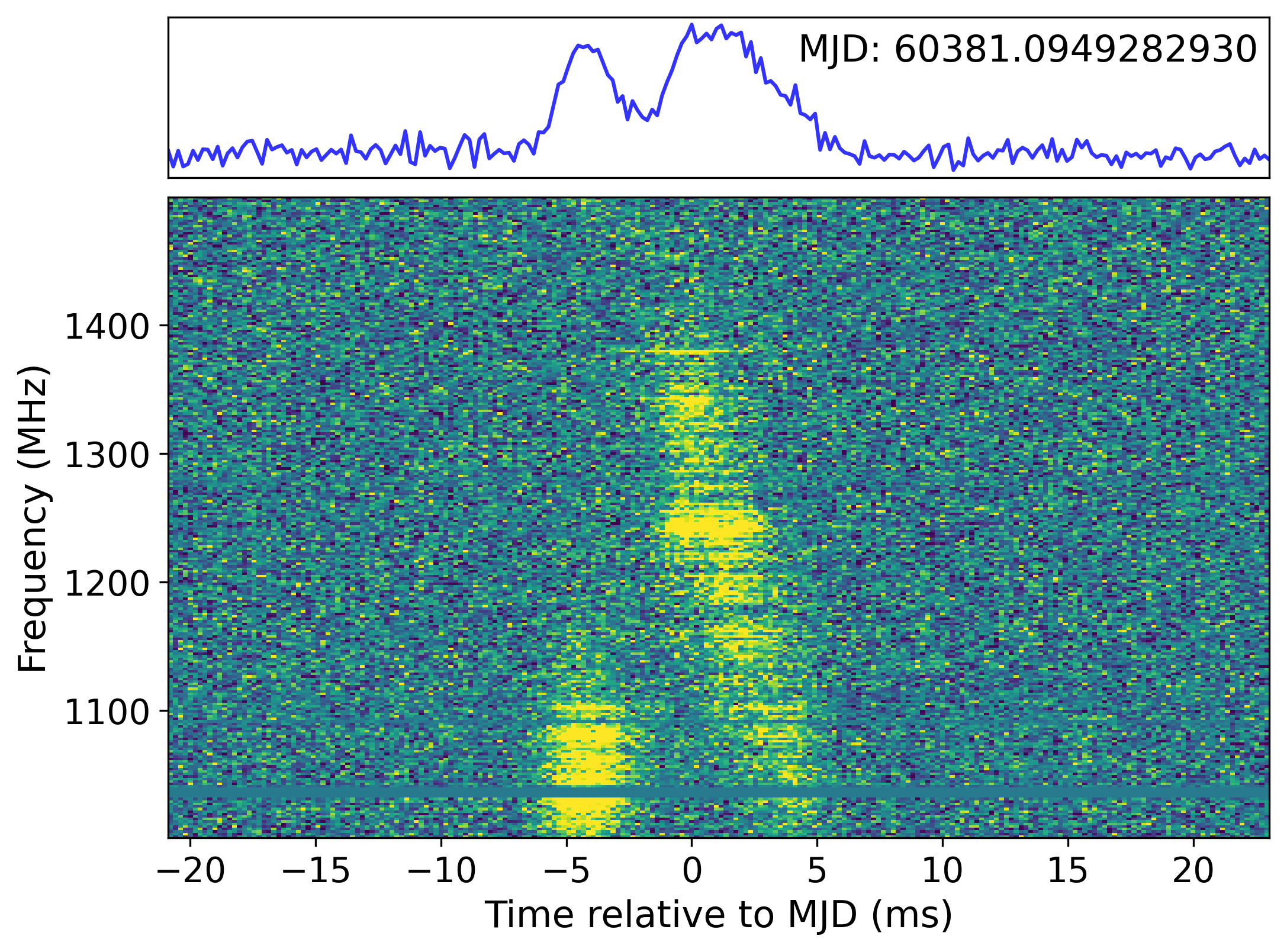}} & 
\adjustbox{valign=t}{\includegraphics[width=0.3\textwidth]{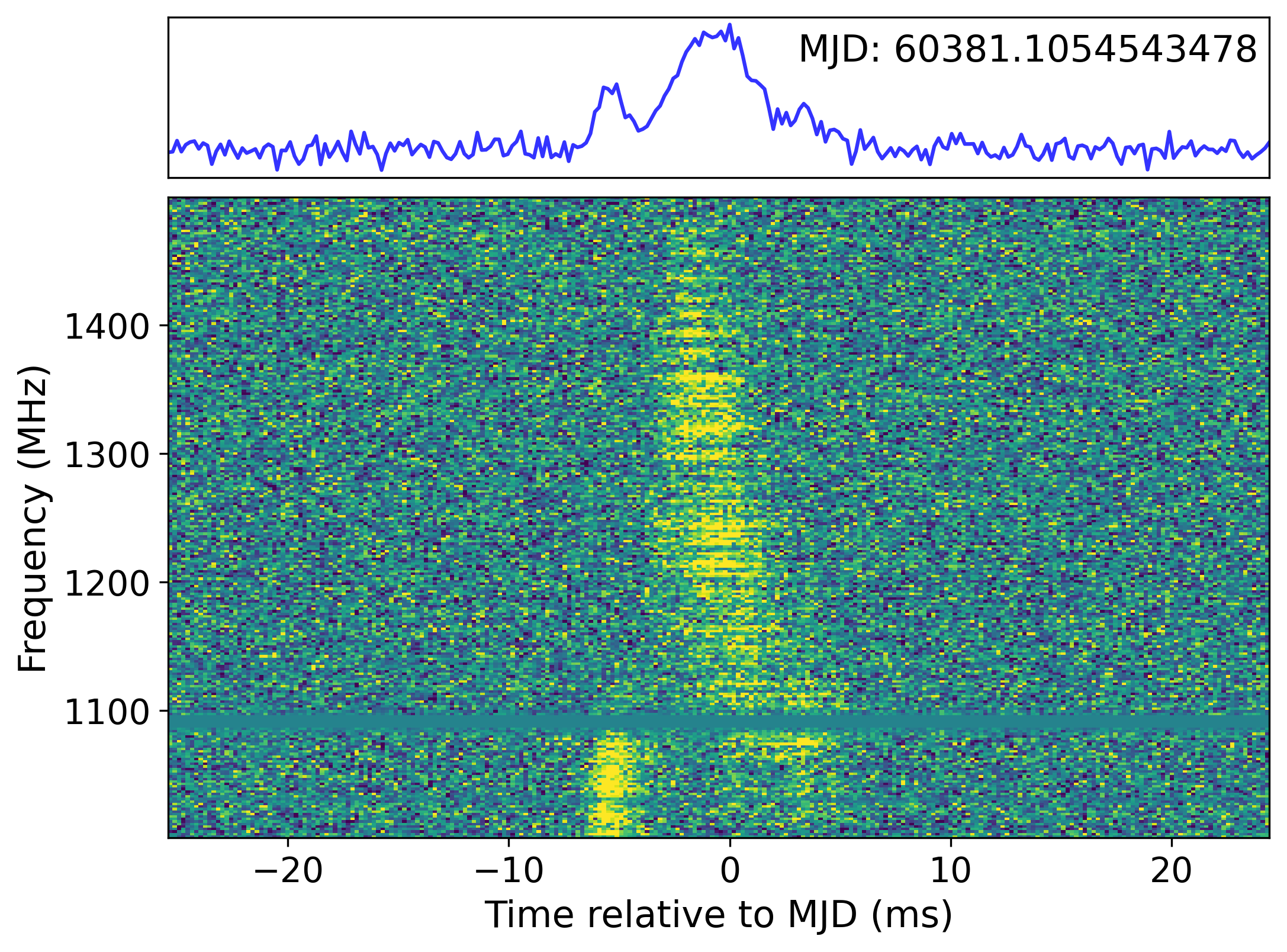}} & 
\adjustbox{valign=t}{\includegraphics[width=0.3\textwidth]{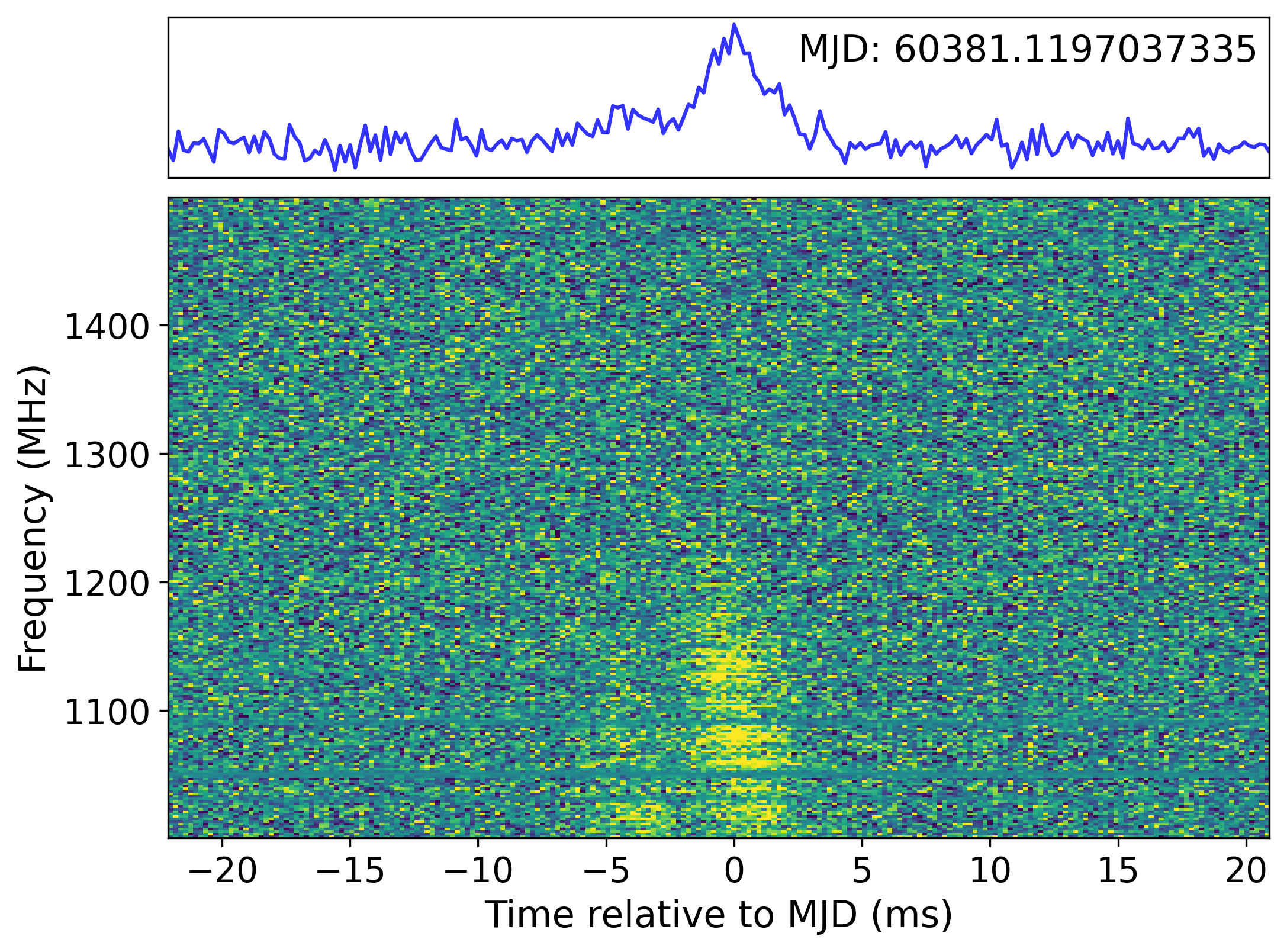}} \\[2mm]

\adjustbox{valign=t}{\includegraphics[width=0.3\textwidth]{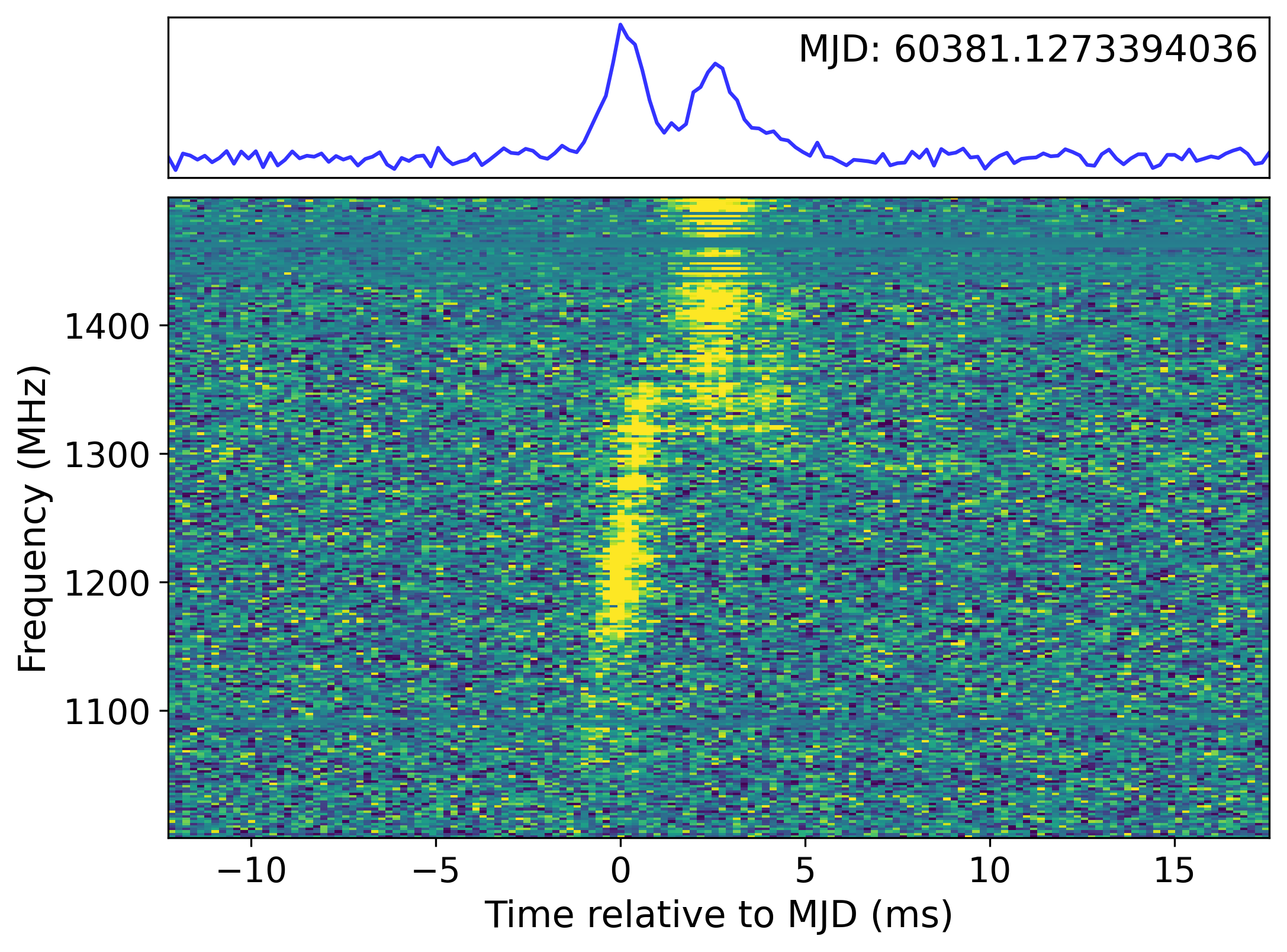}} & 
\adjustbox{valign=t}{\includegraphics[width=0.3\textwidth]{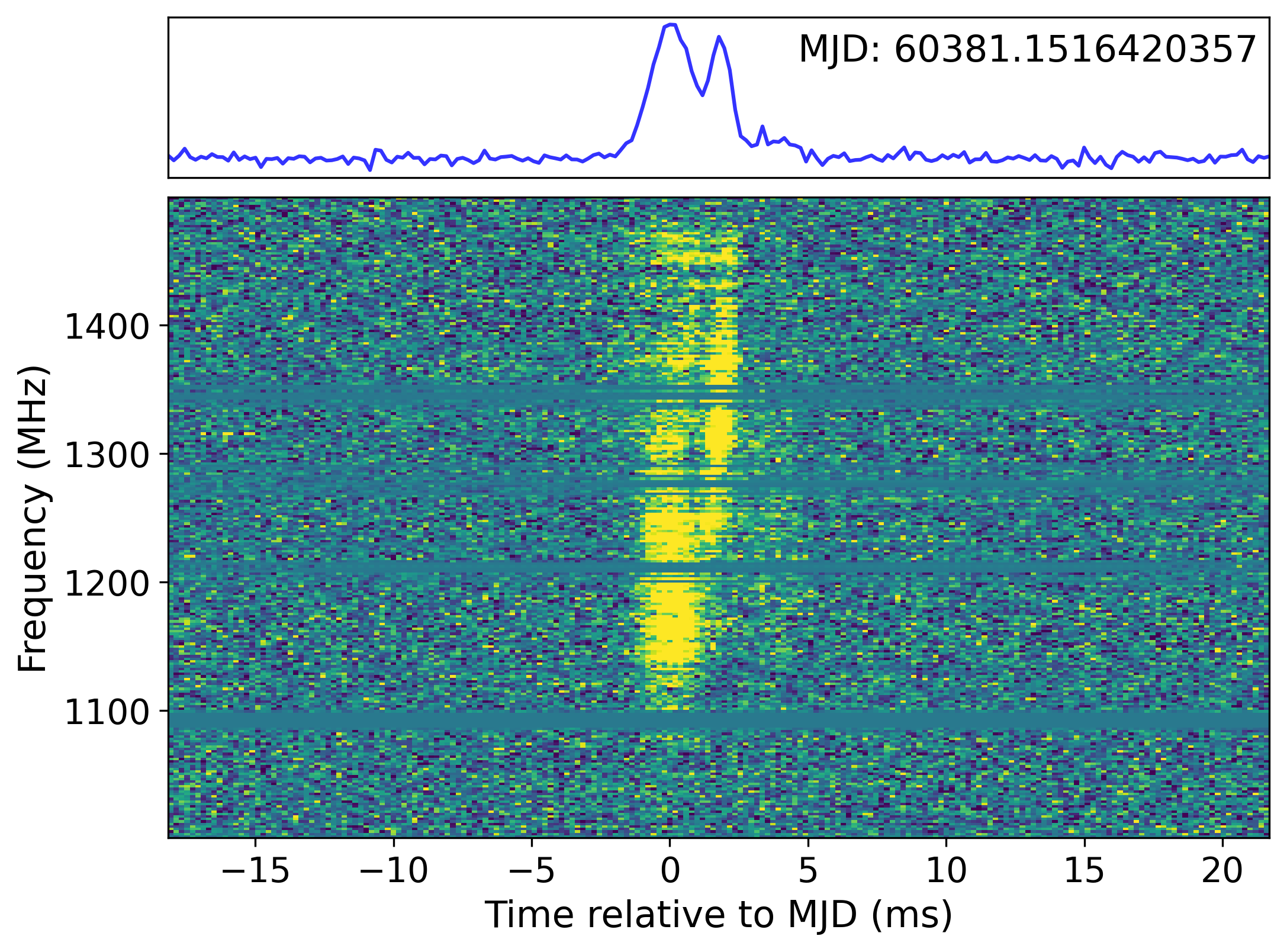}} & 
\adjustbox{valign=t}{\includegraphics[width=0.3\textwidth]{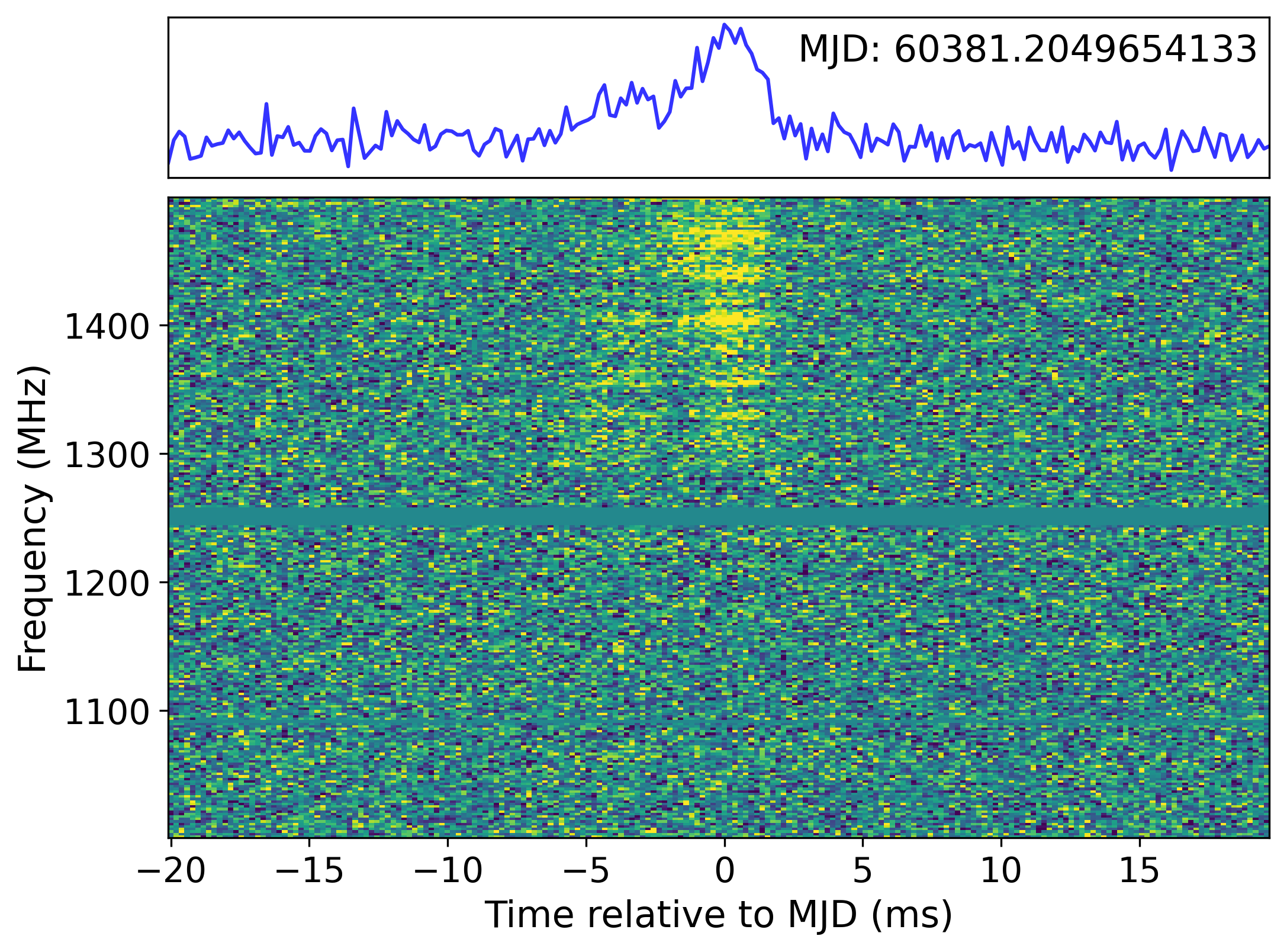}}
\end{tabular}

\caption{The 9 consecutive upward drifting burst-clusters (or upward drifting bursts), shown here, represent the entire dataset used to analyze consecutive time intervals for upward drifting burst-clusters, which is a key focus of this study.}
\label{consecutive_upward_drifting_burst-clusters}
\end{figure*}

Upward drifting burst-clusters refer to cases where the lower-frequency components of the burst-cluster precede the higher-frequency components. These burst-clusters are classified based on the number of components (or sub-bursts) observed in their dynamic spectrum and the position of the frequency bands: high-frequency band (H), middle-frequency band (M), low-frequency band (L), and wide-frequency band (W), as recorded by the FAST. For single-component burst-clusters, this means that the lower-frequency sub-band arrives earlier than the higher-frequency one. In the case of double-component burst-clusters, the sub-burst with the lower peak frequency is detected earlier than the one with the higher peak frequency. For multi-component burst-clusters, the sub-burst with the lowest peak frequency arrives first, while the sub-burst with the highest peak frequency arrives last, allowing for the measurement of the drifting rate. It should be noted that some categories, such as U1-W and Um-M, currently lack samples due to insufficient data. An illustration of these burst-clusters can be seen in Figure~\ref{upward_drifting_burst-clusters}.

As shown in Table \ref{tab:stat_table}, a total of 233 upward drifting burst-clusters were detected. Among them, there were 142 single-component burst-clusters (60.9\%), with the same caveat regarding DM averaging effects as mentioned above for downward drifting cases. Additionally, there were 86 double-component burst-clusters (36.9\%), and 5 multi-component burst-clusters (2.1\%). The distribution across frequency bands is as follows: 45 burst-clusters (19.3\%) in the high-frequency band (H), 47 burst-clusters (20.2\%) in the middle-frequency band (M), 117 burst-clusters (50.2\%) in the low-frequency band (L), and 27 burst-clusters (11.6\%) in the wide-frequency band (W). The number of consecutive time intervals for upward drifting burst-clusters was 90, and the number of intermittent time intervals for upward drifting burst-clusters was 9.

In \cite{2022RAA....22l4001Z}, FRB~20201124A exhibited 263 downward drifting bursts (including single-, double-, and multi-component bursts) alongside 3 upward drifting bursts, all of which were double-component bursts. In contrast, our classification of FRB~20240114A shows that upward drifting burst-clusters account for 23.82\% (233 out of 978) of the total burst-clusters exhibiting drifting behavior. However, the drifting mode of single-component burst-clusters is significantly affected by the choice of DM, while double- and multi-component burst-clusters are not subject to this influence. If we exclude single-component burst-clusters with upward drifting, the number of upward drifting double- and multi-component burst-clusters still reaches 91, accounting for 10.89\% of the total sample of 836 burst-clusters exhibiting drifting behavior (excluding single-component burst-clusters with upward drifting). Furthermore, if we consider only upward drifting bursts, specifically those upward drifting burst-clusters consisting solely of consecutive time intervals, this number drastically decreases to just 9, as illustrated in Figure~\ref{consecutive_upward_drifting_burst-clusters}. This proportion is comparable to that observed in \cite{2022RAA....22l4001Z}. Upward drifting bursts, though rare, have theoretical grounding in magnetospheric trigger dynamics. Sparks triggered earlier may occur along field lines with a larger radius of curvature \citep{2020ApJ...899..109W}. Propagation effects such as interstellar scintillation can enhance the detection of upward drifting features through temporary intensity amplification \citep{1990ARA&A..28..561R}, although intrinsic emission mechanisms remain dominant.

\subsection{No Drifting burst-clusters}

\begin{figure*}[ht]
\centering
\begin{tabular}{ccc}
\adjustbox{valign=t}{
  \begin{overpic}[width=0.3\textwidth]{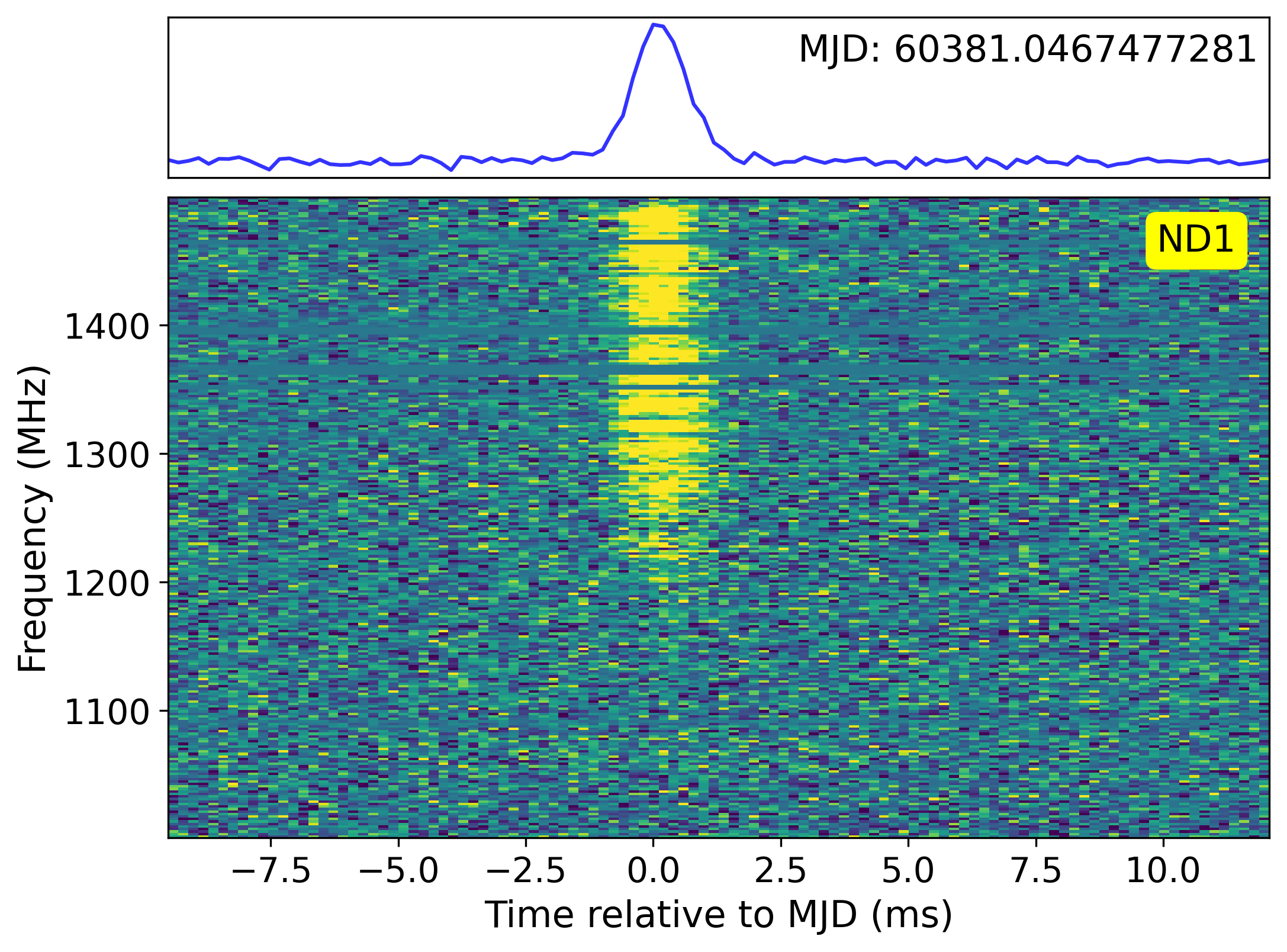}
    \put(4,65){\small (a)}
  \end{overpic}
} &
\adjustbox{valign=t}{
  \begin{overpic}[width=0.3\textwidth]{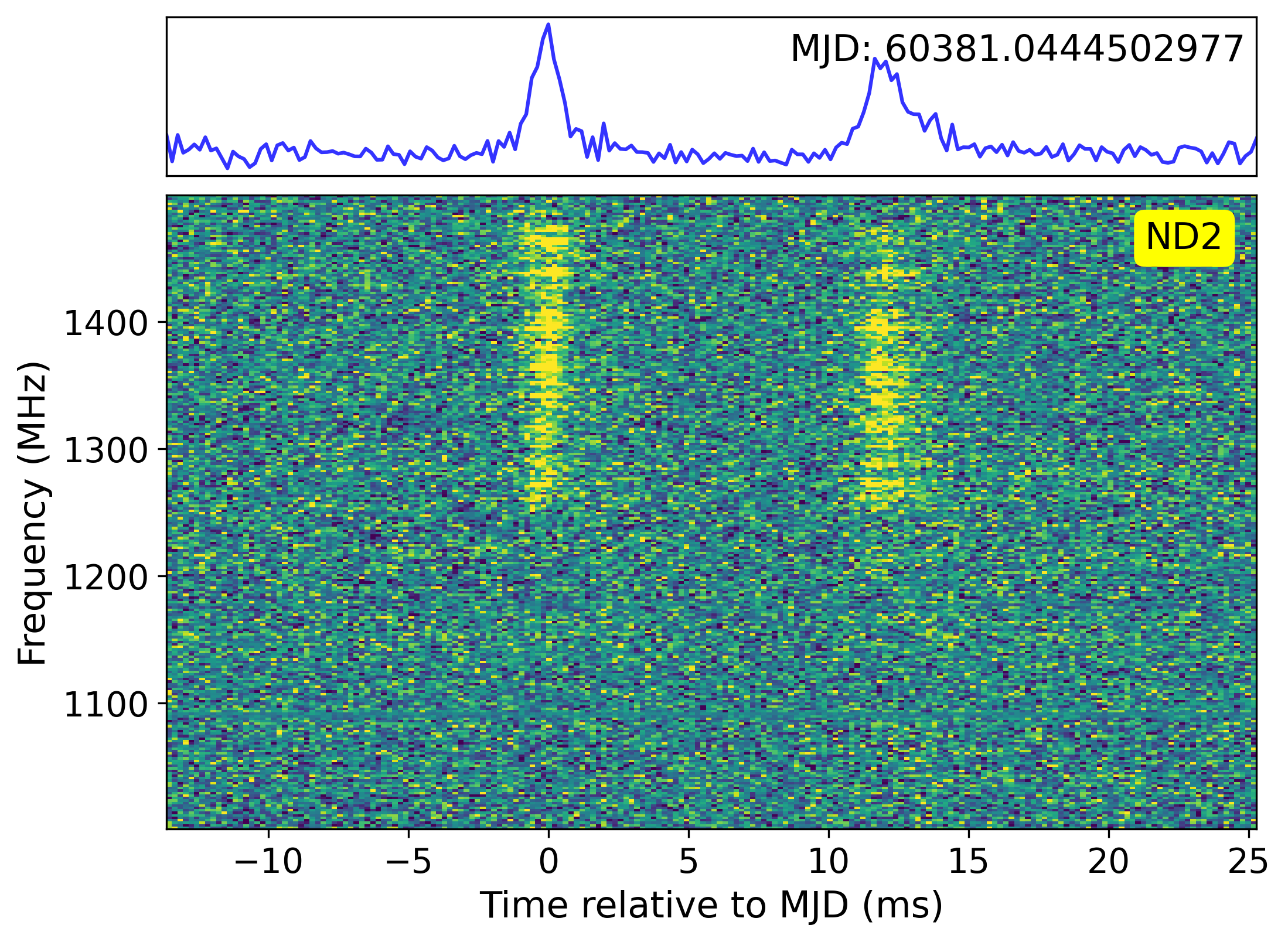}
    \put(4,65){\small (b)}
  \end{overpic}
} &
\adjustbox{valign=t}{
  \begin{overpic}[width=0.3\textwidth]{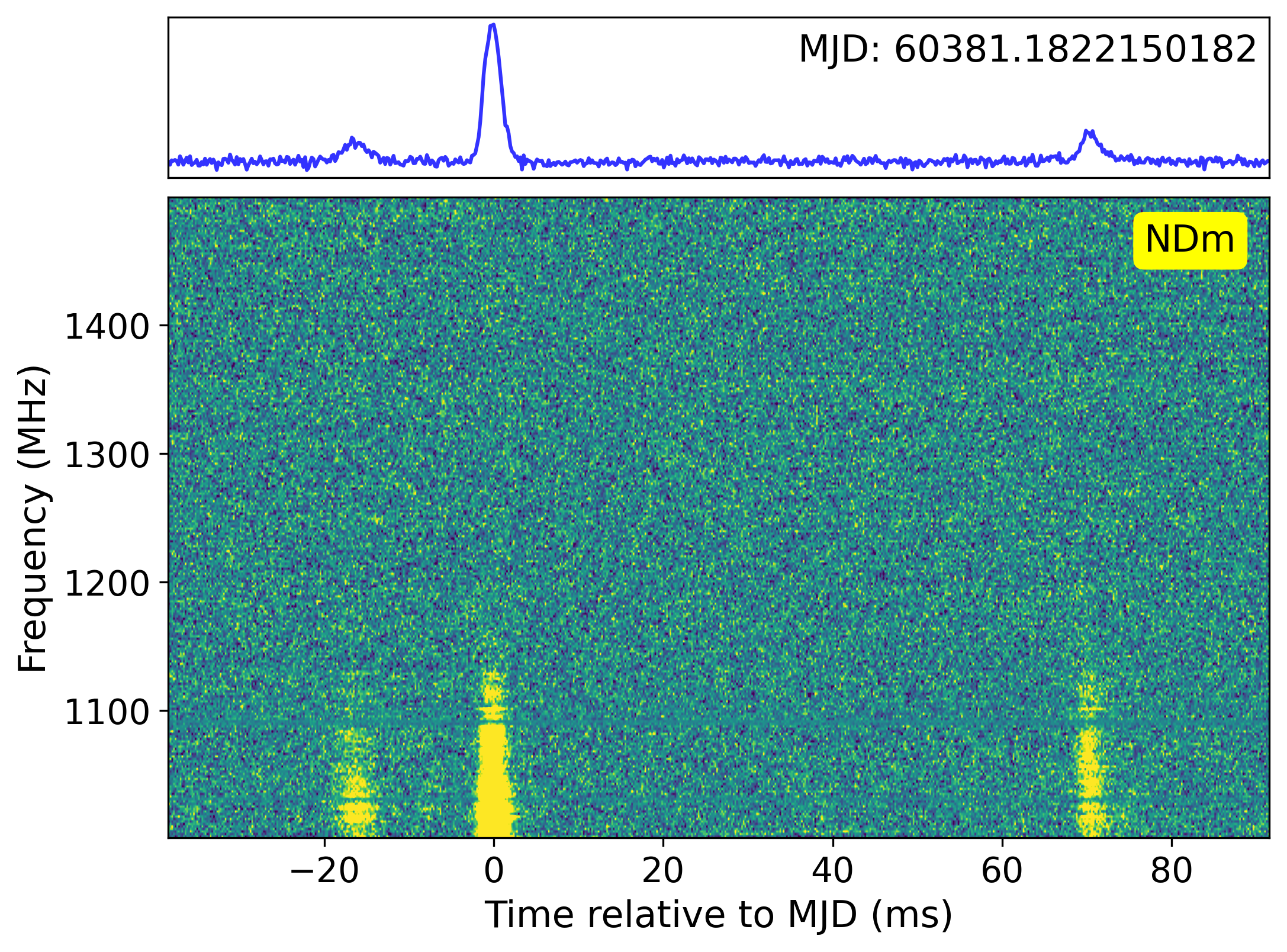}
    \put(4,65){\small (c)}
  \end{overpic}
} \\[2mm]

\adjustbox{valign=t}{
  \begin{overpic}[width=0.3\textwidth]{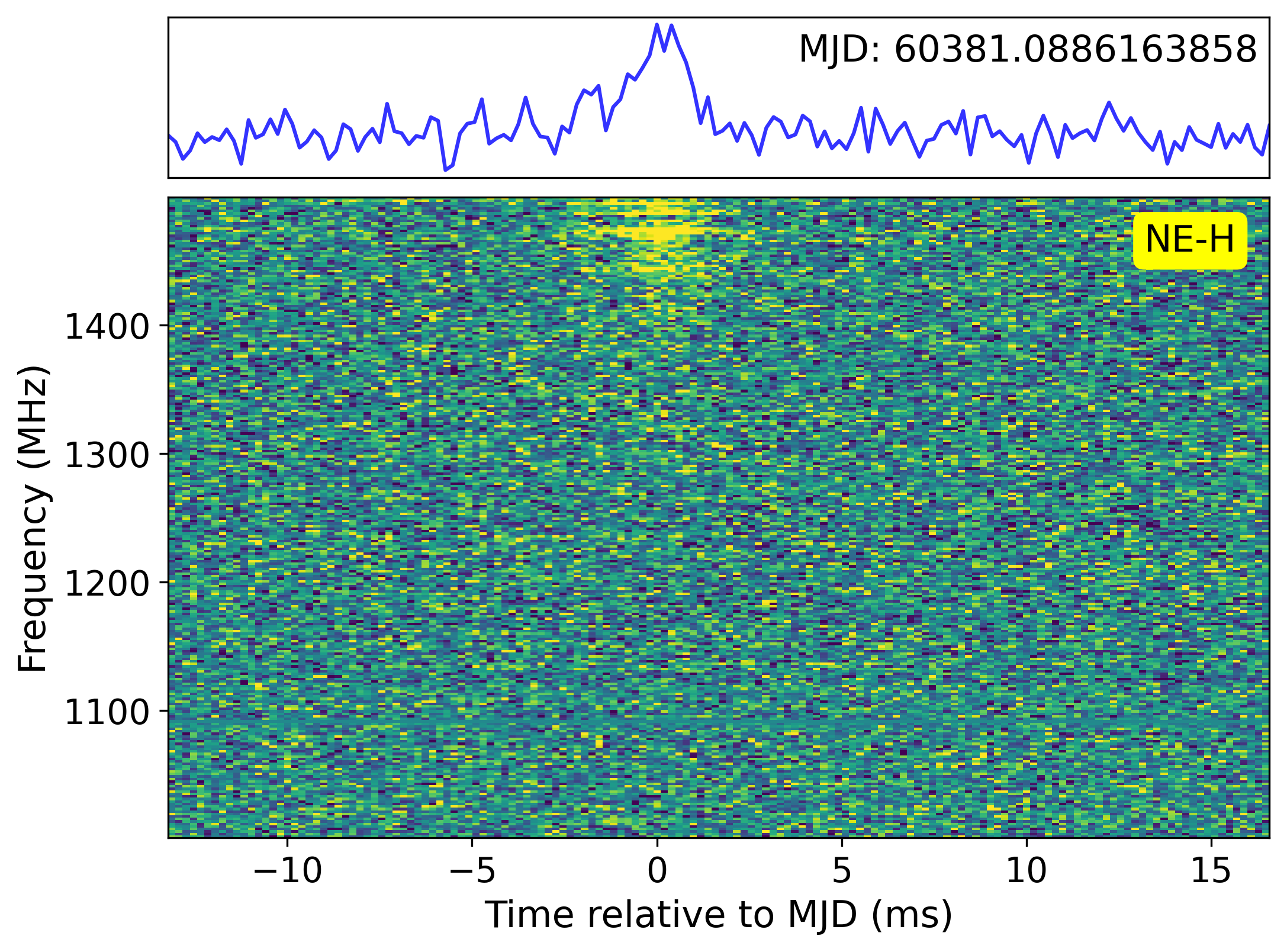}
    \put(4,65){\small (d)}
  \end{overpic}
} &
\adjustbox{valign=t}{
  \begin{overpic}[width=0.3\textwidth]{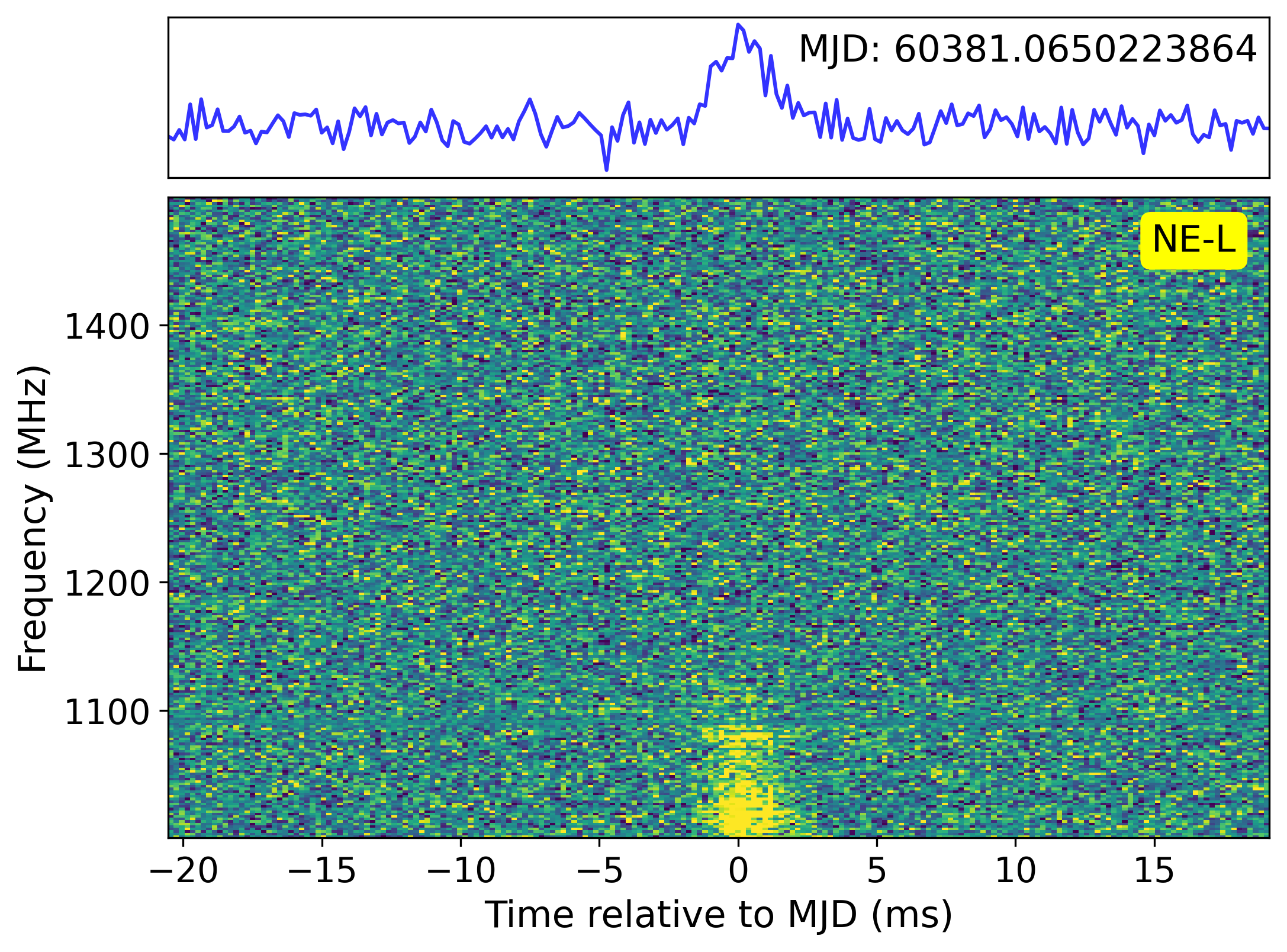}
    \put(4,65){\small (e)}
  \end{overpic}
} &
\adjustbox{valign=t}{
  \begin{overpic}[width=0.3\textwidth]{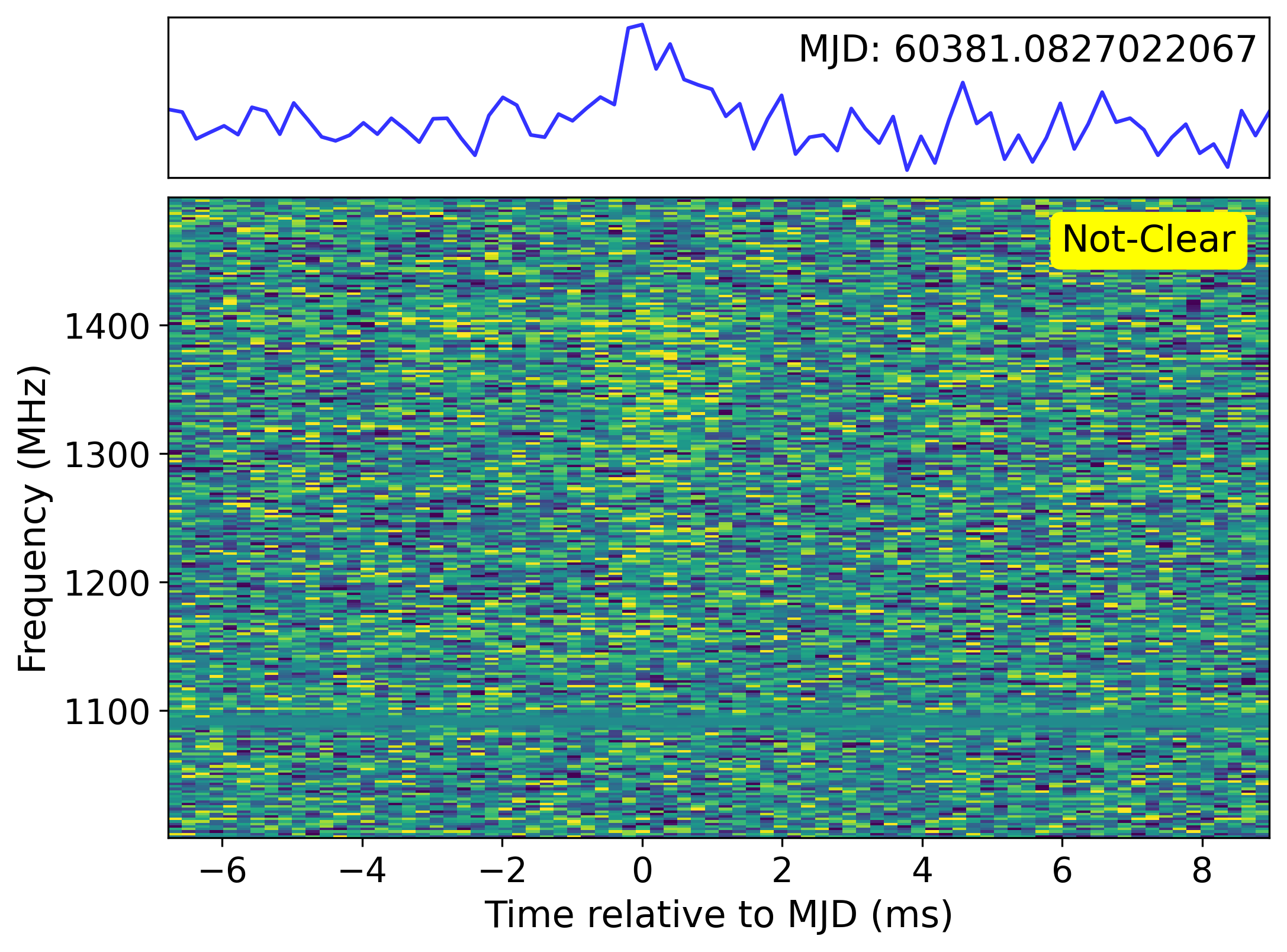}
    \put(4,65){\small (f)}
  \end{overpic}
} \\[2mm]
\end{tabular}

\caption{Representative examples of burst-cluster classifications. The first row shows No Drifting burst-clusters: (a) ND1, a single-component burst-cluster with a stable emission band and no frequency drift, (b) ND2, a double-component burst-cluster with two components showing no drift, and (c) NDm, a multi-component burst-cluster with parallel components. The second row illustrates burst-clusters with no evidence for drifting: (d) NE-H, a high-frequency burst-cluster with insufficient bandwidth to determine drift, (e) NE-L, a low-frequency burst-cluster with insufficient bandwidth to determine drift, and (f) Not-Clear, a burst-cluster with no clear pattern due to low signal-to-noise ratio or interference.}
\label{no drifting burst-clusters}
\end{figure*}

ND burst-clusters are burst-clusters with dynamic spectra that show a distinctly vertical pattern, indicating that the frequency does not drift with a stable emission band, regardless of whether they occur in the high-, middle-, or low-frequency regions. This absence of drifting makes them easily distinguishable from drifting burst-clusters.

Further classification of ND burst-clusters has led to the identification of three sub-classes. ND1 burst refers to single-component burst-cluster with a clear vertical pattern and a stable emission band, without frequency drifting ($\Delta \nu / \Delta t \approx \infty$). 
However, as the drifting rate is affected by the value of DMs, a slightly different DM could cause the ND1 to transition to either downward or upward drifting with a high absolute drifting rate, and vise versa. One should be cautious regarding the statistics of the ND1 bursts. ND2 and NDm burst-clusters represent double- and multi-component burst-clusters, respectively, which exhibit similar behavior but consist of two or more emission components that are arranged in parallel with approximately equal bandwidths ($\Delta \nu / \Delta t \approx 0$). The total number of ND burst-clusters is 720. Figure~\ref{no drifting burst-clusters} illustrates examples of these sub-classes, where the first row showcases representative cases of ND1, ND2, and NDm. The number of sub-classes is as follows: ND1 (670 burst-clusters, 93.1\%), ND2 (49 burst-clusters, 6.8\%), and NDm (1 burst-cluster, 0.1\%). Additionally, there are 9 consecutive time intervals and 42 intermittent time intervals for No Drifting burst-clusters.

\subsection{burst-clusters with No Evidence for Drifting}
Certain burst-clusters are observed near the edges of the FAST frequency band, either at the lower or upper boundary. Due to the narrow emission bandwidth of these burst-clusters, it is challenging to identify any distinct frequency drifting patterns. As a result, these burst-clusters are classified as having ``No Evidence" for drifting. To provide a more refined categorization, they are further divided into two subcategories: NE-L (Low-frequency burst-clusters with no evidence of drifting), where the emission is observed in the lower portion of the FAST frequency band, and NE-H (High-frequency burst-clusters with no evidence of drifting), where the emission occurs in the upper part of the band. As shown in Figure~\ref{no drifting burst-clusters}, the left and middle columns of the second row represent NE-H and NE-L burst-clusters, respectively.

\subsection{burst-clusters with Not-Clear}
Burst-clusters with not clear patterns exhibit the following characteristics: either the signal-to-noise ratio (SNR) is too low, causing excessive background noise fluctuations that obscure the burst-cluster's structure, or the RFI is too strong, severely affecting the signal quality and preventing morphological classification or drifting rate measurement. As shown in Figure~\ref{no drifting burst-clusters}, the right column of the second row illustrates a burst-cluster with Not-Clear pattern.

\subsection{Complex burst-clusters}
These burst-clusters display fascinating features with complex structures in their waterfall plots, such as combinations of upward and downward drifts, or some sub-bursts within certain bursts exhibiting very intricate shapes. The diversity and intricacy of these burst-clusters make it worthwhile to review all the plots in extended material.

\section{Statistical relations Between Drifting Rate and Other Physical Quantities}
\label{sec:drift}

Downward drifting is a common characteristic of repeating FRBs observed across different telescopes and frequency bands. In FRB~20121102A, this phenomenon has been consistently reported in bursts spanning a wide frequency range \citep{Hessels_2019, 2019ApJ...882L..18J, 2021MNRAS.505.3041P}. Similar drifting patterns were identified in FRB~20180814A, where sub-bursts exhibited progressive shifts to lower frequencies within the CHIME band \citep{2019Natur.566..235C,2020ApJ...896L..41C}. Observations of FRB~20180916B using simultaneous LOFAR (110–190\,MHz) and Apertif (1220–1520\,MHz) data confirmed that its characteristic sub-burst downward drifting behavior persists for a decade of radio frequency, with detections down to at least 120\,MHz \citep{2021ApJ...911L...3P, 2021Natur.596..505P}, while CHIME observations revealed similar drifting structures in the 400–800\,MHz band \citep{2020ApJ...891L...6F}. While early studies of FRB~20190711 suggested downward drifting features \citep{2020MNRAS.497.3335D}, subsequent analyses revealed ambiguities in burst classification, mainly due to the burst's complex temporal and spectral features, including multiple sub-bursts with varying polarization and scattering effects, which made it difficult to clearly classify it as a repeating or non-repeating burst. FRB~20201124A displayed complex time-frequency structures with consistent downward drifting across multiple observational campaigns \citep{2021MNRAS.508.5354H, 2022RAA....22l4001Z}.

These studies demonstrate significant variations in drifting rates among different FRBs, potentially correlated with other burst parameters. Notably, \cite{2021MNRAS.507..246C} identified an inverse relation between drifting rates and sub-burst durations, suggesting a unified mechanism governing these observed properties. Previous studies reported drifting rates of order $-200$ MHz ms$^{-1}$ in repeating FRB~20121102A \citep{2019ApJ...876L..23H}, ranging from $-1$ to $-60$ MHz ms$^{-1}$ in FRB~20180916B \citep{2021ApJ...911L...3P}, and from $-5$ to $-166$ MHz ms$^{-1}$ in FRB~20201124A \citep{2022RAA....22l4001Z}. Our study of FRB~20240114A demonstrates that its drifting rates span an even broader range than previously reported. Part of this can be attributed to the fact that our drifting rate measurements are based on burst-clusters, which may extend the range of drifting rates toward values closer to zero for the double- and multi-component burst-clusters, as discussed in Section~\ref{Drifting_Rate}. However, this approach does not affect the extremes of positive or negative infinity for all the burst-clusters.

\subsection{Drifting Rate Analysis}
\label{Drifting_Rate}

\begin{figure}[ht]
    \centering
    \begin{minipage}{0.35\textwidth}  
        \centering
        \includegraphics[width=\textwidth]{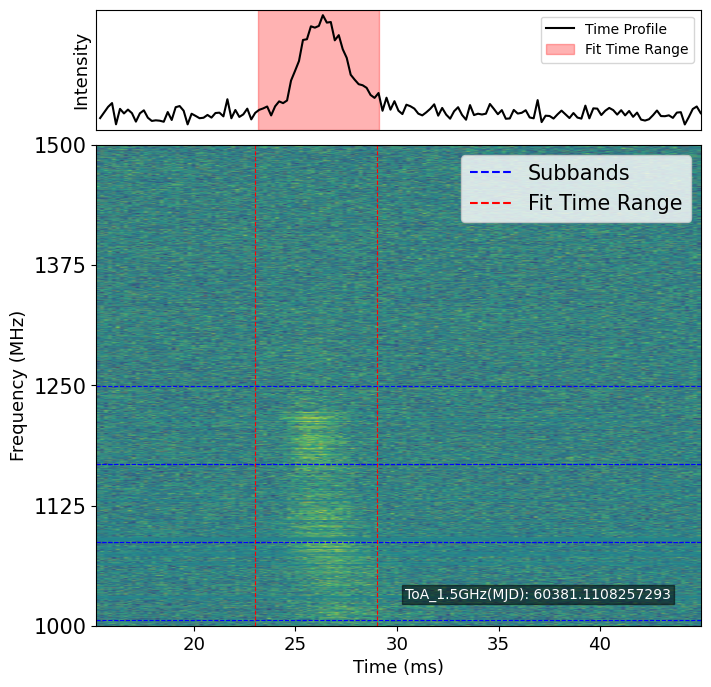}
    \end{minipage}  
    \hspace{0.0\textwidth}  
    \begin{minipage}{0.6\textwidth}  
        \centering
        \includegraphics[width=\textwidth]{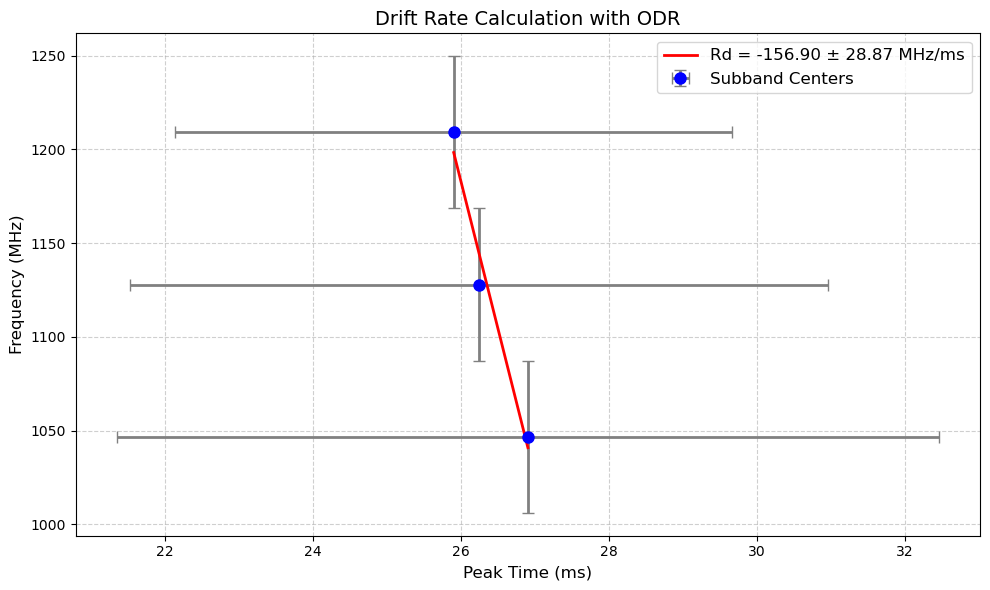}
    \end{minipage}
    \caption{
    Example of the drifting rate calculation for a single-component burst-cluster. The left panel displays the dynamic spectrum with the emission bandwidth evenly divided into three sub-bands. Gaussian fits are applied separately to each sub-band to determine the corresponding peak times and peak frequencies. The drifting rate is then computed using ODR, a method that minimizes the sum of the squared orthogonal distances between the data points and the fitted model, thereby accounting for uncertainties in both time and frequency measurements. In this example, the measured drifting rate is $R_d = -156.90 \pm 28.37\,\mathrm{MHz\,ms}^{-1}$.
    }
    \label{single_pulse_drifting}
\end{figure}

\begin{figure}[ht]
    \centering    \includegraphics[height=0.5\textwidth]{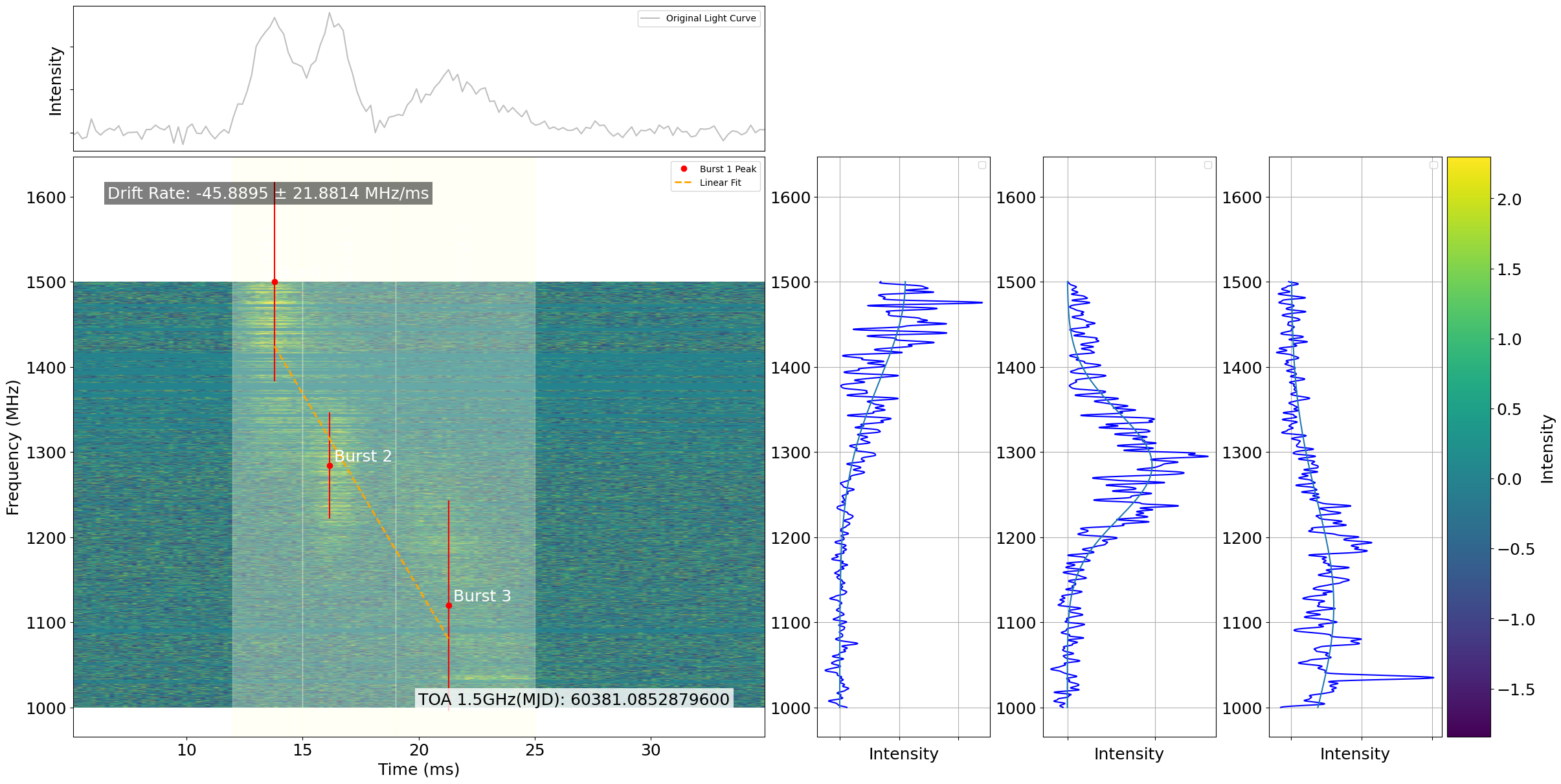}
    \caption{Drifting rate calculation for a multi-component burst-cluster. For each sub-burst, an optimal time window is manually selected. Within this window, time integration and Gaussian fitting are performed to determine the peak frequency, and the drifting rate is subsequently derived by combining the peak times of the sub-bursts. In this example, the measured drifting rate is $R_d = -45.89 \pm 21.88\,\mathrm{MHz\,ms}^{-1}$. A similar procedure is applied to double-component burst-clusters.}    \label{more_than_single_pulse_drifting}
\end{figure}

Following \cite{2022RAA....22l4001Z}, we define the drifting rate as:
\begin{equation}
    R_{\rm d}=\frac{{\rm d} \nu}{{\rm d} t}.
\end{equation}
We categorize burst-clusters into single-, double-, and multi-component burst-clusters, applying different methodologies to determine their drifting rates. 
The drifting of single-component bursts can be considered as intra-burst drifting, while the drifting of both the double- and multi-component burst-clusters can be considered as inter-burst drifting.

For single-component burst-clusters, the emission bandwidth is evenly divided into three sub-bands as a sampling strategy to obtain multiple independent frequency–time centroids. Within each sub-band, Gaussian fitting is applied to determine both the peak time and peak frequency. The drifting rate is then derived from these values using Orthogonal Distance Regression (ODR), as illustrated in Figure~\ref{single_pulse_drifting}. For double- and multi-component burst-clusters, we manually select an appropriate time window for each sub-burst. Within the chosen window, time integration is performed, followed by a Gaussian fitting to extract the peak frequency. The drifting rate is then determined by combining the peak times of each sub-burst, as shown in Figure~\ref{more_than_single_pulse_drifting}. It is important to note that the drifting rate $R_d$ is most intuitive for double- and multi-component burst-clusters, where the unit is MHz\,ms$^{-1}$ and a larger $|R_{\rm d}|$ corresponds to a more pronounced drifting behavior. In contrast, for the single-component burst-clusters, when the drifting rate approaches infinity, it indicates that there is no drifting.

\subsection{Statistical relations Between Peak Frequency, Drifting Rate, Bandwidth, Effective Width and Energy}
\label{Statistical_relations}

\begin{figure}[ht]
    \centering
    \includegraphics[width=\textwidth]{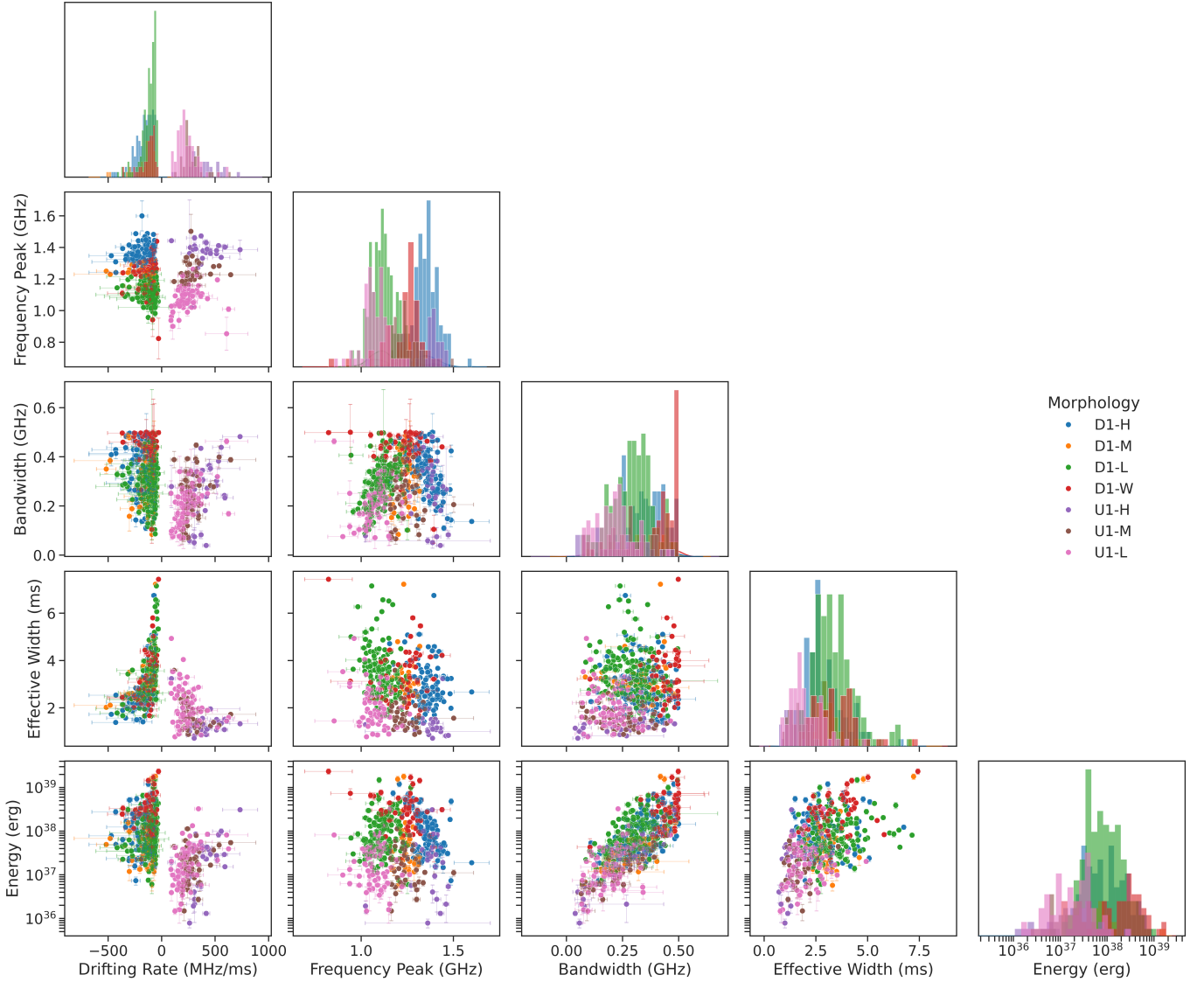}
    \caption{
        This scatter matrix visualizes the relations between the peak frequency, drifting rate, bandwidth, effective width, and energy for the single-peak dataset. The lower triangle contains pairwise scatter plots, and the diagonal displays the distribution of each variable. The different colors represent various morphological types (D1-H, D1-M, D1-L, D1-W for downward drifting and U1-H, U1-M, U1-L, U1-W for upward drifting), helping to differentiate the data samples.
    }
    \label{scatter_plot_Y1}
\end{figure}

\begin{figure}[ht]
    \centering    \includegraphics[width=0.8\textwidth]{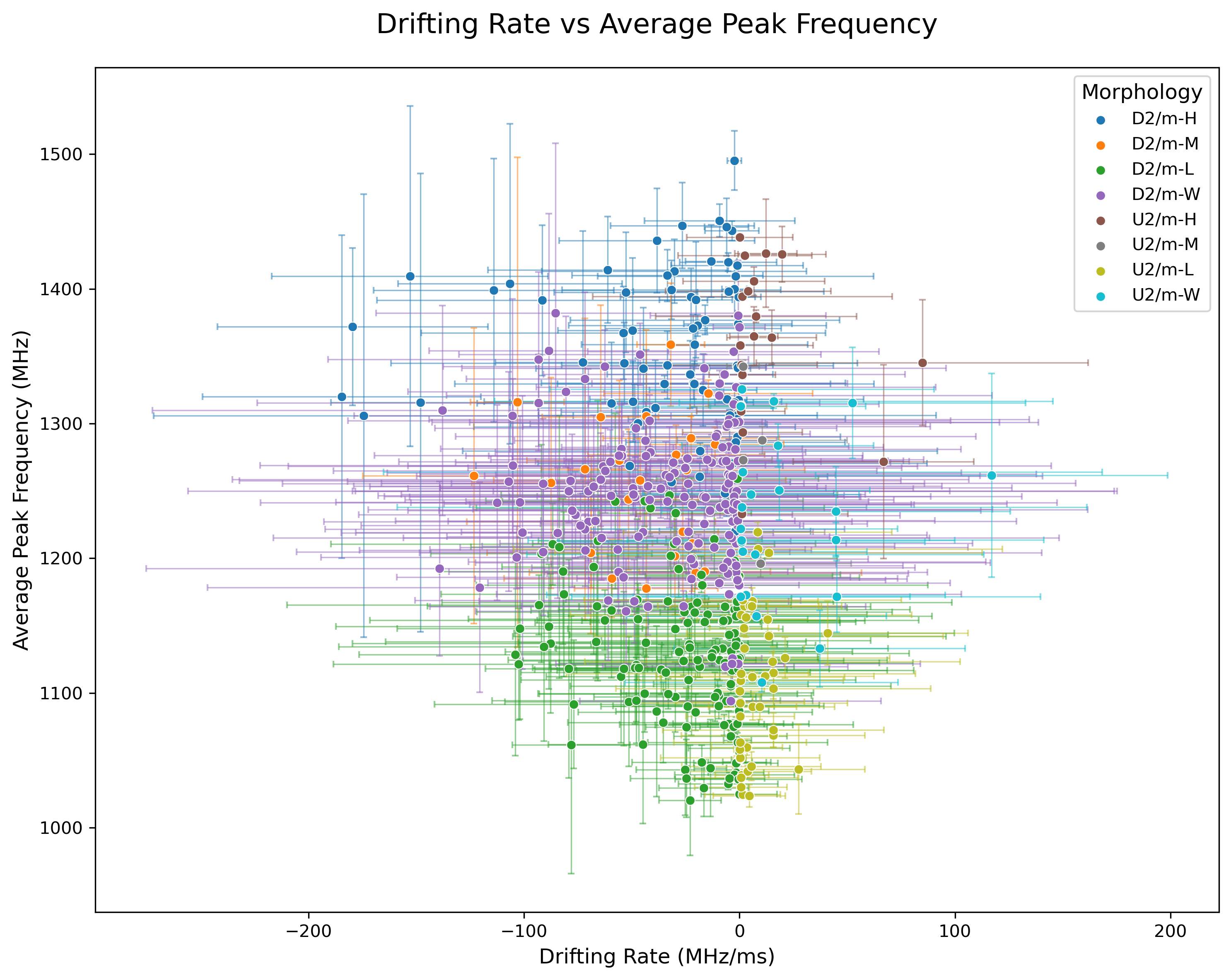}
    \caption{
    Correlation between drifting rate and average peak frequency for composite burst-clusters (Y = 2/m). Colored markers encode morphology parameters in XY-Z format: drifting direction (X: D and U), component count (Y: 2/m), and spectral band (Z: H, M, L and W). Average peak frequencies represent arithmetic means of sub-burst components.}
    \label{FreqPeak_vs_Rd}
\end{figure}

\begin{figure}[ht]
    \centering
    \includegraphics[width=1\textwidth]{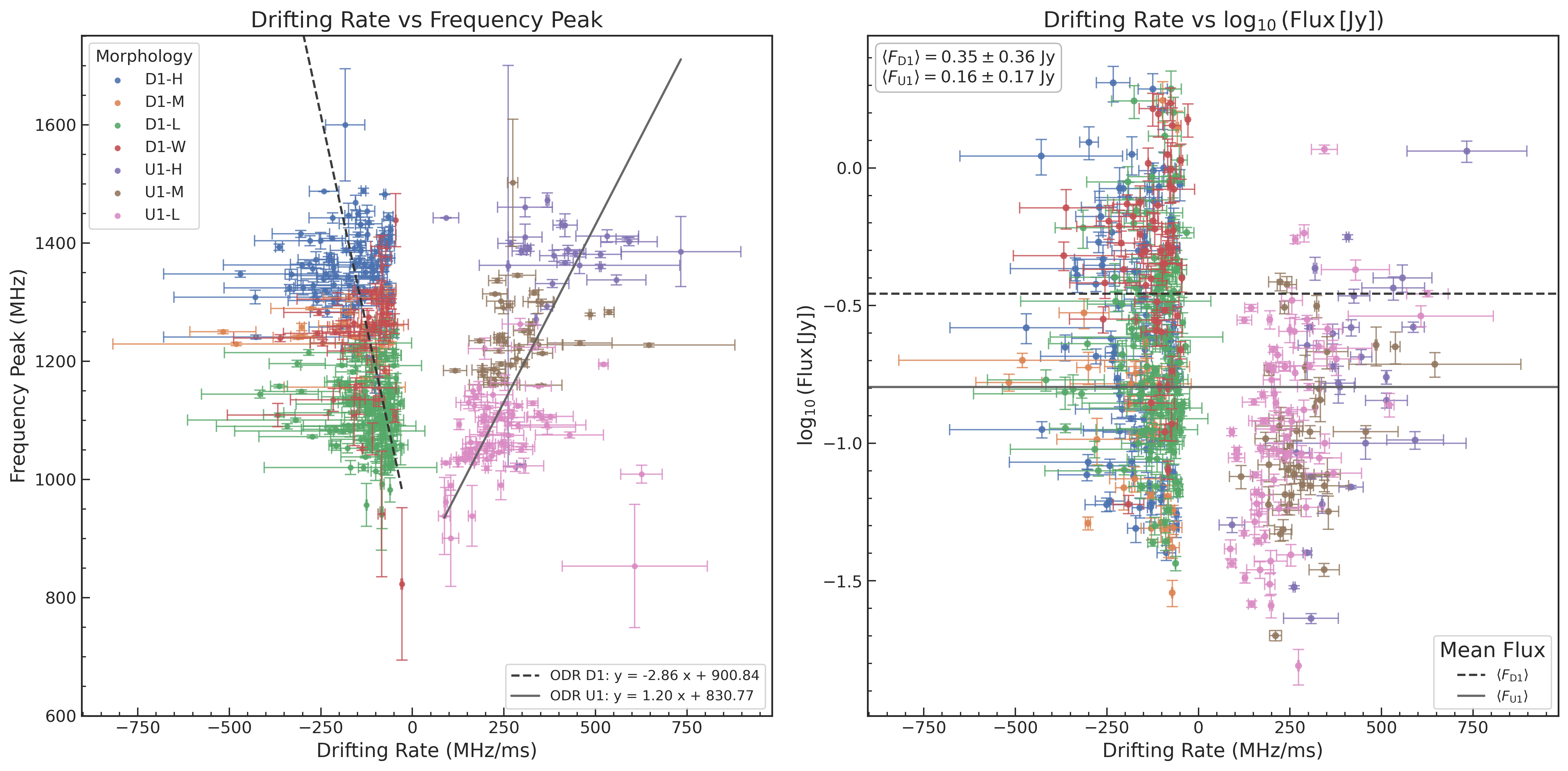}
    \caption{
        Scatter plots of the drifting rate with frequency peak (left panel) and flux (right panel) for single-component burst-clusters exhibiting measurable frequency drifts. Each point represents an individual burst-cluster, color-coded by morphological subtype (D1-H, D1-M, D1-L, D1-W for downward drifting and U1-H, U1-M, U1-L for upward drifting). ODR fits for D1 and U1 are plotted in gray (left panel). In the right panel, horizontal gray lines mark the mean fluxes of the two classes, $\langle F_{\mathrm{D1}}\rangle = 0.35 \pm 0.36~\mathrm{Jy}$ and $\langle F_{\mathrm{U1}}\rangle = 0.16 \pm 0.17~\mathrm{Jy}$, indicating that D1 burst-clusters are on average about 2.2 times brighter. Error bars are plotted only for bursts with frequency peak and drifting rate uncertainties below 500~MHz and 400~MHz\,ms$^{-1}$, respectively.
    }
    \label{Rd_vs_FreqPeak_and_LogFlux}
\end{figure}

In this section, we explore the statistical relations between various physical parameters: (average) peak frequency, drifting rate, bandwidth, effective width, and isotropic energy. The format used is XY-Z, where X denotes the drifting type (D for downward drifting, U for upward drifting), Y represents the number of components (or sub-bursts) in the burst-cluster, with "m" used if the number exceeds two, and Z indicates the observation band: L (low frequency), M (middle frequency), H (high frequency), W (wide frequency). For all the single-component burst-clusters exhibiting drifting behavior, these relations are visualized using scatter matrices, which show pairwise relations between variables along with the distribution of each individual variable along the diagonal as shown in Figure~\ref{scatter_plot_Y1}, where only single-component burst-clusters data (Y = 1) are presented, with different values of X (D and U) and Z (H, M, L and W) represented by points of varying colors. Extending this analysis to double- and multi-component burst-clusters, the relation between drifting rate and average peak frequency is visualized in Figure~\ref{FreqPeak_vs_Rd}, with distinct color coding representing different combinations of the X and Z parameters. For the double- and multi-component burst-clusters datasets, the average peak frequency refers to the average of the individual peak frequencies for each sub-burst. 

The U1 and D1 burst-clusters (single-component with drifting behavior) exhibit a clear correlation between peak frequency and drifting rate, as shown in Figures~\ref{scatter_plot_Y1} and~\ref{Rd_vs_FreqPeak_and_LogFlux}, while the Y=2/m burst-clusters also lack such a correlation, as shown in Figure~\ref{FreqPeak_vs_Rd}.

We make the simplifying assumption that the measured drifting rate is entirely due to the deviation between the adopted average DM and the true per-burst DM, as described by the cold-plasma dispersion relation \citep{2023RvMP...95c5005Z}:
\begin{equation}
    \Delta t = t(\nu_1) - t(\nu_2)
    = \frac{e^2}{2\pi m_e c}\!\left(\frac{1}{\nu_1^2}-\frac{1}{\nu_2^2}\right)\!\mathrm{DM}
    \approx 4.15\,\mathrm{ms}\!
    \left(\frac{1}{\nu_{1,\mathrm{GHz}}^2}-\frac{1}{\nu_{2,\mathrm{GHz}}^2}\right)
    \frac{\mathrm{DM}}{\mathrm{pc}\,\mathrm{cm}^{-3}}.
\end{equation}

\noindent where \( \Delta t \) represents the time delay between two frequencies \( \nu_1 \) and \( \nu_2 \), with \( e \) being the electron charge, \( m_e \) the electron mass, and \( c \) the speed of light. The term \( \frac{1}{\nu_1^2} - \frac{1}{\nu_2^2} \) accounts for the difference in arrival times at different frequencies, while DM quantifies the integrated free electron density along the line of sight. 

Taking the derivative shows that, under the assumption that DM deviations are the only source of the observed slope, the measured drifting rate should increase in magnitude with frequency. In our analysis the daily average DM is used rather than the exact DM of each burst (assuming such measurements could be accurate). In this case, both upward and downward drifting burst-clusters are expected to follow the same frequency-dependent scaling in $|R_d|$. However, our results reveal that the two classes exhibit significantly different slopes: U1 burst-clusters show a strong positive correlation (ODR slope $b = 1.20 \pm 0.07~\mathrm{ms}$; Pearson $r = 0.47$, $p = 3.5\times10^{-9}$; Spearman $\rho = 0.54$, $p = 2.7\times10^{-12}$), while D1 burst-clusters show a weaker dependence (ODR slope $b = -2.86 \pm 0.24~\mathrm{ms}$; Pearson $r = -0.21$, $p = 9.0\times10^{-5}$; Spearman $\rho = -0.23$, $p = 7.6\times10^{-6}$), as shown in the left panel of Figure~\ref{Rd_vs_FreqPeak_and_LogFlux}. The substantial difference in slope magnitude between U1 and D1 indicates that DM deviations alone cannot account for the observed trend, indicating that additional, yet unidentified physical mechanisms may contribute to the frequency dependence of the drifting rate. Furthermore, for single-component burst-clusters, the D1 burst-clusters exhibit significantly larger bandwidth, effective width, and isotropic energy than U1, as shown in Figure~\ref{scatter_plot_Y1}. It is interesting to see that the energy for downward drifting bursts is statistically higher than the energy for the upward drifting ones (lower-left panel). Considering that the isotropic energy of detected bursts can be calculated from the emission bandwidth and effective width, we also plotted the relation between the flux and the drifting rate. The average flux of D1 burst-clusters ($\langle F_{\mathrm{D1}}\rangle = 0.35 \pm 0.36~\mathrm{Jy}$) is about 2.19 times higher than that of U1 burst-clusters ($\langle F_{\mathrm{U1}}\rangle = 0.16 \pm 0.17~\mathrm{Jy}$), as shown in the right panel of Figure~\ref{Rd_vs_FreqPeak_and_LogFlux}.

Another physically meaningful statistical parameter is $\Delta\nu/\nu_0$, which characterizes the degree of spectral confinement of the emission. Here, $\Delta\nu$ denotes the full width at half-maximum (FWHM) of the emission spectrum and $\nu_0$ the central frequency, following the definition in \citet{2024ApJ...974..160K}. In our analysis, the bandwidths are adopted from the catalog of \citet{2025arXiv250714707Z}, where the frequency boundaries were determined using a CDF-based edge-finding method, with Gaussian extrapolation applied to band-truncated bursts and manual selection for a few complex or low-S/N cases. These dataset bandwidths are primarily defined as $2\sigma$ (or CDF-based equivalent) bandwidths, and the Gaussian and CDF results are found to be generally consistent within the measurement uncertainties. For Gaussian profiles, the $2\sigma$ width can be approximately converted to the equivalent FWHM by a factor of $\sqrt{2\ln2}\approx1.18$.

\begin{figure}[ht]
    \centering
    \includegraphics[width=0.8\textwidth]{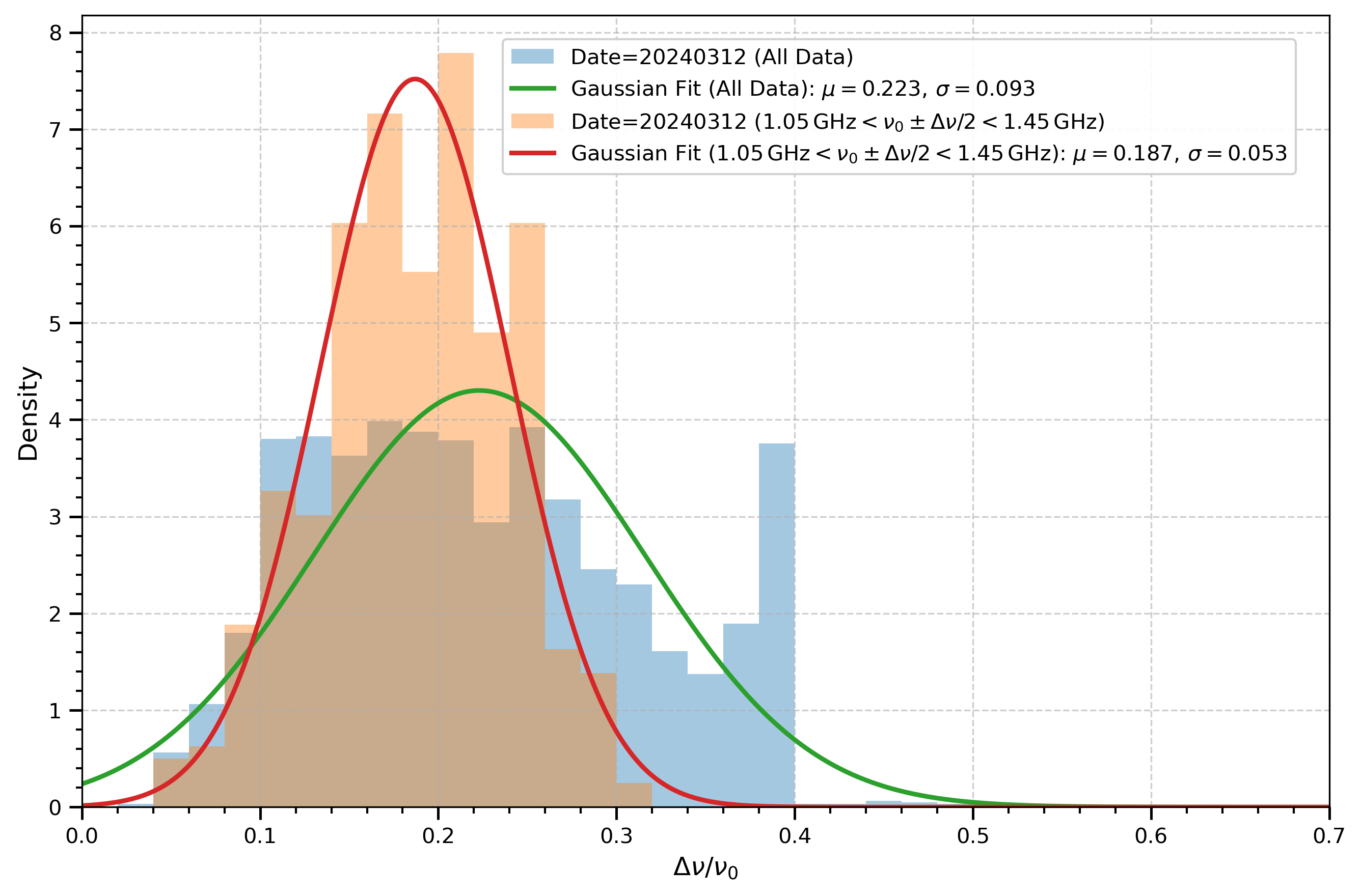}
    \caption{
    Distribution of $\Delta\nu/\nu_0$ for FRB~20240114A bursts detected on March~12, 2024. 
    The blue and orange histograms represent the full dataset and the frequency-limited subset ($1.05~\mathrm{GHz}<\nu_0\pm\Delta\nu/2<1.45~\mathrm{GHz}$), respectively.
    The solid green and red curves show the corresponding Gaussian fits,
    yielding $(\mu,\,\sigma)=(0.223,\,0.093)$ and $(0.187,\,0.053)$. After converting the dataset $2\sigma$ bandwidths to the equivalent FWHM ($\mathrm{FWHM}/(2\sigma)=\sqrt{2\ln2}\approx1.18$), the typical fractional bandwidths become $\Delta\nu/\nu_0\approx0.22$–$0.27$, still well below the theoretical lower limit of about $\Delta\nu/\nu_0\sim0.58$ expected for emission outside the magnetosphere.
    }
    \label{delta_v_over_v0_distribution}
\end{figure}

For the bursts detected on March~12, 2024, the histogram of $\Delta\nu/\nu_0$ is well fitted by a Gaussian, with the full dataset yielding $\mu = 0.223$ and $\sigma = 0.093$, and the frequency-limited subset ($1.05~\mathrm{GHz}<\nu_0\pm\Delta\nu/2<1.45~\mathrm{GHz}$) fitted by a Gaussian with $\mu = 0.187$ and $\sigma = 0.053$, as shown in Figure~\ref{delta_v_over_v0_distribution}. Since many bursts exhibit spectral components extending beyond the 1–1.5,GHz FAST band, focusing on those fully enclosed within the band ensures that the measured bandwidths are not limited by the instrumental frequency coverage. After applying the FWHM conversion, the typical fractional bandwidths become $\Delta\nu/\nu_0 \approx 0.22$-$0.27$, still well below the theoretical lower limit of about $\Delta\nu/\nu_0 \sim 0.58$, which is expected for emission outside the magnetosphere due to high-latitude broadening \citep{2024ApJ...974..160K}. This lower bound arises because, for an FRB source outside the magnetosphere whose angular size exceeds the Doppler beaming angle, high-latitude emission causes spectral broadening that sets a minimum fractional bandwidth. Such high-latitude effects are therefore difficult to reconcile with the markedly confined spectra observed here. This strong spectral confinement, therefore, favors a magnetospheric emission origin; however, additional propagation effects, such as plasma lensing or scintillation, may still influence the observed spectral morphology. Similar narrow-band behavior has also been noted in previous studies of repeating FRBs, including FRB~20220912A, FRB~20230607A, and FRB~20240114A, whose fractional bandwidth distributions typically peak around $\Delta\nu/\nu_0\sim0.1$-$0.2$ \citep{2025ApJ...988...41Z,2023ApJ...955..142Z,2024ApJ...977..177K}.

\subsection{Analysis of consecutive and intermittent for downward and upward drifting}\label{Sec:drifting4.3}

\label{Analysis_of_consecutive_and_intermittent}

\begin{figure}[ht]
    \centering
    \includegraphics[width=\textwidth]{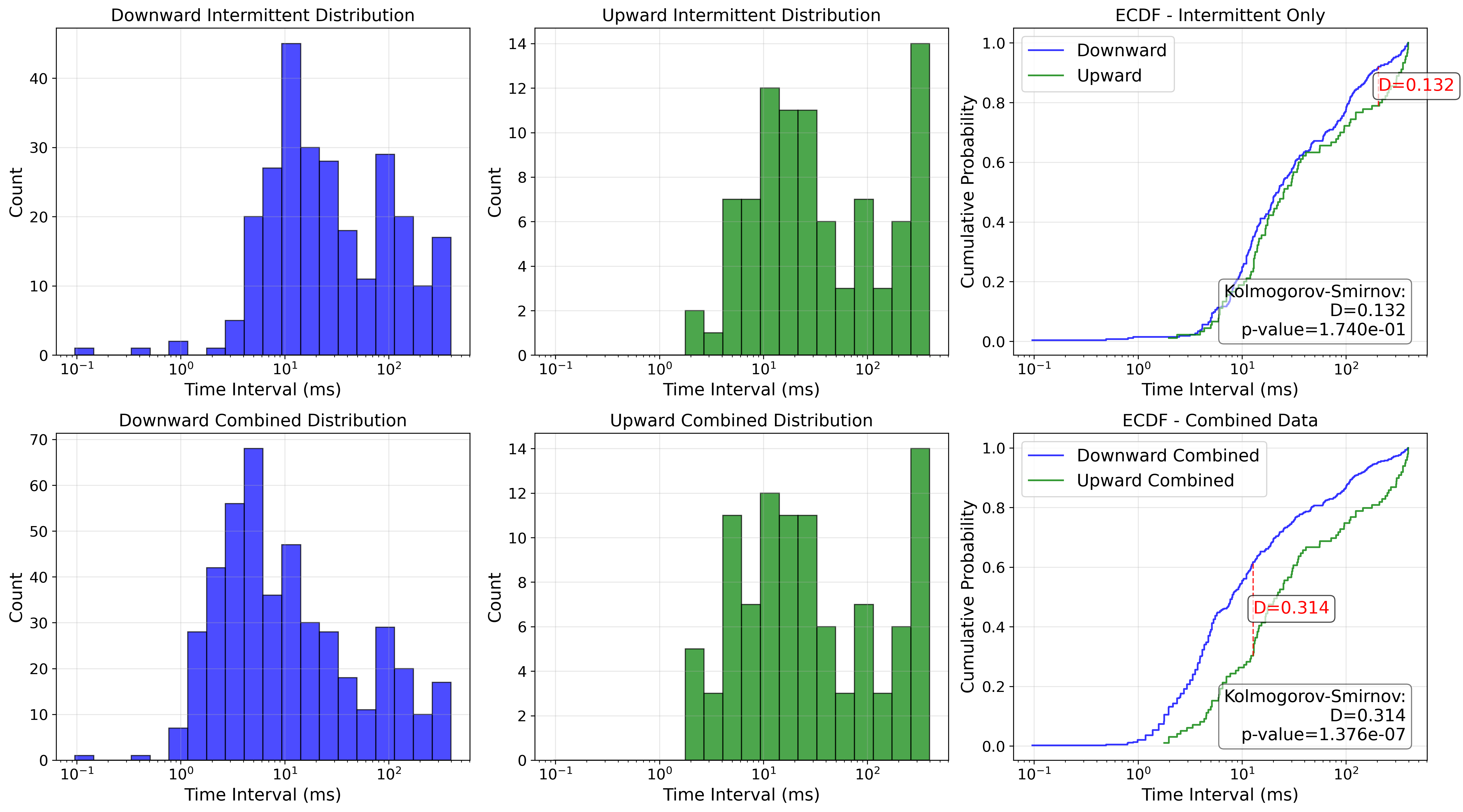}
    \caption{Overall view of the sub-burst interval distributions in left four panels. Empirical CDFs of downward and upward sub-burst intervals with the K-S distance indicated in the right two panels. The top row displays the histogram and cumulative distribution for intermittent sub-burst intervals, whereas the bottom row shows the histogram and cumulative distribution for both intermittent and consecutive sub-burst intervals. These left four panels together illustrate the statistical properties of sub-burst intervals for both downward and upward drifting directions.}
    \label{ks_test}
\end{figure}

In this section, we present the statistical distribution of consecutive and intermittent time intervals for upward and downward drifting burst-clusters, derived from double- and multi-component burst-clusters. First, we focus on the intermittent intervals for upward and downward drifting burst-clusters separately. For the downward drifting burst-clusters, the sample size is 265, with a mean of 64.48 ms and a standard deviation of 89.40 ms, while for upward drifting burst-clusters, the sample size is 90, with a mean of 93.55 ms and a standard deviation of 124.91 ms. A Kolmogorov-Smirnov (K-S) test was performed on the intermittent only data, resulting in a test statistic D = 0.132 and a p-value of $1.740 \times 10^{-1}$, as shown in the first row of Figure~\ref{ks_test}. This high p-value indicates that we cannot reject the null hypothesis, meaning the intermittent intervals for the two types of drifting burst-clusters are statistically consistent with being drawn from the same distribution.

Due to the scarcity of consecutive time intervals for upward drifting burst-clusters, with only 9 such intervals observed (shown in Figure~\ref{consecutive_upward_drifting_burst-clusters}), we were unable to perform a separate statistical analysis for consecutive upward and downward drifting burst-clusters. As a result, we combined consecutive and intermittent intervals for a comprehensive analysis. For the downward drifting burst-clusters, the combined sample size is 449, with a mean of 39.45 ms and a standard deviation of 74.94 ms, and for upward drifting burst-clusters, the combined sample size is 99, with a mean of 85.36 ms and a standard deviation of 121.85 ms. The K-S test for the combined data, shown in the second row of Figure~\ref{ks_test}, yielded a test statistic D = 0.314 and a p-value of $1.376 \times 10^{-7}$. This very small p-value provides strong evidence to reject the null hypothesis, suggesting that the combined intervals for upward and downward drifting burst-clusters are likely drawn from different distributions.

The analyses above indicate a statistically significant difference ($p < 0.05$) in the combined (consecutive and intermittent) time intervals between upward and downward drifting burst-clusters. Since the intermittent upward and downward drifting burst-clusters come from the same distribution and significant differences were observed in the combined data of both intermittent and consecutive intervals, it suggests that the mechanisms driving consecutive upward and downward drifting burst-clusters are different. Specifically, upward drifting burst-clusters exhibit a much longer mean interval, which suggests distinct physical processes or factors governing these burst-clusters compared to downward drifting burst-clusters. Further investigation is necessary to better understand the nature of these processes and their impact on drifting behavior.

\section{Conclusion and Discussion}
\label{sec:Discussion}

This study presents the morphological classification and drifting rate measurements of the repeating fast radio burst source FRB~20240114A, utilizing observational data obtained from FAST on March~12, 2024. As detailed in Section~\ref{DM_Optimized_for_Structure}, we employed the average DM optimized for the burst's finest temporal structures during the March~12, 2024, observations to de-disperse the data. This enabled comprehensive morphological classification (Section~\ref{sec:morphology}) and drifting rate measurements (Section~\ref{Drifting_Rate}), followed by a systematic exploration of statistical relations between drifting rates and various physical parameters, as well as the narrow-band spectral properties that are more naturally explained by magnetospheric emission (Section~\ref{Statistical_relations}). Key concepts, including sub-bursts, bursts, burst-clusters, and consecutive/intermittent time intervals, were rigorously defined and visually illustrated (Figure~\ref{burst_definition}), with consecutive time intervals of downward/upward drifting burst-clusters (i.e., time intervals of downward/upward drifting bursts) also being specified. Notably, as discussed in Section~\ref{Analysis_of_consecutive_and_intermittent}, a K-S test was implemented to statistically compare whether the intermittent and consecutive time intervals of upward/downward drifting burst-clusters originate from identical distributions.

In our study of FRB~20240114A, we detected a small fraction of upward drifting burst-clusters (23.82\%), decreasing to 10.89\% when single-component upward drifting burst-clusters are excluded. Moreover, when only upward drifting burst-clusters consisting solely of consecutive time intervals are considered, this number further drops to just 9, as illustrated in Figure~\ref{consecutive_upward_drifting_burst-clusters}. Notably, these 9 consecutive upward drifting burst-clusters represent a proportion nearly the same as that reported for FRB~20201124A in \cite{2022RAA....22l4001Z}. Considering that the mechanism of upward drifting is still unclear (though a few models have been proposed, e.g. \citet{2020ApJ...899..109W}), such rare upward series in multiple FRBs might hint at analogous conditions in their sources.

In the analysis of the single-component burst-clusters, we observe a positive correlation between the drifting rate and the peak frequency (left panel of Figure~\ref{Rd_vs_FreqPeak_and_LogFlux}). This trend may be attributed to deviations between the adopted average DM and the true per-burst DM. The theoretical derivation shows that, under the assumption that DM deviations are the only source of the observed slope, the measured drifting rate should increase with frequency. However, both U1 and D1 burst-clusters exhibit different linear correlations. This indicates that DM deviations alone cannot fully account for the observed drifting behavior, implying that additional, possibly unknown processes may contribute to the observed frequency dependence.

The analysis presented in this work reveals an inverse correlation between the drifting rate and the sub-burst effective widths (see Figure ~\ref{scatter_plot_Y1}). This inverse relation corroborates the findings of \cite{2021MNRAS.507..246C}, which may indicate a triggered scenario involving a compact object (e.g. a pulsar or magnetar) and a separate FRB-emitting region along the line of sight.
For single-component burst-clusters, our results show that downward drifting events exhibit larger bandwidths, effective widths, and flux compared to upward drifting events (see Figures~\ref{scatter_plot_Y1} and \ref{Rd_vs_FreqPeak_and_LogFlux}). However, the physical origin underlying these observational differences remains unclear, and further investigations are necessary to fully understand the mechanisms responsible for these drifting behaviors.
In addition, the fractional bandwidths remain narrowly confined, with FWHM-corrected values of $\Delta\nu/\nu_0 \approx 0.22$-$0.27$ (see Figures~\ref{delta_v_over_v0_distribution}), well below the $\Delta\nu/\nu_0 \sim 0.58$ limit expected for emission outside the magnetosphere \citep{2024ApJ...974..160K}. This trend is more naturally accommodated by magnetospheric scenarios, though propagation effects may still contribute. Similar narrow-band behavior has also been reported for other repeaters \citep{2023ApJ...955..142Z,2025ApJ...988...41Z,2024ApJ...977..177K}.

As shown in Section \ref{Sec:drifting4.3}, the consecutive time intervals of upward drifting burst-clusters (mean value 93.35 ms) are significantly longer than those of downward drifting burst-clusters (mean value 64.48 ms). In contrast, the intermittent time intervals of upward/downward drifting burst-clusters are drawn from the same distribution (as indicated by the K-S test shown in Figure~\ref{ks_test}), suggesting that while the intermittent time intervals of upward/downward drifting burst-clusters may result from random triggering, the consecutive drifting behaviors reflect distinct physical mechanisms, although it is not clear what those mechanisms should be. Consequently, we concentrate on the drifting patterns of individual bursts. Fully understanding these mechanisms will require high-time-resolution follow-up observations and the development of more comprehensive theoretical models to explain the frequency drifting phenomena observed in FRB~20240114A.

Most bursts exhibit downward drifting patterns, consistent with many repeating FRBs previously studied \citep[e.g.][]{2021ApJ...923....1P, 2022RAA....22l4001Z}. These downward drifting bursts are often interpreted within models invoking radius-to-frequency mapping in magnetospheric emission regions \citep{2019ARA&A..57..417C, 2019ApJ...876L..15W} or plasma lensing effects \citep{2017ApJ...842...35C, 2021arXiv210713549T}, where higher frequencies arrive prior to lower frequencies. This interpretation is supported by the even shorter and more numerous consecutive time intervals of downward drifting burst-clusters, as well as the well-defined negative drifting rates observed. Nevertheless, the morphological asymmetries between upward and downward drifting bursts could arise from the geometric effect of ordered bunch emission along different magnetic field lines. As proposed by \citet{2022ApJ...927..105W}, the drifting direction (upward or downward) is determined by the sequence in which bunches cross the line of sight, with the curvature of field lines and the observer's viewing angle playing crucial roles in shaping the distinct spectro-temporal patterns. The observed downward drifting patterns also find theoretical grounding in magnetospheric curvature radiation models, where relativistic particle bunches propagating along neutron star open field lines naturally exhibit downward drifting due to increasing curvature radii at higher altitudes \citep{2019ApJ...876L..15W}. Notably, \citet{2020ApJ...899..109W} extended this framework to incorporate upward drifting scenarios, demonstrating that earlier-triggered sparks at larger emission heights can reverse the drifting direction under specific geometric configurations and the temporal sequencing of magnetospheric perturbations. Our detection of upward drifting burst-clusters (23.82\% overall, with only nine exceptional consecutive cases) aligns with this extended framework, suggesting rare instances where spark triggering sequences or non-dipolar field geometries dominate over the standard downward drifting mechanism.

\section*{Data Availability}
\label{sec:Data Availability}

The dataset underlying this work is publicly available at the Science Data Bank \url{https://doi.org/10.57760/sciencedb.Fastro.00033}. It includes dynamic spectra of FRB~20240114A in archival format (\texttt{.ar} files), widely used in radio astronomy, along with their corresponding visualization plots (\texttt{.png}). The dataset also provides summary tables in \texttt{.xlsx} format, covering morphological classifications, drifting rate measurements, and other relevant physical quantities. In addition, curated data subsets are included for the K–S test. All files are provided in open formats to support transparent and reproducible research.

\section*{Acknowledgments}
This work made use of the data from FAST FRB Key Science Project. This work is supported by National Natural Science Foundation of China (NSFC) Programs Nos. 12588202, 12041306, 11988101, 11725313, 11690024, 12041303, 12421003, U1731238, 12233002, 12303042, 12403100, W2442001, 12203045, 12447115 and U1931203; CAS International Partnership Program No. 114-A11KYSB20160008; CAS Strategic Priority Research Program No. XDB23000000; the National Key R\&D Program of China (No. 2017YFA0402600); the National SKA Program of China Nos. 2022SKA0130100, 2020SKA0120200, and 2020SKA0120100.
W.W.Z. is supported by CAS Project for Young Scientists in Basic Research, YSBR-063.
D.L. is a New Cornerstone investigator.
P.W. acknowledges support from the CAS Youth Interdisciplinary Team, the Youth Innovation Promotion Association CAS (id. 2021055), and the Cultivation Project for FAST Scientific Payoff and Research Achievement of CAMS-CAS.
Y.F. is supported by the Leading Innovation and Entrepreneurship Team of Zhejiang Province of China grant No. 2023R01008, and by Key R\&D Program of Zhejiang grant No. 2024SSYS0012.
Y.F.H. acknowledges the support from the Xinjiang Tianchi Program.
Q.W. is supported by the China Postdoctoral Science Foundation (CPSF) under Grant Nos. GZB20240308, 2025T180875.
C.W.T. is supported by CAS project No. JZHKYPT-2021-06.
J.R.N. is supported by the Postdoctoral Fellowship Program of CPSF under Grant No. GZB20250737.
This work made use of data from FAST, a Chinese national mega-science facility built and operated by the National Astronomical Observatories, Chinese Academy of Sciences. 
This research has made use of the CHIME/FRB VOEvent Service, BlinkVerse and TransientVerse.
The cartoon plot was created with the help of Ms. Yan-Dong Wang and Mr. Tong-Cai Zhang.
The computation was partially completed on the HPC Platform of Huazhong University of Science and Technology. 

\bibliographystyle{aasjournal}
\bibliography{sample631}



\end{document}